\documentclass[letterpaper]{JHEP3}
\usepackage{epsfig}

\def\ba{\begin{eqnarray}}
\def\ea{\end{eqnarray}}
\def\be{\begin{equation}}
\def\ee{\end{equation}}
\def\nn{\nonumber}
\def\exd{{\rm d}}

\makeatletter
\def\x@arrow{\DOTSB\Relbar}
\def\xlongequalsignfill@{\arrowfill@\x@arrow\Relbar\x@arrow}
\newcommand{\xlongequal}[2]{%
    \ext@arrow 0099\xlongequalsignfill@{#1}{#2}}
\makeatother

\newcommand{\roughly}[1]{\mathrel{\raise.3ex\hbox{$#1$\kern-0.85em
\lower1ex\hbox{$\sim$}}}}

\def\endignore{}
\def\ignore #1\endignore{} % use to "comment out" text
\def\be{\begin{equation}}
\def\beq\begin{equation}
\def\ee{\end{equation}}
\def\bea{\begin{eqnarray}}
\def\eea{\end{eqnarray}}

\def\nn{\nonumber}

\def\pref#1{(\ref{#1})}

\def\dddot#1{#1 \hspace{-3.3mm} \ddot{\phantom{#1}} \hspace{-1.9mm} \dot{\phantom{#1}}}
\def\ddddot#1{#1 \hspace{-3.3mm} \ddot{\phantom{#1}} \hspace{-1.7mm} \ddot{\phantom{#1}}}
\def\beq{\begin{equation}}
\def\eeq{\end{equation}}
\def\beqa{\begin{eqnarray}}
\def\eeqa{\end{eqnarray}}

\def\cA{{\cal A}}

\def\cD{{\cal D}}

\def\cG{{\cal G}}
\def\cH{{\cal H}}
\def\cI{{\cal I}}
\def\cJ{{\cal J}}

\def\cL{{\cal L}}
\def\cM{{\cal M}}
\def\cN{{\cal N}}
\def\cO{{\cal O}}

\def\cR{{\cal R}}

\def\cX{{\cal X}}
\def\cY{{\cal Y}}
\def\cZ{{\cal Z}}

\def\ssN{{\scriptscriptstyle N}}

\def\ssT{{\scriptscriptstyle T}}
\def\ssU{{\scriptscriptstyle U}}
\def\ssV{{\scriptscriptstyle V}}

\def\Va{{V_{,\,a}}}

\def\p{\varphi}
\def\e{\ell}

\newcommand{\bmat}{\left(\begin{array}}
\newcommand{\emat}{\end{array}\right)}

\def\-{\hphantom{-}}

\def\s2{\frac{1}{2}}

\def\IF{\relax{\rm I\kern-.18em F}}
\def\II{\relax{\rm I\kern-.18em I}}
\def\IP{\relax{\rm I\kern-.18em P}}
\def\IC{\relax{\rm I\kern-.48em C}}
\def\IR{\relax{\rm I\kern-.18em R}}
\def\IK{\relax{\rm I\kern-.20em K}}
\def\IM{\relax{\rm I\kern-.25em M}}

\def\Dsl{\,\raise.15ex\hbox{/}\mkern-13.5mu D}
\def \one{\relax{\rm 1\kern-.26em I}}

\def\exd{{\rm d}}

\def\nn{\nonumber}

\def\({\left(}
\def\){\right)}

\title{Inflating in a Trough: Single-Field Effective Theory\\ from Multiple-Field Curved Valleys}
\preprint{CPHT-RR 021.0512\\ CERN-PH-TH/2012-246}

\author{C.P.~Burgess,${}^{1,2}$ M.W.~Horbatsch${}^{~1,3}$ and Subodh.P.~Patil${}^{~4,5}$\\
\\
$^1$ Dept. of Physics \& Astronomy, McMaster University \\
 \qquad 1280 Main St. W, Hamilton, Ontario, Canada, L8S 4L8.\\
\\
$^2$ Perimeter Institute for Theoretical Physics \\
 \qquad 31 Caroline St. N, Waterloo, Ontario, Canada  N2L 2Y5.\\
\\
$^3$ The University of Mississippi, P. O. Box 1848, University, MS 38677, U.S.A.\\
\\
$^4$ Theory Division, Case C01600, CERN, CH-1211, Gen\'eve 23, Switzerland.\\
\\
$^5$ Centre de Physique Th\'eorique,  Ecole Polytechnique and CNRS\\ \qquad Paiaiseau cedex 91128, France.
}

\date{}
\abstract {We examine the motion of light fields near the bottom of a potential valley in a multi-dimensional field space. In the case of two fields we identify three general scales, {\em all} of which must be large in order to justify an effective low-energy approximation involving only the light field, $\ell$. (Typically only one of these -- the mass of the heavy field transverse to the trough -- is used in the literature when justifying the truncation of heavy fields.) We explicitly compute the resulting effective field theory, which has the form of a $P(\ell,X)$ model, with $X = - \frac{1}{2}(\partial \ell)^2$, as a function of these scales. This gives the leading ways each scale contributes to {\em any} low-energy dynamics, including (but not restricted to) those relevant for cosmology. We check our results with the special case of a homogeneous roll near the valley floor, placing into a broader context recent cosmological calculations that show how the truncation approximation can fail. By casting our results covariantly in field space, we provide a geometrical criterion for model-builders to decide whether or not the single-field and/or the truncation approximation is justified, identify its leading deviations, and to efficiently extract cosmological predictions.
}

\begin{document}
\section{Introduction}
\label{sec:introduction}

Scalar fields have long been posited by particle physicists and cosmologists, although experimental evidence for their existence has come only very recently \cite{Higgs}. Their discovery has likely taken so long because in the absence of any symmetries that prevent them, quantum corrections often make it difficult to make scalars very light compared with other particles, an observation that is called the `hierarchy problem' when applied to scalars associated with electroweak symmetry breaking.

Cosmologists also frequently invoke scalar fields because certain features of their classical dynamics are known to be useful for describing the very early universe. For instance, although Hot Big Bang cosmology provides an excellent account of current observations \cite{WMAP}, it also provides evidence for there being two separate epochs during which the expansion of the universe accelerated rather than decelerated with time. We appear to have entered one of these epochs comparatively recently (dominated by Dark Energy), while another (possibly Inflationary \cite{Inflation}) epoch of primordial acceleration seems to have taken place at a much earlier time.
Scalar fields are usually proposed to provide the dynamics that could drive such accelerated expansion, though the propensity of scalars to be heavy has made it difficult to embed these models convincingly into a fundamental theory. At present, string theory provides the most precise framework for doing so \cite{StrInfRevs}, although not yet with decisive success \cite{BuMc:2011fa}.

Yet one lesson does emerge from attempts to marry cosmology with fundamental physics-- fundamental theories contain many scalars in their low-energy spectrum, and although it is hard to make them light enough to be interesting for cosmologists, once a mechanism succeeds in doing so for one it usually also does so for others. Furthermore, heavy fields can sometimes play surprisingly large roles in low-energy dynamics \cite{TW, NoTrunc, OtherTroughs0, OtherTroughs1}, requiring a refined understanding of how decoupling is operative in multiple-field and time-dependent contexts \cite{EFTrevs, dec}.

In this paper we embrace the point of view that multiple scalars are likely to be relevant to cosmology (and elsewhere), and explore as systematically as possible the dynamics of light scalar fields in the presence of other, heavier scalars. To this end we start with multi-scalar interactions whose scalar potential has the shape of a trough or ditch: shallow in the general direction of the light fields, but steeply rising in the transverse, heavy directions. By explicitly integrating out the heavy scalars, we identify which parameters control its decoupling and which features of the heavier scalars influence low-energy dynamics in potentially observable ways.

We find, as already noted in \cite{NoTrunc}, that when the trough is not straight\footnote{More precisely, when the trough bottom is not a geodesic of the target-space metric (see below).} the low-energy theory generally is {\em not} well-described by the `truncation approximation', within which the heavy fields are simply set to vanish. Perhaps more surprisingly, this can remain so even as the mass, $m$, of the heavy field goes to infinity. This is possible because potentials that support curved troughs necessarily involve multiple scales, including the radius of curvature of the trough's bottom (relative to a target-space geodesic) and the curvature scale of the target space's Riemann tensor, as well as how quickly these quantities vary along the trough. Generically {\em all} of these scales must be large to ensure that heavy fields decouple, and so justify a low-energy effective theory.

Concretely, we explicitly construct the leading effective couplings within the effective theory for the case of one heavy ($h$) and one light ($\ell$) scalar, defined by the eigenbasis of the mass matrix at a particular point at the trough's bottom.\footnote{As we see below, this basis need not coincide with the tangent and the normal to the trough at this point.} Even in this simple case there are at least three important scales in the low-energy potential: the heavy mass, $m$; and both the target-space radius of curvature, $\rho$, and the curvature, $\kappa$, of the trough's bottom (relative to a target-space geodesic). Because these are geometrical, they are covariant under field redefinitions and so can be computed equally well in any coordinate system that is convenient.

Our main result is the effective description that captures all of the low-energy effects of the heavy field. This is given by a single-scalar theory with the following action, out to 4-derivative level:
\be \label{eq:Seff0}
 S_{\rm eff}(\ell) = - \int \exd^4 x \sqrt{-g} \; \left[ V_{\rm eff}(\ell) + \frac12 (\partial \ell)^2 + \cH_{\rm eff}(\ell) \,  (\partial \ell)^4 + \cdots \right] \,,
\ee
where (expanding out to quartic order in $\ell$) the effective coupling functions are
\be
V_{\rm eff} = V_0 + V_1 \ell + \frac12 \, \mu_{\rm eff}^2 \ell^2 + \frac16 \, g_{\rm eff} \ell^3 + \frac{1}{24} \, \lambda_{\rm eff} \ell^4 + \cdots,\ee
and $\cH_{\rm eff} = \cH_0 + \cdots$ and so on. More generally, were we to work to higher order in fields and derivatives, we would arrive at a low-energy effective theory that would be a higher-order polynomial function of $X := -\frac{1}{2}(\partial\ell)^2$ (with $\ell$-dependent coefficients, in general): a so-called $P(\ell, X)$ model --- or $k-$inflationary theory \cite{k-inf} in a cosmological context\footnote{Such an effective description was previously advocated in \cite{TW}, where the effective coupling $\mathcal H_{eff}$ was generated by non-canonical kinetic couplings in the parent theory (see also \cite{CLT}, which studied the regimes of validity of this effective description). In what follows we generalize and give context to these findings in a manner that is invariant under field redefinitions.}. The regime of validity of the expansions made in obtaining the effective theory \pref{eq:Seff0} are discussed in \S3. 

What is important is that the effective couplings of the low energy theory are explicitly calculable as functions\footnote{Strictly speaking, at this order in $1/m$ the parameter $\rho$ turns out to appear among interactions involving more than 4 powers of $\ell$, although it can arise in quartic (or lower) powers of $\ell$ at higher order in $1/m$.} of $m$, $\kappa$ and $\rho$, evaluated as an expansion about a particular point, $\varphi$, on the trough's bottom. The leading contributions are given by
\ba \label{eq:gefflambdaeff0}
 &&\qquad\qquad \mu^2_{\rm eff} \simeq  U'' - \frac{ U^{'2}}{\kappa^2 m^2}
 \,, \qquad
 g_{\rm eff} \simeq U''' - \frac{2 U^{'2} m'}{ \kappa^2 m^3} \,, \nn\\
 &&\lambda_{\rm eff} \simeq U'''' - \frac{3 U' \kappa' }{\kappa^3} + \frac{4 U' U'' }{\kappa^2 m^2} \left( \frac{4 m'}{m} + \frac{3 \kappa'}{\kappa} \right)  - \frac{16 U^{'2} \kappa' m'}{\kappa^3 m^3} - \frac{6 U' U'''}{\kappa^2 m^2} \\
 && \qquad\qquad\qquad\qquad- \frac{2 U'^2}{\kappa^2 m^2}\frac{m'^2}{m^2} + \frac{ U^{'2}}{\kappa^4 m^2}  \left(6+3 \lambda_{nnn}-11 \kappa^{'2} + 4 \kappa \kappa'' \right) \,, \nn
\ea
and
\be \label{eq:Heff0}
 \cH_{0} \simeq \frac{1}{2\kappa^2m^2}  \,,
\ee
where $U(\varphi)$ is the value of the scalar potential at the trough's bottom, and primes denote differentiation with respect to arc length (as measured by the target-space metric) along the trough. The quantity $\lambda_{nnn}$ measures how the walls of the trough deviate from a perfect parabola.

Provided $\kappa$, $\rho$ and $m$ are sufficiently large, these effective interactions describe {\em any} low-energy process, including (but not restricted to) predictions for --- and fluctuations about --- cosmological evolution. Because the low-energy theory is a single-scalar model, these predictions are very easily obtained by specializing well-known formulae for single-field inflationary models to the above couplings, thereby extending these single-field predictions to a broader class of multi-field models.

In particular, the implications for fluctuations about cosmological solutions --- such as for non-gaussianity --- can be obtained in either of two equivalent ways. When the above theory is directly viewed as the effective theory of inflation -- in the spirit of Weinberg \cite{SFeft1} -- predictions for fluctuations can be simply extracted using existing single-field calculations \cite{Malda, PXphiInf0, PXphiInf} for general $P(\ell, X)$ models. Alternatively, one can use the effective theory for single-field cosmological fluctuations\footnote{The authors of ref.~\cite{NoTrunc} compute the effective theory for the fluctuations directly, without passing through the intermediate step of eqs.~\pref{eq:Seff0} through \pref{eq:Heff0}.} \cite{SFeft1, SFeft2}, for which we provide the leading contribution to the effective coefficients, $M_n(t)$, as functions of $g_{\rm eff}$, $\lambda_{\rm eff}$ and $\cH_{\rm eff}$.

The remainder of the paper is organized as follows. The next section, \S2\ (with details in Appendix \ref{appssec:troughgeometry}) shows how to characterize shallow troughs geometrically in order to identify the relevant scales in a way that is covariant under field redefinitions (see Appendix \ref{app:covariantcharact}). \S3\ (with details in Appendix \ref{app:integratingout}) then (classically) integrates out the heavy field in the trough to derive the low-energy effective theory, eqs.~\pref{eq:Seff0} through \pref{eq:Heff0} in terms of trough properties and examines its domain of validity. Next, \S4\ tests this effective theory by applying it to several non-gravitational situations where dynamics can be compared between the full multi-scalar system and its effective description. \S5\ then describes the applications to cosmology, illustrating the simplicity of the effective theory's use by giving explicit formulae for inflationary primordial fluctuations and non-gaussianity. Finally, \S6\ briefly summarizes our conclusions.

\section{Covariant characterization of multi-field troughs}
\label{sec:covariantcharact}

This section defines the multi-scalar action of interest and quantifies what it means for the scalar potential to have a trough along which the potential is constant or slowly varying. The goal is to characterize covariantly the geometrical properties of the slowly varying directions of the potential in terms of derivatives of the potential $V$.

\subsection{General form for multi-scalar actions}

Consider the following general action describing $N$ mutually interacting scalar fields, $\phi^a$ within a curved spacetime:\footnote{Conventions: our metric is 'mostly plus' and we adopt Weinberg's curvature conventions \cite{Wbg}, that differ from MTW conventions \cite{MTW} only by an overall sign in the definition of the Riemann tensor.}
\be
 S = - \int \exd^4 x \sqrt{-g} \; \left\{ V(\phi)
 + g^{\mu\nu} \left[ \frac12 \, \cG_{ab}(\phi) \,
 \partial_\mu \phi^a \partial_\nu \phi^b
 + \frac{M_p^2}{2} \, R_{\mu\nu} \right] + \cdots
 \right\} \,.
\ee
This describes the most general Lorentz-invariant interactions possible amongst these scalars at the two-derivative level,\footnote{We do not write the non-minimal coupling $F(\phi) \, R$, where $R = g^{\mu\nu} R_{\mu\nu}$ is the spacetime Ricci scalar, because this can be removed by transforming to Einstein frame through an appropriate Weyl rescaling: $g_{\mu\nu} \to A(\phi) \, g_{\mu\nu}$.} and is completely characterized by the interaction potential, $V(\phi)$, and the target-space metric, $\cG_{ab}(\phi)$ (which is a positive-definite symmetric matrix). Here $M_p$ is the reduced Planck mass defined in terms of Newton's constant by $M_p^2 = (8 \pi G_\ssN)^{-1}$ which only plays a role for applications where couplings to gravity are important (such as to cosmology).

Our interest is in making perturbative (typically semi-classical) predictions in the immediate vicinity of a field-point, $\varphi^a$, and so usually at this juncture we would expand the action in powers of $\phi^a - \varphi^a$. However it is useful to emphasize the invariance of physical predictions under field redefinitions, and this is not well-served by such a linear split between $\phi^a$ and $\varphi^a$. A nonlinear alternative exists --- $\delta \phi^a = \delta \phi^a(\phi, \varphi)$ with $\delta \phi^a \to 0$ as $\phi^a \to \varphi^a$ --- that preserves covariance under field redefinitions, where $\delta \phi^a$ geometrically represent Gaussian normal coordinates in field space. A brief review of this formalism is given in Appendix \ref{app:covariantcharact}.

Recall that under a generic local field redefinition, $\phi^a \to \phi^a + \zeta^a(\phi)$ (for $\zeta^a(\phi)$ an arbitrary, infinitesimal, locally invertible collection of functions), $V(\phi)$ transforms as a scalar: $\delta V = V_{,\,a} \, \zeta^a$, while $\cG_{ab}(\phi)$ transforms as a metric tensor: $\delta \cG_{ab} = \cG_{ab,\,c} \, \zeta^c + \cG_{ac} \, \zeta^c_{,\,b} + \cG_{cb} \, \zeta^c_{,\,a}$, where commas denote differentiation ($V_{,\,a} := \partial V/\partial \varphi^a$ and so on). When expanded in terms of the covariant quantity $\delta \phi^a$, the Lagrangian can be written in terms of covariant derivatives and curvatures of the metric $\cG_{ab}$. For instance, the expansion of the scalar potential gives
\be \label{Vexpnv0}
 V(\phi)
 = V(\varphi) + \Va (\varphi) \, \delta \phi^a
 + \frac{1}{2} \, V_{;\,ab} (\varphi) \, \delta \phi^a \delta \phi^b + \frac{1}{3!} V_{;\,abc} (\varphi) \, \delta \phi^a \delta \phi^b \delta \phi^c + \cdots \,,
\ee
where semicolons denote covariant derivatives constructed using the Christoffel symbols, $\gamma^a_{bc}(\varphi)$, built from first derivatives of the target space metric, $\cG_{ab}(\varphi)$. An expansion of the metric to quartic order similarly gives the standard normal-coordinate expression \cite{GNCexp}
\be \label{eq:kinexpansion}
 \cG_{ab} \, \partial_\mu \phi^a
 \, \partial^\mu \phi^b = \left[ \cG_{ab}(\varphi) + \frac{1}{3} \cR_{acbd}(\varphi) \, \delta \phi^c \delta \phi^d \right] (\partial_\mu \delta \phi^a) (\partial^\mu \delta \phi^b)
 + \cdots \,,
\ee
where ${\cR^a}_{bcd}$ is the Riemann tensor built from $\cG_{ab}$.

In the special case where there are only two fields --- a case we explore in more detail below --- the target-space curvature tensor is particularly simple:
\be \label{eq:2Dcurv}
 \cR_{abcd}
% = \frac12 \, \cR  \left( \cG_{ac} \, \cG_{bd} - \cG_{ad} \, \cG_{bc} \right)
 = \frac{1}{2 \, \rho^2} \,  \left( \cG_{ad} \, \cG_{bc} - \cG_{ac} \, \cG_{bd} \right)  \,,
\ee
characterized by a single function, the target-space radius of curvature,\footnote{In our conventions if the target space were a two-sphere, then $\rho^2 > 0$.} $\rho(\varphi)$.

We next suppose the scalar $V$ has a trough-like shape with a local minimum in several strongly varying directions, but varying slowly along others. For simplicity we describe in detail here a system involving only $N=2$ fields, but the generalization to more than two is straightforward. We first characterize more precisely what it means for the potential $V$ to have a trough. Because this is most easy to do when the trough is perfectly level --- {\em i.e.} when $V$ is perfectly constant along its bottom --- we first do so in this simpler case.

\subsection{Perfectly level troughs}
\label{ssec:leveltroughs}

Given any smooth potential it is always possible to define equipotential curves, {\em i.e.} trajectories in field space, $\phi^a = \chi^a(\sigma)$, along which $V$ is constant: $V[\chi(\sigma)] = V_0$ for all $\sigma$. We define a level trough as an equipotential curve, $\chi^a(\sigma)$, with two additional properties:
\begin{enumerate}
\item[$(i)$] The potential gradient vanishes everywhere along the curve: $V_{,\,a}[\chi(\sigma)] = 0$ for all $\sigma$ and for all $a$;
\item[$(ii)$] All eigenvalues of the `mass' matrix ${\cA^a}_b := \cG^{ac} \, V_{;\,cb}$ are non-negative, and at least one eigenvalue is strictly positive. This condition is required to distinguish troughs from ridges. Notice that because the eigenvalue condition, ${\cA^a}_b e^b = \lambda \, e^a$, is a tensor equation the eigenvalues $\lambda$ are scalars under field redefinitions.
\end{enumerate}

As is shown in detail in appendix \ref{appssec:troughgeometry} these conditions imply that the two independent eigenvectors of ${\cA^a}_b$ can be cleanly identified. First, there is a zero eigenvector given by the tangent, $\dot \chi^a$, to the trough's bottom. Here, and in what follows, over-dots denote differentiation with respect to $\sigma$, where $\exd \sigma^2 := \cG_{ab} \, \exd \chi^a \exd \chi^b$ denotes target-space proper distance along the trough's bottom. The nonzero eigenvector is proportional to the covariant directional derivative of $\dot \chi^a$ along the trough:
\be
 \frac{D \dot \chi^a}{\exd \sigma} := \ddot \chi^a + \gamma^a_{bc} \dot \chi^b \dot \chi^c \,.
\ee

Because $D \dot \chi^a/\exd \sigma$ is orthogonal\footnote{Using the target-space metric, $\cG_{ab}$.} to $\dot \chi^a$ --- see appendix \ref{appssec:troughgeometry} for details --- it is convenient to define the orthonormal basis, $\{ \dot \chi^a, n^b \}$, in field space, with
\be \label{eq:troughcurvaturedef}
 \frac{D \dot \chi^a}{\exd \sigma} := \frac{n^a}{\kappa} \,,
\ee
defining the radius of curvature, $\kappa(\varphi)$, of the bottom of the trough\footnote{In decomposing field excitations with respect to the basis defined by the tangent and normal to the {\it trough} of the potential, we derive independent Frenet-Serret relations \cite{FS} to those introduced in \cite{GNVT}, who define excitations tangent and normal to a background solution (in the context of inflation). We do so as we are interested in understanding how the scales of the parent theory enters the effective theory that describes all low energy processes, and not just those corresponding to perturbations around cosmological evolution.}. (Notice that $\kappa \to \infty$ corresponds to the case of a `straight' trough, for which the valley floor defines a target-space geodesic, $D \dot \chi^a/\exd \sigma = 0$.) In terms of $\dot \chi_a := \cG_{ab} \dot \chi^b$ and $n_a := \cG_{ab} n^b$ we therefore have
\be \label{eq:2rdordermatrix}
 V_{;\,ab} = m^2(\varphi) \, n_a n_b \,,
\ee
everywhere along the trough's bottom, where $m^2(\varphi) = V_{;\, ab} \, n^a n^b > 0$ is the nonzero eigenvalue of ${\cA^a}_b$ required by condition $(ii)$ above.

As shown in detail in Appendix \ref{appssec:troughgeometry}, differentiating eq.~\pref{eq:2rdordermatrix} with respect to $\sigma$ along the bottom of the trough, gives the following expression for the potential's third covariant derivatives,
\bea \label{eq:3rdderivmatrix}
 V_{;\,abc} &=&  %\left( \frac{\exd M^2}{\exd \sigma} \right)
 2 \, m \dot m \, \Bigl( n_a n_b \dot \chi_c + n_b n_c \dot \chi_a + n_c n_a \dot \chi_b \Bigr)  \nn\\
 && \qquad - \frac{m^2}{\kappa} \Bigl(
 n_a \dot \chi_b \dot \chi_c + n_b \dot \chi_a \dot \chi_c + n_c \dot \chi_a \dot \chi_b \Bigr) + V_{nnn} \; n_a n_b n_c \,,
\eea
where $V_{nnn} := V_{;\,abc} \, n^a n^b n^c$ and $\dot m := \exd m/\exd \sigma$. This uses that $V_{;\,abc}$ is completely symmetric when evaluated along the trough's bottom, since
\be \label{eq:v3_curv}
 V_{;\,cba} - V_{;\,cab} =  \cR^{d}_{\phantom{d}cab}V_{,\,d}
\ee
vanishes because $V_{,\,d}$ also vanishes there, using condition $(i)$ above. Among other things, eq.~\pref{eq:3rdderivmatrix} gives the quantities $\dot m$ and $\kappa$ in terms of derivatives of $V$, with
\be \label{eq:kappasimple}
 \frac{1}{\kappa} =  - \frac{1}{m^2} \, V_{;\,abc} \, n^a \, \dot\chi^b \dot\chi^c :=  - \frac{V_{ntt}}{m^2}
 \quad \hbox{and} \quad
 \dot m = \frac{1}{2\,m} \,
  V_{;\,abc} \, n^a n^b \dot\chi^c := \frac{V_{nnt}}{2\,m} \,.
\ee

Expressions for higher derivatives of $V$ are similarly obtained by repeated differentiation, with explicit expressions for the fourth derivatives given in appendix \ref{appssec:troughgeometry}. Notice that these higher derivatives need not be completely symmetric in their indices if the target-space metric is not flat, since (for example)
\be \label{eq:v4_curvcom}
  V_{;\,dcba} - V_{;\,dcab} =
  \cR^{e}_{\phantom{e}cab}V_{;\, de}
  + \cR^{e}_{\phantom{e}dab}V_{;\, ec} \,,
\ee
and so on.

\subsection{Tilted troughs}
\label{ssec:tiltedtroughs}

Of more interest, particularly in cosmology, is the situation where the trough is not completely level, but with a slope along the trough that is much shallower than the directions up the trough's sides.

This situation is handled as above, but with the generalization that derivatives along the trough direction are parametrically small rather than zero. Defining $U(\sigma)$ as the value of the potential along the trough bottom, we have
\be
 U(\sigma) := V[ \chi(\sigma)] \,,
\ee
and because the curve $\chi^a(\sigma)$ runs along the bottom of the (no-longer level) trough, its tangent is parallel to the potential gradient along the bottom: $\dot\chi^a \propto \cG^{ab} V_{,\,b}$. Because of this we replace condition $(i)$ of the flat trough with the following conditions for the potential gradient
\be \label{eq:udoteqn}
 \Va \, \dot \chi^a = \dot U
 \quad \hbox{and} \quad
 \Va \, n^a = 0 \,,
\ee
everywhere along the valley floor.

Successive differentiation --- see appendix \ref{appssec:troughgeometry} --- of these equations again allows the derivation of expressions for higher derivatives of the potential. In particular, differentiating eqs.~\pref{eq:udoteqn} gives the following expression for the second-derivative matrix
\be \label{eq:2ndderivmatrixU}
 V_{;\,ab} = \ddot U \, \dot \chi_a \dot\chi_b +
 \frac{\dot U}{\kappa} \, \Bigl( n_a \dot \chi_b + n_b \dot \chi_a \Bigr) + m^2 \, n_a n_b \,,
\ee
where, as before, we define $m^{2}(\sigma) := V_{; \, ab} \, n^a \, n^b$ and the radius of curvature by $D \dot \chi^a /\exd \sigma = n^a/\kappa$, with $\dot \chi^a$ and $n^b$ being the orthonormal basis adapted to the trough bottom.

In particular, eq.~\pref{eq:2ndderivmatrixU} shows
that $n^a$ and $\dot\chi^a$ need no longer be eigenvectors of the matrix ${\cA^a}_b$, and $m^2$ need no longer be an eigenvalue. Explicit diagonalization gives the eigenvalues
\be \label{eq:evals}
  M^{2}_\pm = \frac{1}{2}
  \left( m^{2} + \ddot{U} \pm (m^{2}-\ddot{U}) \sqrt{1
    +\beta^{2}} \right)  \,,
\ee
with corresponding (orthonormal) eigenvectors
\ba \label{eq:evecs}
  e_{+}^{a}   &=& n^{a}\cos\theta + \dot\chi^a \sin \theta \nn\\
  e_{-}^{a}   &=& - n^{a}\sin\theta + \dot\chi^a \cos \theta \,,
\ea
where
\bea \label{eq:angledef}
  \sin \theta &:=& ({\rm sgn} \, \dot{U}) \cdot \frac{\sqrt{1+\beta^{2}} -1}{\sqrt{2 (1 + \beta^2 - \sqrt{1+\beta^2} )}} \nn\\
  \cos \theta &:=& ({\rm sgn} \, \dot{U}) \cdot \frac{\beta}{\sqrt{2 (1 + \beta^2 - \sqrt{1+\beta^2} )}} \,,
\eea
and
\be \label{eq:betadef}
  \beta = \frac{2 \dot{U}}{\kappa(m^{2} - \ddot{U})}  \,.
\ee

These simplify once restricted to the regime of interest: $m^2$ much bigger than derivatives of $U$. In particular, in this limit
\be
  \beta \simeq \frac{2\, \dot{U}}{\kappa\, m^{2}} \left[
   1 + \frac{\ddot{U}}{m^2} + \cdots \right] \ll 1\,,
\ee
and so the `heavy' eigenvalue becomes
\be  \label{eq:heavy_eval}
  M^2 := M_+^2 \simeq m^2
  + \frac{\dot{U}^{2}}{\kappa^{2} m^{2}}
  + \mathcal{O}\left( \frac{1}{m^{4}}\right) \,,
\ee
while the `light' one is
\be
  \mu^2 :=M_-^2 \simeq \ddot{U}
  -\frac{\dot{U}^{2}}{\kappa^{2} m^{2}}
   + \mathcal{O}\left( \frac{1}{m^{4}}\right) \,.
\ee
The mixing angle is similarly small in this limit,
\be
  \tan \theta = \frac{\sqrt{1+\beta^{2}} -1}{\beta}
  \simeq \frac{\beta}{2} \simeq
  \frac{\dot{U}}{\kappa\, m^{2}} \,,
\ee
and so the corresponding eigenvectors take the approximate forms: $e_{+}^{a} \simeq n^{a} + \frac12 \, \beta \dot\chi^a + \cO(\beta^2)$ and $e_{-}^{a} \simeq \dot\chi^a - \frac12 \, \beta n^a + \cO(\beta^2)$.

Formulae for the third derivatives of $V$ are obtained by successive differentiation, and are derived in detail in appendix \ref{appssec:troughgeometry}. Because the trough is not precisely flat, the third derivatives are in general no longer completely symmetric. Specializing eq.~\pref{eq:v3_curv} to tilted troughs gives (for two fields)
\ba
  \label{eq:tilt_v3sym_1}
  V_{; \, abc} \, \dot\chi^a \, \dot\chi^b \, n^c
  - V_{; \, abc} \, \dot\chi^a \, n^b \, \dot\chi^c &=& 0 \,,
  \nn\\
  \label{eq:tilt_v3sym_2}
  V_{; \, abc} \, n^a \, \dot\chi^b \, n^c
  - V_{; \, abc} \, n^a \, n^b \, \dot\chi^c &=& \frac{\dot{U}}{2 \rho^2} \,,
\ea
which shows that ordering only matters when $V_{;\, abc}$ is contracted with two $n$'s and one $\dot\chi$. Appendix \ref{appssec:troughgeometry} shows that the third derivatives evaluate to
\ba
  V_{;\,abc} \, \dot \chi^a \dot \chi^b \dot \chi^c &=& \dddot{U} - \frac{2 \, \dot U}{\kappa^2} \nn\\
  V_{;\,abc} \, \dot \chi^a \dot \chi^b  n^c =
   V_{;\,abc} \, \dot \chi^a n^b \dot \chi^c =
    V_{;\,abc} \, n^a \dot \chi^b \dot \chi^c &=& - \frac{m^{2}}{\kappa}+\frac{2\, \ddot{U}}{\kappa}  - \frac{\dot \kappa \, \dot U}{\kappa^2}\\
  V_{;\,abc} \, n^a n^b \dot \chi^c &=& 2\, m \dot m +
  \frac{2 \, \dot{U}}{\kappa^{2}} \nn\\
    V_{;\,abc} \, \dot \chi^a n^b n^c =
    V_{;\,abc} \, n^a \dot \chi^b n^c &=& 2\, m \dot m +
  \frac{2 \, \dot{U}}{\kappa^{2}} + \frac{\dot U}{2 \rho^2} \,,\nn
\ea
where
\be
 - \frac{\dot \kappa}{\kappa^2} = \frac{\exd }{\exd \sigma}
 \left( \cG_{ab} \, n^a \frac{D \dot \chi^b}{\exd \sigma}
 \right) = \cG_{ab} \, n^a \frac{D^2 \dot \chi^b}{\exd \sigma^2} \,,
\ee
and $V_{;\,abc} \, n^a n^b n^c$ is not in general related to $\kappa$, $m$ and derivatives of $U$.

For later purposes it is useful also to have expressions for the completely symmetrized derivatives:
\ba
  V_{(ttt)} := V_{;\,(abc)} \, \dot \chi^a \dot \chi^b \dot \chi^c &=& \dddot{U} - \frac{2 \, \dot U}{\kappa^2} \nn\\
  V_{(ttn)} := V_{;\,(abc)} \, \dot \chi^a \dot \chi^b  n^c
  &=& - \frac{m^{2}}{\kappa} + \frac{2\, \ddot{U}}{\kappa}  - \frac{\dot \kappa \, \dot U}{\kappa^2}\\
  V_{(tnn)} := V_{;\,(abc)} \, n^a n^b \dot \chi^c &=& \frac23
  \; V_{;\,abc} \, \dot \chi^a n^b n^c + \frac13
  \; V_{;\,abc} \,  n^a n^b \dot \chi^c \nn\\
  &=& 2\, m \dot m +
  \frac{2 \, \dot{U}}{\kappa^{2}} + \frac{\dot U}{3 \rho^2} \,,\nn
\ea
as well as the contractions of the symmetrized derivative, $V_{;\,(abc)}$, with the eigenvectors $e^a_\pm$, in the small-$\beta$ limit. For instance $V_h := \Va e^a_+ \simeq \frac12 \beta \dot U \simeq (\dot U^2/\kappa m^2)$ and $V_\e := \Va e^a_- \simeq \dot U$, $V_{hh} = M^2$, $V_{\e\e} = \mu^2$ and $V_{\e h} = V_{h\e} = 0$. For small $\beta$ the third derivatives are
\be \label{Vlll}
  V_{\e\e\e} := V_{;\,(abc)} \, e^a_- e^b_- e^c_-
  \simeq V_{ttt} - \frac{3\beta}{2} \, V_{(ttn)}
  \simeq \dddot{U} + \frac{\dot U}{\kappa^2} \,,
\ee
\be \label{Vllh}
  V_{\e\e h} := V_{;\,(abc)} \, e^a_- e^b_- e^c_+
  \simeq V_{(ttn)} + \frac{\beta}{2} \, V_{ttt} - \beta \, V_{(tnn)}
  \simeq - \frac{m^{2}}{\kappa} + \frac{2\, \ddot{U}}{\kappa} - \frac{\dot{U}}{\kappa} \left[ \frac{4 \dot m}{ m} + \frac{\dot \kappa }{\kappa}   \right] \,,
\ee
\be \label{Vlhh}
  V_{\e hh} := V_{;\,(abc)} \, e^a_- e^b_+ e^c_+
  \simeq V_{(tnn)} - \frac{\beta}{2} \, V_{nnn} + \beta \, V_{(ttn)}
  \simeq 2\, m \dot m -  \dot{U} \left[  \frac{ \lambda_{nnn}}{\kappa^{2}}
   - \frac{1}{3 \rho^2} \right] \,,\nn
\ee
\be \label{Vhhh}
  V_{hhh} := V_{;\,(abc)} \, e^a_+ e^b_+ e^c_+
  \simeq V_{nnn} + \frac{3\beta}{2} \, V_{(tnn)}
  \simeq \frac{m^{2}}{\kappa}\lambda_{nnn} + \frac{6 \, \dot U \dot m}{\kappa m} \,,
\ee
where the quantity $\lambda_{nnn}$ defined by
\be \label{Vnnnbardef}
 V_{nnn} := \left( \frac{m^{2}}{\kappa} \right) \, \lambda_{nnn} \,,
\ee
typically remains bounded as $m^2$ gets large. Notice that these reduce to the usual expressions for straight troughs, $\kappa \to \infty$, with a flat target space, $\rho \to \infty$. For some extensions of these expressions to higher derivatives and to $1/m^{2}$ corrections, see appendix \ref{appssec:troughgeometry}.

\section{The low-energy effective theory}
\label{sec:lowenergyEFT}

The previous section shows that there are three separate, possibly large, scales that instantaneously characterize the properties of a trough-shaped potential along its bottom: the scale $m^2(\varphi)$ defining the trough's transverse steepness; the scale $\kappa(\varphi)$ defining the radius of curvature of the trough's valley floor; and the Riemann radius of curvature, $\rho(\varphi)$, of the target-space geometry. There are also the derivatives of these quantities along the trough, as well as third and higher derivatives of $V$ in the direction(s) normal to the trough.

\subsection{Light and heavy states in a trough}

We now assume all of these scales to be much larger than the energy scales of interest, such as the fractional rates of change of quantities along the trough's bottom. We wish to identify the low-energy effective theory that governs the dynamics along the trough in this limit. Our goal is to trace the leading way that each of these scales shows up in the low-energy effective interactions once heavy degrees of freedom are integrated out (at the classical level). In particular, we wish to see how their presence alters the naive truncation approximation, in which the heavy fields are simply set to zero.

In order to do so we must identify the heavy and light degrees of freedom, and integrate out the heavy one. To this end we expand the expansion field, $\delta \phi^a$ in a basis that diagonalizes the mass matrix, ${\cA^a}_b = \cG^{ac} \, V_{;\,cb}$, writing
\be \label{eigeqn}
 \delta \phi^a = \e \, e^a_- + h \, e^a_+ \,.
\ee
By virtue of the above definitions the expansion, eq.~\pref{Vexpnv0}, of the scalar potential becomes:
\ba \label{Vexpnv1}
 V (\phi)
 &=& V(\varphi) + \Bigl( V_\e \; \e + V_h \, h \Bigr)
 + \frac12 \Bigl( M^2 \, h^2 + \mu^2 \, \e^2
 \Bigr) \nn\\
 && \qquad\qquad + \frac16 \Bigl( V_{\e\e\e} \, \e^3
 + 3 V_{h\e\e} \, h \e^2 + 3 V_{hh\e} \, h^2 \e
 + V_{hhh} \, h^3 \Bigr) + \cdots \,,
\ea
where $M^2(\p) \simeq  m^2 + ({\dot{U}^{2}}/{\kappa^{2} m^{2}})$ and $\mu^2 (\p) \simeq  \ddot{U} - ({\dot{U}^{2}}/{\kappa^{2} m^{2}})$ while $V_{\e\e\e}(\p) = V_{;\,(abc)} \, e^a_- e^b_- e^c_-$, $V_{\e \e h} (\p) = V_{;\,(abc)} \, e^a_- e^b_- e^c_+$ and so on are the symmetric derivatives of $V$ as evaluated at the end of the previous section. In terms of these fields the expansion of the kinetic term, eq.~\pref{eq:kinexpansion}, similarly is
\ba \label{eq:kinexpansion2}
 - \frac12 \, \cG_{ab}(\phi) \, \partial_\mu \phi^a
 \, \partial^\mu \phi^b &=&
 - \frac12 \left[ \cG_{ab}(\varphi) + \frac{1}{3} \cR_{acbd}(\varphi) \, \delta \phi^c \delta \phi^d \right] \partial_\mu \delta \phi^a \, \partial^\mu \delta \phi^b
 + \cdots \nn\\
 &=& - \frac{1}{2} \, (\partial \delta \phi)^2 + \frac{1}{12 \, \rho^2} \, \Bigl[
  \delta \phi^2 (\partial \delta \phi)^2 - (\delta \phi \cdot \partial \delta \phi)^2
  \Bigr]  + \cdots \\
 &=& - \frac12 \, \Bigl[ (\partial\e)^2
 + (\partial h)^2  \Bigr] \nn\\
 && \qquad\qquad + \frac{1}{12 \rho^2} \,
 \Bigl[ \e^2 (\partial h)^2 + h^2 (\partial \e)^2
 - 2 \, h \e (\partial \e) (\partial h)
  \Bigr]  + \cdots \,, \nn
\ea
which uses eq.~\pref{eq:2Dcurv} for the target-space curvature.

\subsection{Integrating out the heavy fields}
\label{ssec:integratingout}

The next step is to integrate out the heavy field to obtain the low-energy effective theory of the light field along the bottom of the trough. In the classical approximation the heavy field is integrated out by eliminating it from the action using its equations of motion:\footnote{For time-dependent solutions there generically is more than one such solution, in which case it is the solution corresponding to having $h$ in its adiabatic vacuum that should be used \cite{EFTrevs}.} $S_{\rm eff}(\e) = S[\e, h(\e)]$, where $h(\e)$ is the adiabatic ground state satisfying $\delta S/\delta h = 0$ \cite{EFTrevs}. We summarize the main steps here, with more details given in Appendix \ref{app:integratingout}.

To start, it is useful to integrate by parts in order to write the classical action as follows,
\be \label{act}
 \frac{\mathcal L}{\sqrt{-g}}  = - \frac12 \, \partial_\mu \e \, \partial^\mu \e - V_{\rm tr}(\p, \e) + \frac12 \, h \, \Delta_h h - J_{(1)} h - \frac13 \, J_{(3)} h^3 - \frac14 \, J_{(4)} h^4 \,,
\ee
where the `truncated' potential is
\be
 V_{\rm tr}(\varphi, \ell) := V (\varphi, \ell,{h=0}) = U(\p) + (j + V_\e) \e + \frac{\mu^2}{2} \, \e^2 + \frac16 \, V_{\e\e \e} \, \e^3 + \frac{1}{24} \, V_{\e\e\e\e} \, \e^4
\ee
and we couple an external current, $j$, to the light field, $\e$. The kinetic operator for $h$ is $\Delta_h := \Omega - \cM^2$, where
\begin{eqnarray}  \label{eq:otherdefs}
 \Omega &:=& \left( 1 - \frac{\e^2}{6 \rho^2} \right) \square \nn\\
 \hbox{and} \quad
 \cM^2 &:=& M^2 + V_{hh\e} \, \e + \frac12 \, V_{h h \e\e} \, \e^2 - \frac{1}{2 \rho^2} \, \partial_\mu\e \partial^\mu \e - \frac{1}{3\rho^2} \, \e \, \square \e \,.
\end{eqnarray}
Finally, the $J_{(i)}$ are given by
\bea
 J_{(1)} &:=& V_h + \frac12 \, V_{h\e\e} \, \e^2 + \frac16 \, V_{h\e\e\e} \, \e^3 \nn\\
 J_{(3)} &:=& \frac12 \, V_{hhh} + \frac12 \, V_{h h h \e} \, \e \\
 \hbox{and} \quad J_{(4)} &:=& \frac16 \, V_{hhhh} \,. \nn
\eea

The equation of motion of the field $h$ then is,
\be
 \Delta_h h = J_{(1)} + J_{(3)} h^2 + J_{(4)} h^3 \,,
\ee
which can be solved iteratively to give $h = \Delta_h^{-1} J_{(1)} + \cdots$, where the ellipses involve powers of $J_{(3)}$ and $J_{(4)}$. We insert this back into the classical action, and expand $\Delta_h^{-1}$ in powers of $1/M^2$ to get the following expression (see Appendix \ref{app:integratingout} for details)
\be \label{eq:leff_series v2}
  \mathcal{L}_{\rm eff} =
  \mathcal{L}_{(0)} + \frac{ \mathcal{L}_{(1)}}{M^2} + \frac{ \mathcal{L}_{(2)}}{M^4} + \frac{ \mathcal{L}_{(3)}}{M^6} + \mathcal{O} \left(\frac{1}{M^{8}}\right) \,,
\ee
where
\be \label{Dl1}
 \frac{ \mathcal{L}_{(1)}}{M^2} = \frac{1}{2M^2} \left( V_h + \frac12 \, V_{h\e\e} \, \e^2 + \frac16 \, V_{h\e\e\e} \, \e^3 \right)^2
\ee
\begin{eqnarray} \label{Dl2}
 \frac{ \mathcal{L}_{(2)}}{M^4}   &=& - \frac{1}{2M^4} \left( V_h + \frac{1}{2} \, V_{h\e\e} \, \e^2 + \frac{1}{6} \, V_{h\e\e\e} \, \e^3 \right)^2 \left( V_{hh\e} \, \e + \frac{1}{2} \, V_{hh\e\e} \, \e^2 \right) \\
 && \qquad - \frac{\e^2}{2M^4} \, (\partial\e)^2 \left( V_{h\e\e} + \frac{1}{2} \, V_{h\e\e\e} \, \e \right)^2 + \frac{1}{12 \rho^2 M^4} \, (\partial\e)^2 \left( V_h - \frac{1}{2} \, V_{h\e\e} \, \e^2 - \frac{1}{3} \, V_{h\e\e\e} \, \e^3 \right)^2 \nn
\end{eqnarray}
and
\ba \label{Dl3}
 \frac{ \mathcal{L}_{(3)}}{M^6} &=&  \frac{1}{2M^6}\left(V_h + \frac{\e^2}{2}V_{\e\e h} \right)^2\left(V_{\e h h}\e + \frac{\e^2}{2}V_{\e\e h h}\right)^2 \nn\\
 &+& \frac{1}{6M^6}\left(V_{hhh} + V_{\e hhh}\e\right)\left(V_h + \frac{\e^2}{2}V_{\e\e h} + \frac{\e^3}{6}V_{\e\e\e h}\right)^3 \nn \\
 &+& \frac{V_h^2}{2M^6}\left(\frac{1}{4\rho^4}(\partial\e)^4 + \frac{1}{9\rho^4}\e^2\square\e\square\e + \frac{1}{3\rho^4}(\partial\e)^2\e\square\e\right) \nn\\
 &+& \frac{V_h^2}{2\rho^2 M^6} (\partial\e)^2 \left(\frac{\e}{3}V_{\e h h} + \frac{\e^2}{2}V_{\e\e h h}\right) + \frac{V_h}{2M^6}V_{\e hh}V_{\e\e\e h}\e^2(\partial\e)^2 \\
 &+& \frac{(\partial\e)^2}{M^6}\left(V_hV_{\e\e h} - \frac{V_h^2}{6\rho^2}\right)\left(V_{\e h h }\e + V_{\e\e h h}\e^2\right) \nn\\
 &+& \frac{1}{\rho^2M^6}\left(V_hV_{\e\e h} - \frac{V_h^2}{6\rho^2}\right)\left(\frac{(\partial\e)^4}{2} + \frac{\e^2}{3}(\square\e)^2 + \frac{5}{6}\e\square\e(\partial\e)^2\right) \nn\\
 &+& \frac{1}{2M^6}\left(V^2_{\e\e h} - \frac{V_hV_{\e\e h}}{3\rho^2}\right)\left((\partial\e)^4 + \e^2(\square\e)^2 + 2\e\square\e(\partial\e)^2\right) \,, \nn
\ea
and so on.

Finally, we trade the explicit derivatives of $V$ appearing in these expressions --- including the mass eigenvalues $M^2$ and $\mu^2$ --- in favour of the trough-related quantities $U(\varphi)$, $m(\varphi)$, $\kappa(\varphi)$ and their derivatives along the trough, as well as $\rho(\varphi)$ and transverse derivatives like $V_{hhh}$ and so on, using the following results from earlier sections (and Appendix \ref{appssec:troughgeometry})
\be
 V_\ell = \dot U \,, \quad
 V_h \simeq \frac{\dot U^2}{\kappa m^2} \,, \quad
 V_{hh} = M^2 \simeq  m^2 + \frac{{\dot{U}^{2}}}{{\kappa^{2} m^{2}}} \,, \quad
 V_{\e\e} = \mu^2 \simeq  \ddot U - \frac{{\dot{U}^{2}}}{{\kappa^{2} m^{2}}} \,,
\ee
while $V_{h\ell}=0$. Third derivatives are similarly given by
\ba   \label{eq:tilt_v3_lh}
  V_{\e \e \e} &=& \dddot{U} + \frac{\dot{U}}{\kappa^{2}}
  -\frac{3 \dot{U}}{m^{2}\kappa^{2}}
  \left[ \ddot{U} -\dot{U}
  \left( \frac{2 \dot{m}}{m} + \frac{\dot{\kappa}}{\kappa}  \right) \right]
%  + \mathcal{O}\left(\frac{1}{m^{4}} \right)
 \,, \nn \\
  V_{\e \e h} &=& - \frac{m^{2}}{\kappa}
  + \frac{2 \ddot{U}}{\kappa}
  - \frac{\dot{U}}{\kappa} \left( \frac{4\dot{m}}{m} +\frac{\dot{\kappa}}{\kappa} \right)
  + \frac{\dot{U}}{m^{2} \kappa}
  \left[
  \dddot{U} - \frac{4\ddot{U} \dot{m}}{m}  -  \frac{\dot{U}}{\kappa^{2}} \left(\frac{5}{2}-\lambda_{nnn}\right)
  - \frac{2\dot{U}}{3\rho^{2}}
  \right]
%  \nn \\
%  && + \mathcal{O}\left(\frac{1}{m^{4}} \right)
 \,, \nn \\
  V_{\e h h} &=& 2 m \dot{m} - \dot{U} \left(   \frac{\lambda_{nnn}}{\kappa^{2}} -\frac{1}{3 \rho^{2}} \right)
  + \frac{\dot{U}}{m^{2}\kappa^{2}}
  \left[
  \ddot{U}\Bigl( 2-\lambda_{nnn} \Bigr) - \dot{U} \left(  \frac{7\dot{m}}{m} + \frac{2\dot{\kappa}}{\kappa} \right)
  \right]
 % \nn\\ &&
 % + \mathcal{O}\left(\frac{1}{m^{4}} \right)
 \,,  \\
  V_{h h h} &=& \frac{m^{2}}{\kappa}\lambda_{nnn} +  \frac{6 \dot{U}\dot{m}}{\kappa m}
  + \frac{\dot{U}}{m^{2}\kappa}
  \left[ \frac{ 6 \ddot{U} \dot{m}}{m}
  + \frac{3\dot{U}}{\kappa^{2}} \left( 1 - \frac{1}{2}\lambda_{nnn}\right)
  + \frac{\dot{U}}{\rho^{2}}
  \right]
 % + \mathcal{O}\left(\frac{1}{m^{4}} \right)
 \,, \nn
\ea
which extends the earlier expressions to higher order in $1/m^2$, and where $\lambda_{nnn}$ is as defined in eq.~\pref{Vnnnbardef}. Expressions for fourth derivatives are similarly given in Appendix \ref{app:integratingout}.

The results obtained by substituting these expressions into eqs.~\pref{Dl1} through \pref{Dl3} are most succinctly expressed in terms of an expansion in derivatives of $\ell$. As is shown in detail in Appendix \ref{app:integratingout}, it is always possible to perform a local field redefinition so that the result up to four derivatives has the form
\be \label{eq:leff_deriv_exp}
 \frac{\cL_{\rm eff}}{\sqrt{-g}} = - \hat V_{\rm eff}(\varphi, \e) - \frac12 \, G_{\rm eff}(\varphi, \e) (\partial_\mu \e \, \partial^\mu \e) +  \cH_{\rm eff}(\varphi, \e) (\partial_\mu \e \, \partial^\mu \e)^2 + \cdots \,,
\ee
and so the content of the above calculation is to give expressions for the leading contributions to the functions $\hat V_{\rm eff}$, $G_{\rm eff}$ and $\cH_{\rm eff}$. (The freedom to perform field redefinitions ensures that only two of these functions are independent, as we show in detail below.) We now quote the expressions for these functions that are relevant for terms in $\cL_{\rm eff}$ involving at most four powers of the light field, $\ell$.

The effective scalar potential is given by (see Appendix \ref{app:integratingout} for details)
\ba \label{veffp}
 \hat V_{\rm eff}(\e) &=& U(\varphi + \e) + j \, \ell + \frac{\e^3}{6} \left( \frac{\dot U}{\kappa ^2} \right) \left( 1 + \frac{2 \dot U \dot \kappa}{\kappa m^2} -  \frac{3 \ddot U}{m^2} + \frac{6 \dot U \dot m}{m^3} \right) \nn\\
 && \; + \e^4\left[ -\frac{ \dot U \dot \kappa }{8\kappa ^3}+\frac{ \ddot U}{6\kappa ^2} + \frac{\dot U \ddot U}{\kappa^2 m^2} \left( \frac{2 \dot m}{m} + \frac{5 \dot\kappa}{6\kappa} \right) - \frac{2\dot U^2 \dot \kappa \dot m}{3\kappa^3 m^3} - \frac{\ddot U^2}{2 \kappa^2 m^2}
 \right. \\
 && \qquad\qquad  \left. - \frac{\dot U \dddot U}{4 \kappa^2 m^2}
   + \frac{\dot U^2}{24 \kappa^4 m^2}
 \left(6+3 \lambda_{nnn}-11 \dot \kappa^2+4 \kappa  \ddot \kappa \right) - \frac{2\dot U^2}{\kappa^2 m^2}\frac{\dot m^2}{m^2} \right] + \cO(\ell^5)\,, \nn
\ea
while the kinetic function is
\be \label{eq:Geffnaive}
 G_{\rm eff}(\e) \simeq 1 + \frac{\e^2}{\kappa^2}\left(1 + \frac{2\dot U\dot\kappa}{\kappa m^2} - \frac{3\ddot U}{m^2} + \frac{8\dot U \dot m}{m^3} \right) \,,
\ee
and the 4-derivative term has coefficient
\be \label{eq:Heff}
 \cH_{\rm eff}(\e) = \frac{1}{2\kappa^2m^2} + \cO(\ell) \,.
\ee

As remarked above, since there is only a single light field only two of these three functions are independent. This is usually expressed by performing a field redefinition, $\e \to \hat \e$ to a `canonical' basis chosen to set the kinetic function to unity: $G_{\rm eff}(\ell) (\partial \ell)^2 = (\partial \hat \ell)^2$. The required redefinition satisfies
\be
 \partial_\mu \hat \e := \sqrt{G_{\rm eff}(\varphi, \e)} \; \partial_\mu \e
 \simeq \left[ 1 + \frac{\e^2}{2\kappa^2}\left(1 + \frac{2\dot U\dot\kappa}{\kappa m^2} - \frac{3\ddot U}{m^2} + \frac{8\dot U \dot m}{m^3} \right) + \cdots \right] \partial_\mu \e \,,
\ee
which has as solution
\be
 \hat \e \simeq \e + \frac{\e^3}{6\kappa^2}\left(1 + \frac{2\dot U\dot\kappa}{\kappa m^2} - \frac{3\ddot U}{m^2} + \frac{8\dot U \dot m}{m^3} \right) + \cdots \,.
\ee
Notice that once this is used (and dropping the `caret' over $\ell$) the effective scalar potential changes to
\ba \label{eq:vtrunc2can}
 V_{\rm eff}(\varphi, \e) &\simeq& \hat V_{\rm eff}(\varphi, \e) - (\dot U + j) \frac{\e^3}{6\kappa^2}\left(1 + \frac{2\dot U\dot\kappa}{\kappa m^2} - \frac{3\ddot U}{m^2} + \frac{8\dot U \dot m}{m^3} \right) \nn\\
  && \qquad\qquad\qquad\qquad\qquad\qquad - \frac{\ddot{U} \e^4}{6\kappa^2}\left(1 + \frac{2\dot U\dot\kappa}{\kappa m^2} - \frac{3\ddot U}{m^2} + \frac{8\dot U \dot m}{m^3} \right) + \cdots \nn\\
 &\simeq& U(\varphi + \e) + j \, \ell - \e^3 \left( \frac{\dot U^2 \dot m}{3 \kappa^2 m^3} \right) \nn\\
 && \qquad + \e^4\left[ -\frac{ \dot U \dot \kappa }{8\kappa ^3} + \frac{\dot U \ddot U }{\kappa^2 m^2} \left( \frac{2\dot m}{3m} + \frac{\dot \kappa}{2\kappa} \right)  - \frac{2\dot U^2 \dot \kappa \dot m}{3\kappa^3 m^3} - \frac{\dot U \dddot U}{4 \kappa^2 m^2} \right. \\
 && \qquad\qquad\qquad\qquad + \left. \frac{ \dot U^2}{24 \kappa^4 m^2}  \left(6+3 \lambda_{nnn}-11 \dot \kappa^2+4 \kappa  \ddot \kappa \right)- \frac{2\dot U^2}{\kappa^2 m^2}\frac{\dot m^2}{m^2} \right]  + \cdots \,, \nn
\ea
where the ellipses denote terms involving higher powers of $\hat\e$ or $1/m^2$, and the new term involving $j$ is absorbed into a redefinition of $j$. Expression \pref{eq:Heff} for $\cH_{\rm eff}$ remains unchanged by this field redefinition to the order in $\ell$ to which we work.

Notice that there are two interesting special cases for which all of the differences between $V_{\rm eff}(\varphi, \ell)$ and $U(\varphi + \e)$ vanish. First, they do so (even for finite $m$ and $\kappa$) for a level trough with all derivatives of $U$ vanishing. This is required in order for the full theory and the effective theory to agree on the value of the potential at its minimum (and so also on measurable quantities like the curvature of spacetime, say). Second, they also vanish in the limit of a straight trough, where $\kappa \to \infty$, in which case a truncation of $V(\ell,h)$ to $h=0$ would have been a good approximation. What is perhaps noteworthy is the appearance of terms that are suppressed only by $1/\kappa$ and not by $1/m$, and so which survive even for infinitely steep troughs for which $m^2 \to \infty$ with $\kappa$ fixed.

Of course, we equally well could have made an alternative choice of variables, $\hat \e \to \check \e$ for which $V_{\rm eff}(\varphi, \check \e) = U(\varphi + \check \e)+\check j \check \e$, at the expense of making the kinetic term non-canonical. For troughs that are not flat, what counts physically is neither $V_{\rm eff}$ or $G_{\rm eff}$ separately, but their relative form and we see that generically either $V_{\rm eff} \ne U(\varphi + \ell)$ or $G_{\rm eff} \ne 1$.

In summary, we see that (for two scalar fields) the most general possible effective interactions governing the dynamics of the light field at low energies (and out to quartic order in $\ell$) along a potential trough are given by eq.~\pref{eq:leff_deriv_exp} with $G_{\rm eff}(\ell) = 1$, $\cH_{\rm eff} = 1/(2 \kappa^2 m^2)$ and $V_{\rm eff}$ given by eq.~\pref{eq:vtrunc2can}. What makes this effective theory so useful (as for any low-energy effective theory) is that these interactions can be used to describe {\em all} physical processes involving at most quartic interactions that can appear at low energies in the full theory. In particular, it identifies that only the combinations of $\kappa$, $m$ and $\rho$ that appear in eqs.~\pref{eq:Heff} and \pref{eq:vtrunc2can} can be relevant at low energies for a broad class of physical situations.

\subsection{Domain of validity}
\label{sec:ValidityDomain}

Before applying this effective theory to some simple illustrative examples it is worth recapping the approximations on which its validity relies.

\subsubsection*{Semiclassical limit}

First, because it is derived purely within the classical approximation, the effective field theory implicitly relies on there being small parameters that parametrically suppress quantum corrections. In the full theory this is often assured through the existence of small dimensionless couplings, like gauge or quartic-scalar couplings. It implicitly also relies on a low-energy approximation, both to justify the low-energy, single-field approximation (see below) and to justify semiclassical methods in the full theory. For instance, the energies to which the full two-field theory are applied must be small relative to the higher energy scales being ignored (such as -- but not restricted to -- the Planck scale) in order to suppress loops, and so is a precondition for justifying the semiclassical treatment of gravity.

\subsubsection*{Low energies}

The additional condition required to replace the full two-scalar system with its one-scalar effective theory in the trough requires the energies of interest to be low enough not to dynamically excite any heavy quanta.\footnote{For time-dependent -- such as cosmological -- applications, `energies' here means both adiabatic energies of perturbations and any measures of background time-dependence, such as the Hubble scale, $H$.} In practice, the validity of the derivative expansion used in \pref{eq:leff_deriv_exp} requires all derivatives to be much smaller than the high-energy scales. As we saw when inverting the heavy-field operator $\Delta_h = \Omega - \cM^2$ as a power series in $\Omega / \cM^2$, the relevant scale controlling this low-energy expansion is set by
\be
 \cM^2 = M^2 + V_{hh\e} \, \e + \frac12 \, V_{h h \e\e} \, \e^2 - \frac{1}{2 \rho^2} \, \partial_\mu\e \partial^\mu \e - \frac{1}{3\rho^2} \, \e \, \square \e \,,
\ee
rather than directly by $\kappa$ and $m$. In particular, the low-energy approximation (and the effective field theory description derived here) can fail if the various terms in $\cM^2$ cancel, even if they are separately large. This is the reason the effective single-field approximation fails in explicit examples \cite{OtherTroughs0}, and we see it here as arising for the usual reason: a breakdown of the large hierarchy of scales on which the decoupling of high scales is based.

Furthermore, even when an effective single-field description exists, it need not be the one obtained by simply truncating the heavy fields \cite{NoTrunc}. As we see above, setting $h = 0$ requires $V_h$ to vanish, but because $V_h \simeq (\dot U/\kappa m)^2$ this need not be a good approximation.

For time-dependent problems, since effective theories only capture adiabatic evolution the low-energy limit also requires the time scales for significant changes to low-energy classical fields to be much larger than those, such as $1/m$ and $1/\kappa$, set by high-energy scales.

\subsubsection*{Small fields}

The explicit form given for the effective Lagrangian in eq.~\pref{eq:leff_deriv_exp} also relies on expanding in powers of $\ell$, and in the presence of shallow troughs in the scalar potential this is (by assumption) {\em not} required by the low-energy approximation. In practice the need to expand in powers of $\ell$ arises from the complexity of solving the full field equations, even in the limit where $m$ is very large.

This complexity has two logically different sources. First, for the kinetic energies the small-field limit enters when evaluating the target-space curvature only at the background, $\varphi$, rather than also as a function of $\ell$. This approximation implicitly requires $\ell$ not to be large compared with the target-space radius of curvature: $\ell \ll \rho$.

%The second source is the scalar potential where the small-$\ell$ expansion is also used, and here it is the radius of curvature of the trough that provides the relevant comparison: $\ell \ll \kappa$. The necessity for this requirement can be seen in more detail by re-examining the integration over the heavy field, $h$. In particular we found that the classical solution, $h(\ell)$, approximately satisfies
%
%\be
% \cM^2 h \simeq J_{(1)} \sim \frac{m^2 \ell^2}{\kappa} + \cdots \,,
%\ee
%
%where we use the fact that $V_{h\e\e} \sim m^2/\kappa$. This shows that even if $\cM^2 \sim m^2$, fluctuations in $h$ driven by changes to $\ell$ can be of order $h \sim \ell^2/\kappa$, and so are only negligible relative to those of $\ell$ if $\ell \ll \kappa$.

Secondly, and more generally, because $J_{(1)} \sim m^2 \ell^2/\kappa$ grows with $m^2$ there could be contributions to the effective action to order $1/m^2$ coming from what are formally much higher orders in the $1/m$ expansion, such as those arising from contributions like $\Delta \mathcal L \sim J^n_{(1)} J_{(n)}/\cM^{2 n}_{eff}$. However, these are also higher order in $\e$ -- being at least of order $\e^{2 n}$ -- showing how $\ell \ll \kappa$ is implicitly required to justify their neglect. This of course is an artefact of having expanded around the point $\varphi$. In order to analyse the system far away from $\varphi$ (i.e. for large $\ell$), it suffices to simply shift the expansion base point, $\varphi$.
%the energy changes unsuppressed by $m$ that come with large excursions in $h$.

\section{Some flat examples}
\label{sec:examples}

It is useful to compare the above expressions with concrete examples, to check their validity against known systems before seeking new applications to cosmological models.

\subsection{The mexican hat}
\label{ssec:mexicanhat}

Consider first the most familiar case of a curved trough: two scalar fields with a flat target space mutually coupled through an $O(2)$-invariant `Mexican hat' or `wineglass' potential:
\be
 \frac{\cL}{\sqrt{-g}} = - \partial_\mu \Phi^* \partial^\mu \Phi - V(\Phi^* \Phi)\,,
\ee
where $\Phi = \frac{1}{\sqrt2} (\cX + i \cY) = \frac{1}{\sqrt2} \, \cZ \, e^{i \vartheta}$. The target-space metric for this model is flat, as is explicit when written in terms of $\cX$ and $\cY$, for which the target-space Christoffel symbols vanish. Consequently, in these coordinates $V_{;\,a_1..a_n} = V_{,\,a_1..a_n}$ and so on.

In this section we choose the potential to have the explicit form
\be \label{eq:mexicanV}
 V(\cX, \cY) = V_0+ \frac{\lambda^2}{4} \left( \cX^2 + \cY^2 - \frac{\nu^2}{\lambda^2} \right)^2 \,,
\ee
which has a level trough at $V = V_0$ along the curve $\cZ = \sqrt{ \cX^2 + \cY^2} = \nu/\lambda$. The unit tangent and normal to this trough are
\be
 \vec e_\e = \left( \begin{array}{c}
  - \sin \vartheta \\ \cos \vartheta \\ \end{array} \right)
  \quad \hbox{and} \quad
 \vec e_h = \left( \begin{array}{c}
  -\cos \vartheta \\ -\sin \vartheta \\ \end{array} \right) \,,
\ee
where $\cos \vartheta := {\cX}/{\cZ}$ and $\sin \vartheta := \cY/\cZ$. These are also eigenvectors of the mass matrix, ${\cA^a}_b = \delta^{ac} \, V_{,\,cb}$,
\be
 \cA = - \nu^2 \cI + \lambda^2 \left( \begin{array}{cc}
                   3 \cX^2 + \cY^2 & 2 \cX \cY \\
                   2 \cX \cY & 3 \cY^2 + \cX^2 \\
                 \end{array}
               \right) \,,
\ee
with eigenvalues $\mu^2 = M_-^2 = - \nu^2 + \lambda^2(\cX^2 + \cY^2)$ and $M^2 = M_+^2 = - \nu^2 + 3 \lambda^2 (\cX^2 + \cY^2)$. Evaluated at the bottom of the trough these reduce to
\be
 M_-^2 = 0 \quad \hbox{and} \quad
 m^2 := M^2_+ = 2\nu^2 \,,
\ee
as usual. Clearly $V_h = V_{,\,a} \, e^a_h$ and $V_\e = V_{,\,a} \, e^a_\e$ both vanish everywhere at the bottom of the trough.

Since the trough is level, it follows that $\dot U = \ddot U = 0$ and from equation (\ref{eq:kappasimple}) that the radius of curvature of the bottom of the trough is $1/\kappa = - V_{h\e\e}/m^2$, where $V_{,\,\cX\cX\cX} = 6\lambda^2 \cX$, $V_{,\,\cX\cX\cY} = 2\lambda^2 \cY$, $V_{,\,\cX\cY\cY} = 2\lambda^2 \cX$ and $V_{,\,\cY\cY\cY} = 6\lambda^2 \cY$. Combining definitions,
\ba
 V_{h\e\e} &=& - V_{,\,\cX\cX\cX} \, s^2 c - V_{,\,\cX \cX \cY} (-2 s c^2 + s^3) - V_{,\,\cX \cY \cY} (c^3 - 2 s^2 c) - V_{,\,\cY\cY\cY} \, s c^2 \nn\\
 &=& -\lambda^2 \cX (6 s^2 c + 2 c^3 - 4 s^2 c)
 - \lambda^2 \cY (6sc^2 + 2 s^3 - 4 sc^2) \nn\\
 &=& -2\lambda^2( \cX c + \cY s) = -2\lambda^2
 \sqrt{\cX^2 + \cY^2} \,,
\ea
where $c := \cos \vartheta$ and $s := \sin \vartheta$. Evaluated at the trough's minimum, $\sqrt{\cX^2 + \cY^2} = \nu/\lambda$, this allows $\kappa$ to be simplified to
\be
 \label{eq:kappa_m_reln}
 \kappa = \frac{m^2}{2 \lambda \nu}
  = \frac{\nu}{\lambda} \,,
\ee
as expected. In particular, the $O(2)$ symmetry ensures physical quantities do not vary along the trough, so $\dot \kappa = \dot m = 0$ and so on. For reference, we list all the symmetrized derivatives, $V_{i_{1} \cdots i_{k}}$, (evaluated at the trough minimum) for the mexican hat potential:
\ba
 V_{\e \e \e} = 0 \,, \qquad
 V_{h \e \e} = -2 \lambda \nu \,, \qquad
 V_{h h \e} = 0 \,, \qquad
 V_{h h h} = -6 \lambda \nu \,, \nn\\
 V_{\e \e \e \e} = 6 \lambda^2 \,, \qquad
 V_{h \e \e \e} = 0 \,, \qquad
 V_{h h \e \e} = 2\lambda^2 \,, \qquad
 V_{h h h \e} = 0 \,, \qquad
 V_{h h h h} = 6 \lambda^2 \,,
\ea
and $V_{i_{1} \cdots i_{k}} = 0$ for $k \geq 5$.

Specializing the low-energy effective Lagrangian, eq.~\pref{eq:leff_deriv_exp}, to this case we find
\bea
 \label{eq:EFTM2_mexhat}
 \frac{\cL_{\rm eff}}{\sqrt{-g}}
 &=& - V_0 - \frac12 \, \partial_\mu \e
 \, \partial^\mu \e + \frac{1}{2m^2 \kappa^2} (\partial_\mu \e \partial^\mu\e)^2
 + \mathcal{O}\left(\frac{1}{m^{4}}\right) \nn\\
 &=& -V_0 - \frac12 \, \partial_\mu \e
 \, \partial^\mu \e + \frac{\lambda^2}{4\nu^4} (\partial_\mu \e \partial^\mu\e)^2
 + \mathcal{O}\left(\frac{1}{m^{4}}\right)  \,,
\eea
where the second line uses the above calculations of $m$ and $\kappa$. Notice that the symmetry $\e \to \e + c$ of the low-energy theory ensures the existence of a conserved Noether current,
\be \label{eq:JeffMH}
 J^\mu_{\rm eff} = - \cN \left[ 1 - \frac{\lambda^2}{\nu^4} \left(\partial_\lambda \e \partial^\lambda \e \right) \right] \partial^\mu \e + \cdots \,,
\ee
which corresponds (up to a constant normalization, $\cN$) to the current due to $O(2)$ invariance in the full theory
\be
 J^\mu = - \cZ^2 \, \partial^\mu \vartheta \,.
\ee

%Furthermore, from the form of (\ref{eq:EFTM2_mexhat}) we see that the Lagrangian is of the form
%
%\be
%\mathcal L = P(X,\ell) := X + \frac{2}{m^2\kappa^2}X^2 - V_0,
%\ee
%
%where $X:= -\frac{1}{2}(\partial\ell)^2$, from which one can deduce that the light quanta propagate with a reduced speed of sound given by \cite{k-inf}:
%\be
%c_s^2 = \frac{P_{,X}}{P_{,X} + 2XP_{,XX}} \approx 1 - \frac{8X}{m^2\kappa^2}.
%\ee
%% TOOK THIS OUT SINCE IT ONLY IS USED FOR COSMOLOGY, SO MOVED IT THERE.

\subsubsection*{Slowly rolling solutions}

As an application of this Lagrangian, consider next the energetics of the slowly rolling solution where the field $\Phi$ rotates around the bottom of the potential at constant angular speed: {\em i.e.} $\cZ$ is constant but $\vartheta = \omega t$. In this case the centrifugal force shifts $\cZ$ away from the minimum so that $\cZ^2 = \cX^2 + \cY^2 = (\nu^2 + \omega^2)/\lambda^2$. The potential evaluated at this shifted position is
\be
 V - V_0 = \frac{\nu^4}{4\lambda^2}
 - \frac{\nu^2}{2\lambda^2} (\nu^2 + \omega^2)
 + \frac{1}{4\lambda^2} (\nu^2 + \omega^2)^2
 = \frac{\omega^4}{4\lambda^2}  \,,
\ee
and so the total energy density is
\be
 \varepsilon = \frac12 \, (\cX^2 + \cY^2) \, \omega^2 + V
 = V_0 + \frac{\omega^2 \nu^2}{2\lambda^2} + \frac{3 \, \omega^4 }{4\lambda^2}\,.
\ee
The conserved `angular momentum' of this motion is similarly given by
\be
 J^0 = \cZ^2 \dot \vartheta = (\nu^2 + \omega^2)\frac{\omega}{\lambda^2} \,,
\ee
where in this section we temporarily use dots to denote time derivatives.

We next calculate this same energy density and conserved charge in the effective field theory, to see how it arises there. For the slowly-rolling field configuration in the low-energy theory, we solve $\Box \e = 0$ using the leading-order solution $\e = f \omega t$, for which $\dot \e = \partial_t \e = f \, \omega$ is a constant. Evaluating $\mathcal{L}_{\rm eff}$ at this solution then gives
\be \label{leffatsoln}
  \frac{\mathcal{L}_{\rm eff}}{\sqrt{-g}} = - V_0 +
  \frac{\dot \e^2}{ 2}
  + \frac{\lambda^2 \dot \e^{4}}{4 \nu^{4}}
   + \mathcal{O}\left( \frac{1}{\nu^{6}} \right) \,.
\ee

To find the energy of this solution we compute the effective Hamiltonian density for this system, which is
\be \label{heffexpdef}
  \mathcal{H}_{\rm eff} = \pi_{\rm eff} \, \dot \e - \cL_{\rm eff} \,,
\ee
where the canonical momentum is defined by
\be
  \frac{\pi_{\rm eff}}{\sqrt{-g}} := \frac{1}{\sqrt{-g}} \; \frac{\delta S_{\rm eff}}{\delta \dot \e} = \dot \e  + \frac{\lambda^2 \dot \e^{3}}{ \nu^{4}}
   + \mathcal{O}\left( \frac{1}{\nu^{6}} \right) \,.
\ee
Using this the Hamiltonian density becomes
\be \label{heffexp}
 \frac{\mathcal{H}_{\rm eff}}{\sqrt{-g}}  = V_0+
   \frac{\dot \e^2}{ 2}
  + \frac{3\,\lambda^2 \dot \e^{4}}{4 \nu^{4}}
  + \mathcal{O}\left( \frac{1}{\nu^{6}} \right)  \,,
\ee
and so the energy density obtained by evaluating this at $\dot \e = f \omega $ is
\ba \label{heffsoln}
  \varepsilon_{\rm eff} &=& V_0 +
  \frac{\omega^{2}f^{2}}{2}
  + \frac{3\,\lambda^2 \omega^{4} f^4}{4\nu^4}
  + \cdots \nn\\
  &=& V_0 +
  \frac{\omega^{2}\nu^{2}}{2\lambda^2}
  + \frac{3\,\omega^{4}}{4\lambda^2}
  + \mathcal{O}\left(\frac{1}{\nu^{6}} \right) \,,
\ea
where the second equality uses $f = \nu/\lambda$ to secure agreement of the $\omega^2$ term with its counterpart in the exact result obtained from the full theory. Once this is done the $\omega^4$ term also agrees.

The conserved charge is similarly given by
\be
 J^0_{\rm eff} = \cN \left[ 1 + \frac{\lambda^2\dot \e^2}{\nu^4}  \right] \dot \e = \cN \left( 1 + \frac{\omega^2}{\nu^2}  \right) \frac{\nu \omega}{\lambda} \,,
\ee
which again agrees with the full theory given the normalization $\cN = f = \nu/\lambda$. These examples show how it is the new $\cO(1/\nu^4)$ effective interactions that bring the low-energy theory the news of the energy shift that centrifugal motion brings for slow motion in the full theory.

\subsection{The cowboy hat}

An instructive variation on the previous example is the case of an $O(2)$-breaking potential, wherein the circular trough is deformed to an ellipse.\footnote{And so with the sombrero shape deforming into a cowboy hat, hence the name.} This deformation is simply achieved by deforming the potential of eq.~\pref{eq:mexicanV} to
\be \label{eq:cowboyV}
 V(\cX, \cY) = V_0 + \frac{1}{4} \left( \lambda_x \cX^2 + \lambda_y  \cY^2 - v^2 \right)^2 \,,
\ee
which reduces to the case considered above if $\lambda_x = \lambda_y = \lambda$ and $v^2 = \nu^2/\lambda$.

The trough minimizing $V$ in this case is the ellipse
\be \label{eq:ellipttrough}
 \lambda_x \cX^2 + \lambda_y \cY^2 = \cZ^2 \Bigl( \lambda + \lambda' \cos 2\vartheta \Bigr) = v^2 \,,
\ee
where $\lambda := \frac12(\lambda_x + \lambda_y)$, $\lambda' := \frac12 (\lambda_x - \lambda_y)$ and, as before, $\cX + i \, \cY := \cZ \, e^{i\vartheta}$. The mass matrix along the trough has eigenvalues $M_-^2 = 0$ and $M_+^2 = m^2$, with
\bea \label{eq:msqellipse}
 m^2 &=& 2 \Bigl( \lambda_x^2 \cX^2 + \lambda_y^2 \cY^2 \Bigr)
 = 2 \cZ^2 \Bigl( \lambda^2 + {\lambda'}^2 + 2 \lambda \lambda' \cos 2 \vartheta \Bigr) \nn\\
 &=& \frac{2 v^2 ( \lambda^2 + {\lambda'}^2 + 2 \lambda \lambda' \cos 2 \vartheta )}{\lambda + \lambda' \cos 2\vartheta }  \approx 2 \lambda v^2 \left[ 1 + \frac{\lambda'}{\lambda} \, \cos 2 \vartheta + \cO \left({\lambda'}^2 \right) \right] \,.
\eea
The corresponding eigenvectors are also the tangent and normal to the trough, and are given by
\be
 \vec t = \vec e_\e = \frac{\sqrt2}{m} \left( \begin{array}{c}
 - \lambda_y \cY \\ \lambda_x \cX \\ \end{array} \right)
 \quad \hbox{and} \quad
 \vec n = \vec e_h = -\frac{\sqrt2}{m} \left( \begin{array}{c}
 \lambda_x \cX \\ \lambda_y \cY \\ \end{array} \right) \,.
\ee
Notice in particular that if $\lambda' \ne 0$ then $\dot m \ne 0$ along the trough's bottom.

The trough's radius of curvature is given by $\kappa = -m^2/V_{,ijk} t^i t^j n^k$, where the required third derivatives now are
\be
 V_{\cX\cX\cX} = 6 \lambda_x^2 \cX \,, \quad V_{\cX\cX\cY} = 2 \lambda_x \lambda_y \cY \,, \quad V_{\cX\cY\cY} = 2 \lambda_x \lambda_y \cX \,, \quad V_{\cY\cY\cY} = 6 \lambda_y^2 \cY \,.
\ee
After some algebra this gives
\be \label{eq:kappaellipse}
 \kappa = \frac{m^3}{2\sqrt2 \, \lambda_x \lambda_y v^2 }
  \approx \frac{v}{\sqrt\lambda} \left[ 1 + \frac{3\lambda'}{2\lambda} \, \cos 2 \vartheta + \cO \left({\lambda'}^2 \right) \right] \,,
\ee
which reduces to the mexican-hat expression, eq.~\pref{eq:kappa_m_reln}, when $\lambda' = \frac12(\lambda_x - \lambda_y) \to 0$ and $v = \nu/\sqrt\lambda$. From this we see that $\dot \kappa$ does not vanish along the trough bottom because $\dot m$ does not, and that
\be \label{eq:dkvsdm}
 \frac{\dot \kappa}{\kappa} = \frac{3\, \dot m}{m} \,.
\ee

The low-energy effective Lagrangian derived for physics near the trough's bottom again satisfies $U = V_0$ and so $\dot U = \ddot U = 0$, and because of this variables can be found for which simultaneously $G_{\rm eff} = 1$ and $V_{\rm eff} = V_0$. The leading contribution to the effective theory in these variables is therefore again eq.~\pref{eq:EFTM2_mexhat}:
\bea
 \label{eq:EFTM2_cowhat}
 \frac{\cL_{\rm eff}}{\sqrt{-g}}
 &=& - V_0 - \frac12 \, \partial_\mu \e
 \, \partial^\mu \e + \frac{1}{2m^2 \kappa^2} (\partial_\mu \e \partial^\mu\e)^2
 + \mathcal{O}\left(\frac{1}{m^{4}}, \e^5\right) \nn\\
 &=& - V_0  - \frac12 \, \partial_\mu \e
 \, \partial^\mu \e + \frac{4\lambda^2_x \lambda_y^2 v^4}{m^8} (\partial_\mu \e \partial^\mu\e)^2
 + \mathcal{O}\left(\frac{1}{m^{4}} , \e^5\right)  \,.
\eea
where the second line uses the above calculations of $m$ and $\kappa$, eqs.~\pref{eq:msqellipse} and \pref{eq:kappaellipse}, with $\vartheta \to \vartheta_0$ now regarded as the point about which $\cL_{\rm eff}$ is expanded.

A potential puzzle with this result is that to within the accuracy it is written it shares the shift symmetry, $\ell \to \ell + c$, of the circular case, which implies the existence of the conserved current to within the same level of accuracy
\be \label{eq:JeffMH}
 J^\mu_{\rm eff} = - \cN \left[ 1 - \frac{16\lambda^2_x \lambda_y^2 v^4}{m^8} (\partial \e)^2 \right] \partial^\mu \e + \cdots \,.
\ee
Conservation of this current should only be an artefact of stopping at $\cO(\ell^4)$ when writing the effective Lagrangian, since there is no reason why $\cH_{\rm eff}(\ell)$ should be completely $\ell$-independent. Assuming there to be a term in $\cH_{\rm eff}(\ell)$ of order $\ell^2/m^2 \kappa^4$ we are led to expect failure of current conservation to first arise at the 4-derivative level:
\be
 \partial_\mu J^\mu \propto \left( \frac{\lambda' \cN f^5}{m^2 \kappa^4} \right) \omega^4  = \left( \frac{\lambda' }{\lambda^2} \right) \omega^4 \,.
\ee

The potential puzzle arises once we ask at what level the previously conserved current, $J^\mu$, fails to be conserved in the full theory. This is governed by the $\vartheta$ field equation, which states
\be \label{eq:divJ}
 \partial_\mu J^\mu = -\partial_\mu \Bigl( \cZ^2 \partial^\mu \vartheta \Bigr) = \lambda' \cZ^2 \sin 2\vartheta \Bigl[ \cZ^2 (\lambda + \lambda' \cos 2 \vartheta ) - v^2 \Bigr] \,.
\ee
Notice in particular that the right-hand side vanishes when evaluated along the trough's bottom, which is where eq.~\pref{eq:ellipttrough} is satisfied. Suppose now we take $\lambda' \ll \lambda$ and perturb about the slowly rolling solution of the mexican hat. Then eq.~\pref{eq:divJ} can be linearized in $\lambda'$ and simplifies to
\bea \label{eq:divJapprox}
 \partial_\mu J^\mu &\simeq& \lambda \lambda' \cZ^2 \left( \cZ^2 - \frac{\nu^2}{\lambda^2} \right) \sin 2\vartheta \nn\\
% &\simeq& \frac{\lambda' \omega^2}{\lambda} \left( \frac{\nu^2 + \omega^2}{\lambda^2} \right) \sin 2\vartheta \nn\\
 &\simeq&  \frac{\lambda' \omega^2 \nu^2}{\lambda^3} \sin 2 \vartheta +\cO(\omega^4) \,,
\eea
where $\vartheta \simeq \omega t$. This seems to have the dependence on $f = \nu/\lambda$ and $\omega$ that would come from the contribution to $\partial_\mu J^\mu_{\rm eff}$ of a term like $\lambda' \e^2 (\partial \e)^2$ in the effective Lagrangian.

However, we used the freedom to redefine fields to set $G_{\rm eff} = 1$ in order to find a current conserved up to order $\omega^4$ in the effective theory, so we should see if we can also do so in the full theory. To this end imagine redefining the low-energy angular variable,
\be
 \hat \vartheta := \vartheta + \frac{a}{4} \, \sin 2 \vartheta \,,
\ee
and define
\be
 \hat J^\mu := - \Bigl( \cZ^2 \partial^\mu \hat \vartheta \Bigr)
 \simeq - \cZ^2 \partial^\mu \vartheta \left( 1 + \frac{a}{2} \, \cos 2 \vartheta \right) \,,
\ee
so
\ba
 \partial_\mu \hat J^\mu &\simeq& - \partial_\mu \Bigl( \cZ^2 \partial^\mu \vartheta \Bigr) \left( 1 + \frac{a}{2} \, \cos 2 \vartheta \right) + {a \cZ^2} \, \sin 2\vartheta \; \partial^\mu \vartheta \partial_\mu \vartheta \nn\\
 &\simeq& \left( \frac{\lambda'}{\lambda} - a \right) \frac{\omega^2 \nu^2}{\lambda^2} \; \sin 2 \vartheta \,,
\ea
where the last approximate equality works to linear order in $\lambda'$, assumes $a = \cO(\lambda')$ and linearizes as before about the $\lambda' = 0$ solution $\vartheta \simeq \omega t$ and $\cZ^2 \simeq (\nu^2 + \omega^2)/\lambda^2$. We see that the choice $a = \lambda'/\lambda$ defines a current, $\hat J^\mu$, whose non-conservation first arises at $\cO(\omega^4)$ when linearized in $\lambda'$, just as was the case for the low-energy effective theory.

\section{Applications to inflationary models}

We next consider non-flat troughs and ask whether and how the effective analysis presented here can be used to describe the dynamics of multi-field inflationary models. Our goal is twofold. First, we provide simple criteria for when a given multi-field model with a trough is well-described by our effective Lagrangian. Second, we show how our effective action provides a simple shortcut for calculating inflationary observables for multi-field models using well-known results for single-field models.

Our starting point is the effective field theory computed out to quartic order in $\ell$ and up to order $1/m^2$: eqs.~\pref{eq:leff_deriv_exp}, \pref{eq:Heff} and \pref{eq:vtrunc2can}, which we repeat here for convenience (with $j=0$):
\be \label{eq:leff_deriv_exp2}
 \frac{\cL_{\rm eff}}{\sqrt{-g}} = - V_{\rm eff}(\varphi, \e) - \frac12 \, (\partial_\mu \e \, \partial^\mu \e) +  \cH_{\rm eff}(\varphi, \e) (\partial_\mu \e \, \partial^\mu \e)^2 + \cdots \,,
\ee
with
\be \label{eq:Heff2}
 \cH_{\rm eff}(\varphi,\e) \simeq \frac{1}{2\kappa^2m^2} + \cO(\ell) \,,
\ee
and
\ba \label{eq:veff2}
 V_{\rm eff}(\varphi, \e)  &\simeq& U(\varphi + \e) + \delta V (\varphi, \ell) \nn\\
 &\simeq& U(\varphi) + U' (\varphi) \, \ell + \frac12 \, \mu^2_{\rm eff}(\varphi) \, \ell^2 + \frac{1}{3!} \, g_{\rm eff}(\varphi) \, \ell^3 + \frac{1}{4!} \, \lambda_{\rm eff} (\varphi) \, \ell^4 + \cdots \,,
\ea
with
\ba \label{eq:gefflambdaeff}
 &&\qquad\qquad \mu^2_{\rm eff} \simeq  U'' - \frac{ U^{'2}}{\kappa^2 m^2}
 \,, \qquad
 g_{\rm eff} \simeq U''' - \frac{2 U^{'2} m'}{ \kappa^2 m^3} \,, \nn\\
 &&\lambda_{\rm eff} \simeq U'''' - \frac{3 U' \kappa' }{\kappa^3} + \frac{4 U' U'' }{\kappa^2 m^2} \left( \frac{4 m'}{m} + \frac{3 \kappa'}{\kappa} \right)  - \frac{16 U^{'2} \kappa' m'}{\kappa^3 m^3} - \frac{6 U' U'''}{\kappa^2 m^2} \\
 && \qquad\qquad\qquad\qquad - \frac{2U'^2}{\kappa^2 m^2}\frac{m'^2}{m^2} + \frac{ U^{'2}}{\kappa^4 m^2}  \left(6+3 \lambda_{nnn}-11 \kappa^{'2} + 4 \kappa \kappa'' \right) \,, \nn
\ea
and so on. In this section only we switch to using primes to denote differentiation with respect to trough arc length: {\em e.g.} $\kappa' := \exd \kappa/\exd \sigma = (\exd \kappa/\exd \varphi) \exd \varphi/\exd \sigma$, and reserve over-dots for FRW time derivatives.

For cosmological applications we expect this kind of single-field description to apply whenever all time-dependence scales are smaller than the parameters $m$, $\kappa$, $\rho$ and so on. In particular, we do not expect this type of single-field model to capture the `quasi-single-field models' \cite{QSFinf} that satisfy $m \simeq H$\footnote{See however \cite{PS} for an interesting case study of the regimes that interpolate between those of \cite{QSFinf} and those of the single field effective description.}.

\subsection{Basic inflationary observables}

Suppose we now imagine $\ell$ to be the inflaton, with inflation driven by a slow roll along the trough's bottom. Imagine also choosing $\varphi$ so that $\ell = 0$ denotes the epoch of horizon exit of some reference comoving scale. In this case the action, \pref{eq:leff_deriv_exp2}, is equivalent to a single-field inflationary model, with scalar potential $V_{\rm eff}$ and non-minimal Lagrangian function \cite{k-inf} $P(X,\ell) = - V_{\rm eff}(\ell) + X + 4 \cH_{\rm eff} X^2$. We may therefore use standard single-field formulae for a $P(X, \ell)$ theory \cite{PXphiInf0, PXphiInf} when making inflationary predictions.

In particular, it is clear that the presence of both $\cH_{\rm eff}$ and $\delta V$ imply the inflationary slow-roll differs from a naive analysis that simply uses $U$ as the inflationary potential along the trough's bottom. These differences track the influence of the heavy second field on the low-energy inflationary dynamics. For instance, the slow-roll parameters defined by the scalar potential at horizon exit are
\ba
 \epsilon_\ssV &:=& \frac12 \left( \frac{M_p V_{\rm eff}'}{V_{\rm eff}} \right)^2_{\ell = 0} \simeq \frac12 \left( \frac{M_p \, U'}{U} \right)^2 := \epsilon_\ssU \nn\\
 \eta_\ssV &:=&  \left( \frac{M_p^2 V_{\rm eff}''}{V_{\rm eff}} \right)_{\ell = 0} \simeq \frac{M_p^2 U''}{U} - \frac{M_p^2 U^{'2}}{\kappa^2 m^2 U} := \eta_\ssU - 2 \epsilon_\ssU  \left( \frac{ U}{\kappa^2 m^2} \right) \,,
\ea
showing $\epsilon_\ssV$ agrees with $\epsilon_\ssU$ while $\eta_\ssV$ and $\eta_\ssU$ can differ. Notice that $\eta_\ssV < \eta_\ssU$ because $U > 0$ during inflation, and (if $\epsilon_\ssU$ and $\eta_\ssU$ are comparable in size) the correction is sizeable if $U$ is comparable to $\kappa^2 m^2$.

Furthermore, the presence of $\cH_{\rm eff}$ in $P(X,\ell)$ implies an effective `speed of sound',
\be \label{eq:cs}
 c_s^2 := \frac{P_{,\,X}}{P_{,\,X} + 2X P_{,\,XX}}
 = \frac{1 + 8 \cH_{\rm eff} X}{1 + 24 \cH_{\rm eff} X}
 \simeq 1 - 16 \, \cH_{\rm eff} X
 \simeq 1 - \frac{8 X}{\kappa^2 m^2} \,,
\ee
which is smaller than unity because $X = \frac12 \, \dot \ell^2 > 0$. In terms of the trough and slow roll parameters, using $3H \dot \ell \simeq - U'$ and $3 M_p^2 H^2 \simeq U$ we find that
\be
 c_s^2 \simeq 1 - \frac{8\epsilon_\ssU U}{3\kappa^2 m^2}
 \simeq 1 - \frac{8\epsilon_\ssU H^2M_{pl}^2}{\kappa^2 m^2} \,.
\ee
The Hubble scale as a function of the rolling field $\ell$ is \cite{PXphiInf}
\be
  H^2 = \left( \frac{2 X P_{,\,X} - P}{3M_p^2} \right)_{\ell = 0}
  %\simeq \frac{V_{\rm eff} + X + 12 \cH_{\rm eff} X^2}{3 M_p^2}
  \simeq \frac{1}{3M_p^2} \left[ V_{\rm eff} + X \left( 1 + \frac{6 X}{\kappa^2 m^2} \right) \right] \,,
\ee
whose time-dependence governs the slow-evolution parameters relevant to basic inflationary observables. We imagine this evolution to be slow because of the shallowness of the trough bottom, and so take $X/V_{\rm eff} \ll 1$. We then follow the small corrections from slow roll arising from the effective interactions induced by the heavy field.

The relevant first rate of change of $H$ is given by
\ba
 \epsilon &:=& - \frac{\dot H}{H^2} = \frac{X P_{,\,X}}{M_p^2 H^2} = \frac{3(X + 8 \cH_{\rm eff} X^2)}{V_{\rm eff} + X + 12 \cH_{\rm eff} X^2} \nn \\
 &\simeq& \frac{3X}{V_{\rm eff}} \left( 1 - \frac{X}{V_{\rm eff}}  + 8 \cH_{\rm eff} X  + \cdots \right) \,,
\ea
which may be inverted to give $X$ as a function of $\epsilon$:
\be
 \frac{3 X}{V_{\rm eff}} \simeq \epsilon + \frac{\epsilon^2}{3} \Bigl( 1 - 8 \cH_{\rm eff} V_{\rm eff} \Bigr) \,,
\ee
where to leading order in the slow-roll approximation we would have had $(3X/V_{\rm eff})_{\ell =0} \simeq \epsilon_\ssV = \epsilon_\ssU$. 

A second useful slow roll parameter is given by $\eta := \frac{\dot \epsilon}{\epsilon H}$ which is related to the parameters $\eta_\ssV$ and $\epsilon$ above, and can be rewritten to leading order as \cite{PXphiInf}
\ba
 \eta &=& \frac{\dot \epsilon}{\epsilon H}  = -2\eta_\ssV +4\epsilon \nn \\
  &\simeq& -2\eta_\ssU + \frac{4 U}{\kappa^2 m^2} + \frac{12 X}{V_{\rm eff}} \left( 1 - \frac{X}{V_{\rm eff}} + 8 \cH_{\rm eff} X \right) \,.
\ea
Furthermore, we have $s := \dot c_s/(c_s H) \simeq 0$, which vanishes in our case as $\cH_{\rm eff}$ is $\ell$-independent only as a consequence of our having expanded $\cL_{\rm eff}$ to quartic order in fields. The effective theory obtained to all orders in fields (but to quartic order in derivatives) would in general exhibit a varying speed of sound along the trough.

The utility of these expressions lies in the following general results for properties of the spectra of primordial scalar and tensor fluctuations \cite{PXphiInf}:
\ba
 P_\zeta(k) &\simeq& \left( \frac{H^2}{8\pi^2 c_s \epsilon \, M_p^2} \right)_{k}\nn\\
 P_h(k) &\simeq& \left( \frac{2 H^2}{\pi^2 \, M_p^2} \right)_{k} \,,
\ea
where $(\cdots)_k$ denotes evaluation at horizon exit for mode $k$. These expressions are valid so long as the parameters $\epsilon, \eta$, and in particular $c_s$ vary slowly enough\footnote{We must go beyond quartic order in $\ell$ when the speed of sound varies more rapidly, while remaining within the effective theory and preserving slow roll. (See also \cite{NoTrunc, OtherTroughs1}.)} (to quartic order in fields, the latter is satisfied by default). Of particular observational interest are the following expressions\footnote{The $k$ dependence of the spectral indices and the tensor to scalar ratio can be obtained (accurate up to terms that are second order in the slow roll parameters) by simply evaluating the first order expressions at the instant of horizon crossing.}
\ba
 n_s(k) -1 &:=& \frac{\exd P_\zeta}{\exd \ln k} \simeq \left(-2\epsilon - \eta - s\right)_k \nn\\
 n_{\ssT}(k) &:=& \frac{\exd P_h}{\exd \ln k} \simeq -(2\epsilon)_k \nn\\
 \hbox{and} \quad r &:=& \frac{P_h}{P_\zeta} \simeq (16 c_s \epsilon)_k \simeq -8 (c_s n_\ssT)_k \,.
\ea
We note that were we to compute the effective theory to all orders in fields (alternatively, recompute the effective expansion to quartic order at each instant the mode of interest $k$ crosses the horizon), we could infer from the above the presence of features in the scalar spectrum generated by a varying speed of sound. By current observational constraints \cite{WMAP}: $n_s = 0.968 \pm 0.012$ and $r < 0.2$ with no significant evidence for any spectral running.

\subsection{Nongaussianity}

We note that in addition to gravitational non-linearities, there are three sources of nonlinearity in the action \pref{eq:leff_deriv_exp2} that can give rise to primordial non-gaussianity: the cubic scalar potential term with coupling $g_{\rm eff}$; the quartic scalar potential term with coupling $\lambda_{\rm eff}$; and the quartic derivative interaction with coefficient $\cH_{\rm eff}$. General bispectrum and trispectrum predictions for the multi-scalar trough model are straightforwardly obtained by combining the above expressions for these couplings with existing single-field calculations \cite{Malda, PXphiInf0, PXphiInf}, whose validity relies on the condition that $c_s$ varies sufficiently slowly ($\dot c_s \ll c_s H$).

For example, for the primordial bi-spectrum we quote these as
\be
 \langle \zeta(k_1) \zeta(k_2) \zeta(k_3) \rangle = \frac{(2\pi)^7 [ P_\zeta(K) ]^2}{\prod_i k_i^3} \, \delta^3(\vec k_1 + \vec k_2 + \vec k_3) \Bigl( \cA_\lambda + \cA_c + \cA_o + \cA_\epsilon + \cA_\eta + \cA_s \Bigr) \,,
\ee
where $K := k_1 + k_2 + k_3$ and the coefficients $\cA_i$ are given by
\ba
 && \quad \cA_\lambda = \left\{ \frac{1}{c_s^2} - 1 - \frac{\lambda}{\Sigma} \Bigl[ 2 - (3 - 2 \gamma) \, l \Bigr] \right\}_K \bar \cA_\lambda \,, \qquad
 \cA_c = \left( \frac{1}{c_s^2} - 1 \right)_K \bar \cA_c \,, \\
 \cA_o &=& \left( \frac{1}{c_s^2} - 1 - \frac{2\lambda}{\Sigma} \right)_K \Bigl( \epsilon F_{\lambda \epsilon} + \eta F_{\lambda \eta} + s F_{\lambda s} \Bigr) + \left( \frac{1}{c_s^2} - 1 \right)_K \Bigl( \epsilon F_{c \epsilon} + \eta F_{c \eta} + s F_{c s} \Bigr) \,, \nn\\
 && \qquad\qquad
 \cA_\epsilon = \epsilon \bar \cA_\epsilon \,, \qquad
 \cA_\eta = \eta \bar \cA_\eta \,, \quad \hbox{and} \quad
 \cA_s = s F_s \,, \nn
\ea
where $\gamma = 0.577...$ is the Euler-Mascheroni constant and the $k_i$-dependent functions, $\bar \cA_\lambda$, $\bar \cA_c$, $\cA_\epsilon$, $\cA_\eta$, $F_s$, $F_{\lambda \epsilon}$, $F_{\lambda \eta}$, $F_{\lambda s}$, $F_{c\epsilon}$, $F_{c\eta}$ and $F_{cs}$ and $c_1$ are given explicitly in ref.~\cite{PXphiInf} --- {\em c.f.} eqs.~(4.44) through (4.49) and the Appendices of this reference. The new parameters $\lambda$, $\Sigma$ and $l$ are defined by
\ba
 \lambda &:=& X^2 P_{,\,XX} + \frac{2X^3}{3} \, P_{,\,XXX}
 \simeq 8 \cH_{\rm eff} X^2 \nn\\
 \Sigma &:=& X P_{,\,X} + 2 X^2 P_{,\,XX} \simeq X + 24 \cH_{\rm eff} X^2 \,,
\ea
and
\be
 l := \frac{\dot \lambda}{\lambda H} \,.
\ee
It is clearly a great simplification to be able to use standard single-field results such as these to extract predictions for the broad class of multi-scalar models to which our effective theory applies.

\subsection{Relationship with the EFT of Cheung et al.}

For the simple effective theory we have derived here, there is a direct relation with the effective expansion of \cite{SFeft2}, where it was shown that the most general form for the action for the adiabatic mode (for example, in unitary gauge\footnote{In this gauge, spacetime has been foliated in such a manner as to have gauged away the inflaton field fluctuation.}) can be parametrized as:
\begin{eqnarray*}
 S &=& \int~d^4x\sqrt{-g} \left[\frac{M_p^2}{2}R + M_p^2g^{00}\dot H - M_p^2(3H^2 + \dot H) + \frac{1}{2!}M^4_2(t)(g^{00}+1)^2 +\frac{1}{3!}M^4_3(t)(g^{00}+1)^3 \right. \\
 && \qquad\qquad \left. -\frac{1}{2}\bar M^4_1(t)(g^{00}+1)\delta K^\mu_\mu -\frac{1}{2}\bar M^4_2(t)\delta K^{\mu~2}_{~\mu} - \frac{1}{2}\bar M^4_3(t)\delta K^{\mu}_{~\nu}K^{\nu}_{~\mu}+ \cdots \right] \,,
\end{eqnarray*}
where $\delta K^\mu_\nu$ is the variation of the extrinsic curvature of the constant time hypersurfaces with respect to the background FRW metric. The first three terms in the expansion above ensure tadpole cancellation.

Were we to minimally couple a scalar field with the Lagrangian density $\mathcal L = P(X,\ell)$ to gravity and expand the action around a background homogeneous solution $\ell_0$, we would deduce the co-efficients $\bar M_n \equiv 0$ and
\be
 M^4_n(t) = (-1)^nX^n\frac{\partial^n P}{\partial X^n}\bigg|_{\ell_0}\,,
\ee
and so $M^4_2 \simeq 8 \cH_{\rm eff} X^2$ to the order to which we work in the above. Evidently, our effective expansion to quartic order furnishes the leading $M_n$ co-efficients of the effective theory of \cite{SFeft2}. Proceeding to higher orders in the derivative and field expansion would successively yield the higher order $M^4_n$ coefficients.

\section{Conclusions}
\label{sec:conclusions}

To summarize, in this paper we show how to identify covariantly the effective theory that captures the low-energy limit of a multi-scalar system slowly evolving along a shallow trough in the scalar potential. We illustrate this for a simple two-scalar system by explicitly integrating out the heavy field to obtain the single-scalar low-energy effective theory, \pref{eq:leff_deriv_exp}, with effective couplings,
\pref{eq:Heff} and \pref{eq:vtrunc2can}.

We give explicit covariant expressions for the scales that must be large in order for the truncation approximation to be valid, and see why it is not sufficient for the heavier field merely to be heavy. In particular, it is also necessary for the trough not to be too strongly curved, and for the heavy mass and trough curvature not to vary too strongly along the trough's bottom. Because these criteria are covariant under field redefinitions, they can be computed for specific theories using any convenient field parametrization.

By comparing the effective theory with the full theory in several simple (non-gravitational) examples, we show that $\cH_{\rm eff}$ precisely captures the centrifugal energy caused when slow motion along a curved trough forces the fields to climb a small distance up the trough walls.

Finally, we show how simply inflationary observables can be computed for multi-field models whenever such an effective description applies, by using well-known predictions for single-field models with a quartic effective scalar potential. This extends these single-field predictions by showing that they also apply to a broad class of multi-field models, and identifies which features of the multi-field potential are relevant to observations. In particular, we find that the effective theory contains an effective higher-derivative coupling, $\cH_{\rm eff}$, that contributes to cosmological observables as a contribution to the effective speed of sound of the primodial cosmological fluid.

\section*{Acknowledgements}

We thank Ana Ach\'ucarro, Ed Copeland, Jinn-Ouk Gong, Richard Holman, Gonzalo Palma, Sarah Shandera and Alex Vikman for helpful discussions. The research leading to these results has received funding from the [European Union] Seventh Framework Programme [FP7-People-2010-IRSES] under grant agreement n°269217. Our work was supported in part by funds from the Natural Sciences and Engineering Research Council (NSERC) of Canada. Research at the Perimeter Institute is supported in part by the Government of Canada through Industry Canada, and by the Province of Ontario through the Ministry of Research and Information (MRI). M.W.H. was supported at the University of Mississippi by NSF Grant PHY-1055103. S.P. was supported at the \'{E}cole Polytechnique by funds from CEFIPRA/IFCPAR project 4104-2 and ERC Advanced Investigator Grants no. 226371 ``Mass Hierarchy and Particle Physics at the TeV Scale'' (MassTeV), and at CERN by a Marie Curie Intra-European Fellowship of the European Community's 7'th Framework Programme under contract number PIEF-GA-2011-302817. S.P. wishes to thank the Perimeter Institute for hospitality during the initial stages of this work, and the organizers of COSMO11 at Porto, where the stimulating discussions that led to this investigation commenced.

\appendix

\section{Covariant field expansions}
\label{app:covariantcharact}

As in the main text we consider the action describing $N$ mutually interacting scalar fields, $\phi^a$, written in the Einstein frame
\be
 S = - \int \exd^4 x \sqrt{-g} \; \left\{ V(\phi)
 + g^{\mu\nu} \left[ \frac12 \, \cG_{ab}(\phi) \,
 \partial_\mu \phi^a \partial_\nu \phi^b
 + \frac{1}{16 \pi G_\ssN} \, R_{\mu\nu} \right] + \cdots
 \right\} \,.
\ee

Our interest is in analyzing the theory in the immediate vicinity of a field-point, $\varphi^a$, in a way that emphasizes the invariance of physical predictions under field redefinitions. This section describes how to do so explicitly, but contains only standard material that the cognoscenti should feel free to skip \cite{CovExp}. Recall that under generic infinitesimal local field redefinitions the potential, $V(\phi)$, transforms as a scalar while the kinetic coefficient, $\cG_{ab}(\phi)$, transforms as a symmetric covariant tensor. That is, if $\phi^a \to \phi^a + \zeta^a(\phi)$, the potential, $V(\phi)$, transforms as $V \to V + V_{,\,a} \, \zeta^a$ and $\cG_{ab}(\phi)$ transforms so $\delta \cG_{ab} = \cG_{ab,\,c} \, \zeta^c + \cG_{ac} \, \zeta^c_{,\,b} + \cG_{cb} \, \zeta^c_{,\,a}$. Here commas denote differentiation ($V_{,\,a} := \partial V/\partial \varphi^a$ and so on).

The goal is to define a field expansion of the action, $\phi^a = \varphi^a + \delta \phi^a$, about a particular field point, $\varphi^a$, that makes manifest this target-space covariance. To this end imagine constructing the target-space geodesic, $\psi^a(\sigma)$, that connects $\varphi^a$ to $\phi^a$. It is useful to use as parameter target-space arc-length along the curve,
\be
 \exd \sigma^2 = \cG_{ab}(\phi)\; \exd \phi^a \exd \phi^b \,,
\ee
and so
\be
 \frac{D \dot \psi^a}{\exd \sigma}
 := \ddot \psi^a + \gamma^a_{bc} \,
 \dot \psi^b \, \dot \psi^c = 0 \,,
\ee
where over-dots denote $\exd/\exd \sigma$ and $\gamma^a_{bc}$ are the Christoffel symbols
\be
 \gamma^a_{bc} := \frac12 \, \cG^{ad} \Bigl( \cG_{bd,\,c} + \cG_{cd,\,b} - \cG_{bc,\,d} \Bigr) \,,
\ee
built from the target-space metric $\cG_{ab}$. Defining $\psi^a(0) = \varphi^a$ and $\psi^a (\epsilon) = \phi^a$, we consider the point $\phi^a$ to be near $\varphi^a$ to the extent that $\epsilon$ is small (compared with other scales in the problem).

The covariant formulation of the quantity $\delta \phi^a$ is then $\epsilon \, \xi^a$, where $\xi^a := \dot \psi^a(0)$ is the tangent to this geodesic evaluated at $\varphi^a$. Although in principle any family of curves could be used in this way to define $\delta \phi^a$, the utility of using geodesics can be seen once physical quantities are expanded in powers of $\xi^a$. For instance, expanding $\psi^a$ in powers of $\epsilon$ gives
\ba
 \psi^a(\epsilon) &=& \varphi^a + \epsilon \, \dot \psi^a(0)
 + \frac{\epsilon^2}{2} \, \ddot \psi^a(0)
 + \cO(\epsilon^3) \nn\\
 &=& \varphi^a + \epsilon \, \xi^a
 + \frac{\epsilon^2}{2} \Bigl[ \eta^a -
 \gamma^a_{bc} \, \xi^b \xi^c \Bigr] + \cO(\epsilon^3) \,,
\ea
where $\eta^a := [D \dot\psi^a/\exd \sigma](0)$ vanishes for a geodesic, and so on. Evaluating the scalar potential in the same way then gives
\ba \label{app_Vexpnv0}
 V\left[ \psi\left(\epsilon \right) \right]
 &=& V(\varphi) + \epsilon \, \Va \, \dot\psi^a(0)
 + \frac{\epsilon^2}{2} \, \Bigl[
 V_{,\,ab} \, \dot\psi^a(0) \, \dot\psi^b(0) +
 \Va \, \ddot \psi^a(0) \Bigr] + \cdots \nn\\
 &=&  V(\varphi) + \epsilon \, \Va \, \xi^a
 + \frac{\epsilon^2}{2} \, V_{;\,ab} \, \xi^a \xi^b +
 \frac{\epsilon^3}{3!} V_{;\,abc} \, \xi^a \xi^b \xi^c + \cdots \,,
\ea
where the last line repeatedly uses $D \dot \psi^a/\exd \sigma = 0$. This ensures all coefficients involve only tensor quantities; in this case covariant derivatives built from the target-space metric: $V_{;\,ab} := V_{,\,ab} - \gamma^c_{ab} \, V_{,\,c}$ and so on.

This simplicity arises because the expansion in powers of $\xi^a$ is equivalent to the use of Gaussian normal coordinates for the target space, for which the first derivative of the metric at $\varphi^a$ vanishes. To see this, evaluate the term cubic in $\epsilon$ in the scalar kinetic term using the expansions $\cG_{ab}[\psi(\epsilon)] = \cG_{ab}(\varphi) + \epsilon \, \cG_{ab,\,c}(\varphi) \, \xi^c + \cdots$ and $\partial_\mu \psi^a(\epsilon) = \epsilon \, \partial_\mu \xi^a -  \epsilon^2 \, \gamma^a_{bc} \, \xi^b \partial_\mu \xi^c + \cdots$ (where the last expansion specializes to constant background fields, $\partial_\mu \varphi^a = 0$), to get
\bea
 \cG_{ab}[\psi(\epsilon)] \, \partial_\mu \psi^a(\epsilon)
 \, \partial^\mu \psi^b(\epsilon) &=& \epsilon^2 \, \cG_{ab}(\varphi) \,
 \partial_\mu \xi^a  \partial^\mu \xi^b \nn\\ &&\qquad
 + \epsilon^3 \left[ \cG_{ab,\,c} \, \xi^c \partial_\mu \xi^a
 \partial^\mu \xi^b - 2 \cG_{ab} \gamma^a_{cd} \, \xi^c \partial_\mu \xi^d \partial^\mu \xi^b \right] + \cO(\epsilon^4) \nn\\
 &=& \epsilon^2 \, \cG_{ab}(\varphi) \, \partial_\mu \xi^a  \partial^\mu \xi^b + \cO(\epsilon^4) \,.
\eea
Continuing on to quartic order in the kinetic term gives the standard normal-coordinate expression \cite{GNCexp}
\be \label{app_eq:kinexpansion}
 \cG_{ab}[\psi(\epsilon)] \, \partial_\mu \psi^a(\epsilon)
 \, \partial^\mu \psi^b(\epsilon) = \left[ \epsilon^2 \, \cG_{ab}(\varphi) + \frac{\epsilon^4}{3} \cR_{acbd}(\varphi) \, \xi^c \xi^d \right] \partial_\mu \xi^a \partial^\mu \xi^b
 + \cO(\epsilon^5) \,,
\ee
where ${\cR^a}_{bcd}$ is the Riemann tensor built from $\cG_{ab}$.

In the special case where there are only two fields --- a case we explore in more detail below --- the curvature tensor is particularly simple:
\be \label{app_eq:2Dcurv}
 \cR_{abcd} =  \frac1{2\rho^2}  \left( \cG_{ad} \, \cG_{bc} - \cG_{ac} \, \cG_{bd}\right) \,,
\ee
characterized purely by a single function $\rho$, related to the Ricci scalar \footnote{Given the Weinberg curvature convention \cite{Wbg} in which we work,  the Ricci scalar is negative for a target space two-sphere of radius $\rho$.} as $\cR(\varphi) = \cR^{ab}_{~~ab} = -1/\rho^2$.

\section{Geometry of a trough}
\label{appssec:troughgeometry}

This appendix computes in detail the properties of $V$, assuming it has a trough-like shape for a system involving only $N=2$ fields. Following the main text, we do so first for the case of a perfectly level trough, and then for the general case where the trough is slightly tipped.

\subsection*{Perfectly level troughs}
\label{appsssec:leveltroughs}

As discussed in the main text, a potential with a level trough is one for which there is an equipotential curve, $\chi^a(\sigma)$, with two defining properties. Property $(i)$ states that $V_{,\,a}[\chi(\sigma)] = 0$ for all $\sigma$;
and property $(ii)$ states that all eigenvalues of the `mass' matrix ${\cA^a}_b := \cG^{ac} \, V_{;\,cb}$ are non-negative, and at least one eigenvalue is strictly positive.

To see what these conditions imply, imagine differentiating the condition $V_{,\,a}[\chi(\sigma)] = 0$ with respect to the arc-length, $\sigma$, along the trough. This gives
\be \label{appeq:leveltroughcondition}
 0 = \frac{D V_{,\,a}}{\exd \sigma} :=
 \frac{\exd V_{,\,a}}{\exd \sigma} - \gamma^c_{ab} \,
 V_{,\,c} \, \dot \chi^b = V_{;\,ab} \, \dot\chi^b \,.
\ee
Eq.~\pref{appeq:leveltroughcondition} states that (for all $\sigma$) the vector $\dot \chi^a$ is a zero eigenvector of the mass matrix: ${\cA^a}_b \, \dot \chi^b = 0$, showing that this matrix must have a zero eigenvalue.

Repeatedly differentiating with respect to $\sigma$ gives the further identities involving higher derivatives of $V$:
\be \label{appeq:3rdordereq}
 0 = \frac{D}{\exd \sigma} \Bigl( V_{;\,ab} \, \dot\chi^b
 \Bigr) = V_{;\,abc}  \, \dot\chi^b \dot\chi^c
 + V_{;\,ab} \, \frac{D \dot \chi^b}{\exd \sigma} \,,
\ee
and so on. In general the second term does not vanish, since the direction defined by the bottom of the trough need not be a geodesic of the target-space metric, $\cG_{ab}$.

The radius of curvature, $\kappa(\sigma)$, of the trough's valley floor is also easily computed in terms of derivatives of the potential $V$. This is because the tangent, $\dot \chi^a$, is a unit vector, $\cG_{ab} \, \dot \chi^a \, \dot\chi^b = 1$, provided the parameter, $\sigma$, along the curve is arc-length. This ensures that it must be orthogonal to its derivative along the curve:
\be
 0 = \frac{D}{\exd \sigma} \Bigl( \cG_{ab} \, \dot\chi^a
 \dot \chi^b \Bigr) = 2 \, \cG_{ab} \, \dot \chi^a \,
 \frac{D \dot \chi^b}{\exd \sigma} \,,
\ee
and so defining the unit vector in the $D \dot\chi^a/\exd \sigma$ direction by $n^a$, the radius of curvature of the trough's valley floor is defined by
\be \label{appeq:troughcurvaturedef}
 \frac{D \dot \chi^a}{\exd \sigma} := \frac{n^a}{\kappa(\sigma)} \,.
\ee
Notice that $\kappa \to \infty$ corresponds to the case of a `straight' trough, where the valley floor defines a target-space geodesic, $D \dot \chi^a/\exd \sigma = 0$.
With definition (\ref{appeq:troughcurvaturedef}), equation (\ref{appeq:3rdordereq}) becomes
\be \label{appeq:3rdordereq_n}
  V_{;\,ab} \, n^b / \kappa(\sigma) =  - V_{;\,abc}  \, \dot\chi^b \dot\chi^c \,.
\ee

When there are only two fields the same arguments just given also give a simple expression for $D \, n^a/\exd \sigma$. Since $n^a$ is a unit vector, $\cG_{ab} \, n^a n^b = 1$, its derivative along $\chi^a(\sigma)$ must be perpendicular to itself: $\cG_{ab} \, n^a (D \, n^b/\exd \sigma) = 0$, and so $D \, n^a/\exd \sigma$ must be parallel to $\dot\chi^a$. The coefficient can be found by differentiating the condition $\cG_{ab} \, n^a \dot\chi^b = 0$ along the curve, giving $\cG_{ab} (D \, n^a/\exd \sigma) \dot\chi^b = - \cG_{ab} n^a (D \dot\chi^b/\exd \sigma) = -1/\kappa(\sigma)$, and so
\be \label{appeq:Dndsvskappa}
 \frac{D \, n^a}{\exd \sigma} := - \frac{\dot\chi^a}{\kappa(\sigma)} \,.
\ee

When there are only two fields, let $m^{2}(\sigma)$ denote the strictly positive eigenvalue of the mass matrix that is required by condition $(ii)$ above, and let $e^{a}_+$ be the corresponding normalized eigenvector. Then we have
\be
 0 = V_{;\,ab} \, \dot \chi^a \, e^b_+ =
 m^2(\sigma) \, \cG_{ab} \, \dot \chi^a \, e^b_+ \,,
\ee
where the first equality holds because $\dot \chi^a$ is a zero eigenvector of the mass matrix. Since $m^{2}(\sigma)$ is strictly positive, it follows that $e^a_+$ is orthogonal to $\dot \chi^a$, and thus $e^a_+ = n^a$. Therefore,
\be \label{appeq:massiveevec}
 {\cA^a}_b \, n^b = m^2(\sigma) \, n^a \quad
 \hbox{or, equivalently} \quad V_{;\,ab} \, n^b
 = m^2(\sigma) \, \cG_{ab} \, n^b \,.
\ee

Equation (\ref{appeq:3rdordereq_n}) can now be further simplified to
\be \label{appeq:3rdordereq_n_simple}
  \frac{m^2(\sigma)}{\kappa(\sigma)} \, \cG_{ab} \, n^b =  - V_{;\,abc}  \, \dot\chi^b \dot\chi^c \,.
\ee
Contracting equation (\ref{appeq:3rdordereq_n_simple}) with $n^a$ yields a simple expression for the radius of curvature, $\kappa(\sigma)$, in terms of derivatives of $V$:
\be \label{appeq:kappasimple}
 \frac{1}{\kappa(\sigma)} =  - \frac{1}{m^2(\sigma)} \, V_{;\,abc} \, n^a \,
 \dot\chi^b \dot\chi^c \,.
\ee
On the other hand, contracting equation (\ref{appeq:3rdordereq_n_simple}) with $\dot \chi^a$ yields the identity
\be \label{appeq:vlll}
 V_{;\,abc} \, \dot\chi^a \, \dot\chi^b \, \dot\chi^c  = 0 \,.
\ee
We can obtain another interesting identity by differentiating equation (\ref{appeq:massiveevec}) with
respect to $\sigma$:
\be
 V_{;\,abc} \, n^b \, \dot\chi^c + V_{;\,ab} \, \frac{D n^b}{\exd \sigma} =
 \frac{\exd m^{2}(\sigma)}{\exd \sigma} \, \cG_{ab} \, n^b
 + m^{2}(\sigma) \, \cG_{ab} \, \frac{D n^b}{\exd \sigma} \,.
\ee
Using (\ref{appeq:Dndsvskappa}), and the fact that $\dot\chi^a$ is a zero eigenvector of the mass matrix, this becomes
\be \label{appeq:dm2id_1}
  V_{;\,abc} \, n^b \, \dot\chi^c  =
 \frac{\exd m^{2}(\sigma)}{\exd \sigma} \, \cG_{ab} \, n^b
 - \frac{m^{2}(\sigma)}{\kappa(\sigma)} \, \cG_{ab} \, \dot\chi^b \,.
\ee
Contracting equation (\ref{appeq:dm2id_1}) with $n^a$ yields the identity
\be \label{appeq:dm2id_2}
   V_{;\,abc} \, n^a \, n^b \, \dot\chi^c  =
 \frac{\exd m^{2}(\sigma)}{\exd \sigma} \,,
\ee
whereas contracting equation (\ref{appeq:dm2id_1}) with $\dot\chi^a$ yields equation (\ref{appeq:kappasimple}).

Now, writing the commutator of two covariant derivatives in terms of the
curvature,
\be \label{appeq:v3_curv}
 V_{;\,cba} - V_{;\,cab} =  \cR^{d}_{\phantom{d}cab}V_{,\,d} = 0 \,,
\ee
we find that
\be \label{appeq:v3_sym}
  V_{;\,abc} = V_{;\,(abc)} \,,
\ee
where $(\cdots)$ denotes the normalized completely symmetric product: $V_{;\,(a_1..a_n)} = \frac{1}{n!} \, (V_{;\,a_1..a_n} + \hbox{permutations})$. It is important to note that unlike the identity $V_{;\, ab} = V_{;\, (ab)}$, which holds everywhere, equation (\ref{appeq:v3_sym}) only holds along the curve $\chi^{a}(\sigma)$.

%It will turn out that higher covariant derivatives are not completely symmetric in their indices, so it is useful to define:
%%
%\be
%  V_{i_{1} \cdots i_{k}} :=
%  V_{;\,(a_{1}\cdots a_{k})} \, e^{a_{1}}_{i_{1}} \cdots e^{a_{k}}_{i_{k}} \,,
%\ee
%%
%where $e_{l}^{a} \equiv \dot\chi^a$, and
%$e_{h}^{a} \equiv n^{a}$, and the indices $i_{j}$ take on values $\e$ and
%$h$. The reason for this choice of notation will become clear in the future.

In summary, we have obtained formulas for all possible contractions of
third covariant derivatives of $V$ with $\dot\chi^a$ or $n^a$, in terms of $m$, $\dot{m}$, $\kappa$, and $V_{nnn} \equiv V_{;abc}n^{a}n^{b}n^{c}$.
This last quantity
measures how the walls of the trough deviate from a perfect parabola.

Now we will derive similar formulas for the fourth derivatives of the potential.
Differentiating equation (\ref{appeq:3rdordereq_n_simple}), and using equations (\ref{appeq:troughcurvaturedef}), (\ref{appeq:Dndsvskappa}), (\ref{appeq:v3_sym}), and (\ref{appeq:dm2id_1}), we obtain
\be
  V_{;\,abcd} \, \dot\chi^b \, \dot\chi^c \, \dot\chi^d =
  \frac{m^{2}}{\kappa}\left( \frac{\dot{\kappa}}{\kappa} - \frac{6\dot{m}}{m}\right)n_{a}
  + \frac{3m^{2}}{\kappa^{2}}\dot\chi_a \,.
\ee
Contracting this equation with $\dot\chi^a$, we obtain
\be \label{appeq:vllll}
  V_{;\,abcd} \, \dot\chi^a \, \dot\chi^b \, \dot\chi^c \, \dot\chi^d =
  \frac{3m^{2}}{\kappa^{2}} \,.
\ee
On the other hand, contracting with $n^a$ yields
\be \label{appeq:vnccc}
  V_{;\,abcd} \, n^a \, \dot\chi^b \, \dot\chi^c \, \dot\chi^d =
  \frac{m^{2}}{\kappa}\left( \frac{\dot{\kappa}}{\kappa} - \frac{6\dot{m}}{m} \right) \,.
\ee
Now, differentiating equation (\ref{appeq:dm2id_1}), and using equations
(\ref{appeq:troughcurvaturedef}), (\ref{appeq:Dndsvskappa}), and
(\ref{appeq:3rdordereq_n_simple}), we obtain
\be
  V_{;\,abcd} \, n^b \, \dot\chi^c \, \dot\chi^d =
  - \frac{1}{\kappa}V_{;\,abc} \, n^b \, n^c
  + 2(\dot{m}^{2} + m\ddot{m})n_a
  -\frac{2m^{2}}{\kappa^{2}}n_a
  +\frac{m^{2}}{\kappa}\left( \frac{\dot{\kappa}}{\kappa}
  - \frac{4\dot{m}}{m} \right)\dot\chi^a \,.
\ee
Contracting this equation with $\dot\chi^a$, and using equations
(\ref{appeq:dm2id_2}) and (\ref{appeq:v3_sym}) yields equation (\ref{appeq:vnccc}).
On the other hand, contracting with $n^a$ yields
\be \label{appeq:vnncc}
  V_{;\,abcd} \, n^a \, n^b \, \dot\chi^c \, \dot\chi^d +
  \frac{V_{nnn}}{\kappa} =
  2(\dot{m}^{2} + m\ddot{m}) - \frac{2m^{2}}{\kappa^{2}} \,.
\ee
Finally, another identity is obtained by differentiating $V_{nnn}$ along the trough, and using (\ref{appeq:dm2id_2}):
\be
V_{;\,abcd} \, n^a \, n^b \, n^c \, \dot\chi^d =
\frac{D}{\exd \sigma}(V_{nnn}) + \frac{6m\dot{m}}{k} \,.
\ee
Now, to find the symmetries of $V_{;\,abcd}$ we use
\ba \label{appeq:v4_curvcom}
  V_{;\,dcba} - V_{;\,dcab} &=& R^{e}_{\phantom{e}cab}V_{;\, de}
  + R^{e}_{\phantom{e}dab}V_{;\, ec} \,,
  \\
  \label{appeq:v4_curvcom2}
  V_{;\,dcba} - V_{;\,dbca} &=& R^{e}_{\phantom{e}dbc; \, a}V_{;\, e}
  + R^{e}_{\phantom{e}dbc}V_{;\, ea} \,.
\ea
Contracting (\ref{appeq:v4_curvcom}) with various combinations of $n^a$ and
$\dot\chi^a$, we obtain the relations
\ba
  \label{appeq:v4sym_1}
  V_{;\, abcd} \, \dot\chi^a \, \dot\chi^b \, \dot\chi^c \, n^d
  - V_{;\, abcd} \, \dot\chi^a \, \dot\chi^b \, n^c \, \dot\chi^d &=& 0 \,,
  \nn\\
  \label{appeq:v4sym_2}
  V_{;\, abcd} \, n^a \, n^b \, \dot\chi^c \, n^d
  - V_{;\, abcd} \, n^a \, n^b \, n^c \, \dot\chi^d &=& 0 \,,
  \\
  \label{appeq:v4sym_3}
  V_{;\, abcd} \, \dot\chi^a \, n^b \, \dot\chi^c \, n^d
  - V_{;\, abcd} \, \dot\chi^a \, n^b \, n^c \, \dot\chi^d &=&
  -m^{2} / 2 \rho^{2} \,,\nn
\ea
whereas contracting (\ref{appeq:v4_curvcom2}) with various combinations of $n^a$ and
$\dot\chi^a$ yields
\ba
  V_{;\, abcd} \, \dot\chi^d -
  V_{; \, (abc)d} \, \dot\chi^d &=& 0 \,,
  \nonumber
  \\
  \label{appeq:v4sym_4}
  V_{;\, abcd} \, n^a \dot\chi^b n^c n^d -
  V_{;\, abcd} \, n^a n^b \dot\chi^c n^d &=& 0 \,,
  \\
  V_{;\, abcd} \, \dot\chi^a \dot\chi^b n^c n^d -
  V_{; \, abcd} \dot\chi^a n^b \dot\chi^c n^d &=& -m^2/2\rho^2 \,.
  \nonumber
\ea
Note that the first equation of (\ref{appeq:v4sym_4}) may also be obtained by differentiating equation
(\ref{appeq:v3_sym}) along the trough.

Also, the first equation of (\ref{appeq:v4sym_1}), combined with the first equation of (\ref{appeq:v4sym_4}), implies
that when $V_{;\, abcd}$ is contracted with three $\dot\chi$'s and one $n$,
the ordering of indices does not matter. Similarly, the second equation of (\ref{appeq:v4sym_1}), combined with the
second equation of (\ref{appeq:v4sym_4}), implies that when $V_{;\, abcd}$ is contracted with three $n$'s and one $\dot\chi$,
the ordering of indices also does not matter.
Finally, the third equations of (\ref{appeq:v4sym_1}) and (\ref{appeq:v4sym_4}), in combination with
the first equation of (\ref{appeq:v4sym_4}), relate all possible contractions of
$V_{; \, abcd}$ with two $\dot\chi$'s and two $n$'s to each other.

In summary, it is possible to obtain formulas for all possible contractions of fourth covariant derivatives of $V$ with
$\dot\chi^a$ or $n^a$, in terms of $m$, $\dot{m}$, $\ddot{m}$, $\kappa$, $\dot{\kappa}$, $\rho$,
$V_{nnn}$, $\dot{V}_{nnn}$, and $V_{nnnn} \equiv V_{; \, abcd}n^a n^b n^c n^d$.

\subsection*{Tilted troughs}
\label{appsssec:tiltedtroughs}

We next turn to the situation where the trough is not completely level, but with derivatives along the trough assumed to be parametrically small rather than zero. Define the potential along the trough's bottom as the slowly varying function $U(\sigma) := V[ \chi(\sigma)]$, with the gradient of $V$ along the trough bottom given by $V_{,\,a}[\chi(\sigma)] = u_a(\sigma)$, where $u^2 = \cG^{ab} u_a u_b$ is much smaller than the other scales in the potential.

In this case, we again choose the curve $\chi^a(\sigma)$ to run along the bottom of the (no-longer level) trough, and so by construction its tangent is parallel to the potential gradient along the bottom:\footnote{Notice $\chi^a(\sigma)$ is {\em not} required to be a physical trajectory, in that it need not be a solution to the equations of motion.} $\dot\chi^a \propto \cG^{ab} u_b$. Contracting $V_{,\,a}$ with both $\dot\chi^a$ and $n^a$ then leads to the two equations
\be \label{appeq:udoteqn}
 u_a \, \dot\chi^a = \dot U \quad\hbox{and} \quad
 u_a \, n^a = 0 \quad \hbox{for all $\sigma$} \,.
\ee

Differentiating the first of these with respect to $\sigma$ then gives
\be \label{appeq:uddoteqn}
 \ddot U = V_{;\,ab} \, \dot \chi^a \, \dot \chi^b
 + \Va \, \frac{ D \dot \chi^a}{\exd \sigma}
 = V_{;\,ab} \, \dot \chi^a \, \dot \chi^b + \frac{
 u_a \, n^a}{\kappa} = V_{;\,ab} \, \dot \chi^a \, \dot \chi^b \,.
\ee
Similarly, differentiating the second of eqs.~\pref{appeq:udoteqn} gives
\be \label{appeq:troughcondition}
 0 = \frac{D}{\exd \sigma} \Bigl( V_{,\,a} \, n^a \Bigr)
 = V_{;\,ab} \, n^a \dot\chi^b + V_{,\,a} \, \frac{D \,
 n^a}{\exd \sigma}
 = V_{;\,ab} \, n^a \dot\chi^b
 - \frac{\dot U}{\kappa} \,,
\ee
where the last equality uses both eq.~\pref{appeq:Dndsvskappa} and the first of eqs.~\pref{appeq:udoteqn}. In particular, this shows
that $n^a$ and $\dot\chi^a$ need no longer be eigenvectors of the matrix ${\cA^a}_b$. Instead, we have
\ba
  V_{; \, ab } \, \dot\chi^a &=& \ddot{U} \, \dot\chi_b
  + \frac{\dot{U}}{\kappa} \, n_b \,,
  \nn\\
  V_{; \, ab } \, n^a &=& \frac{\dot{U}}{\kappa} \dot\chi_b
  + m^{2} n_b \,,
\ea
where we define
\be \label{appeq:tilt_mdefn}
  m^{2}(\sigma) := V_{; \, ab} \, n^a \, n^b \,.
\ee
In matrix notation,
\be
  V_{; \, ab} =
  \left( \begin{array}{cc}
  \ddot{U} & \dot{U}/\kappa \\
  \dot{U}/\kappa & m^{2} \end{array} \right) \,.
\ee
Diagonalizing this matrix, we find the heavy eigenvalue
\ba
  M_{+}^{2} &=& \frac{1}{2}
  \left( m^{2} + \ddot{U} + (m^{2}-\ddot{U})\sqrt{1
    +\beta^{2}} \right) \nn \\
  \label{appeq:heavy_eval}
  &=& m^{2} + \frac{\dot{U}^{2}}{\kappa^{2} m^{2}}
  + \mathcal{O}\left( \frac{1}{m^{4}}\right) \,,
\ea
with corresponding eigenvector
\ba \label{appeq:evec_heavy}
   e^a_+ &=&
  ({\rm sgn} \, \dot{U}) \cdot \frac{\beta n^{a} + (\sqrt{1+\beta^{2}}-1)\dot\chi^a}
  {\sqrt{2(1+\beta^{2}-\sqrt{1+\beta^{2}})}} \nn\\
  &=& n^{a}\cos\theta + \dot\chi^a \sin \theta \,,
\ea
where
\be
  \beta = \frac{2\dot{U}}{\kappa(m^{2} - \ddot{U})} \ll 1 \,,
\ee
and
\be
  \tan \theta = \frac{\sqrt{1+\beta^{2}}-1}{\beta} \,.
\ee

The light eigenvalue similarly is
\ba
  M_{-}^2 &=& \frac{1}{2}
  \left( \ddot{U} + m^{2}
     - (m^{2}-\ddot{U}) \sqrt{1
    +\beta^{2}}
   \right) \nn \\
   &=& \ddot{U} -\frac{\dot{U}^{2}}{\kappa^{2} m^{2}}
   + \mathcal{O}\left( \frac{1}{m^{4}}\right) \,,
\ea
and the corresponding eigenvector is
\ba \label{appeq:evec_light}
   e^a_- &=&
  ({\rm sgn} \, \dot{U}) \cdot \frac{\beta \dot\chi^a - (\sqrt{1+\beta^2}-1)n^a}
  {\sqrt{2(1+\beta^2-\sqrt{1+\beta^2})}} \nn\\
  &=& \dot\chi^a \cos \theta - n^a \sin \theta \,.
\ea
It is straightforward to check that these eigenvectors are
orthonormal.
Now we obtain formulas for the third covariant derivatives of $V$ in terms of
derivatives of $U$. Differentiating equation (\ref{appeq:uddoteqn}) yields
\ba
  {\dddot U} &=&
%\frac{\exd}{\exd \sigma}   \Bigl( V_{;\,ab} \, \dot \chi^a \, \dot \chi^b \Bigr) =
   V_{;\,abc} \, \dot \chi^a \, \dot \chi^b \, \dot \chi^c + 2  V_{;\,ab} \, \dot \chi^a \, \frac{D\dot \chi^b}{\exd \sigma} \nn\\
   \label{appeq:tilt_vccc}
   &=& V_{;\,abc} \, \dot \chi^a \, \dot \chi^b \, \dot \chi^c + \frac{2}{\kappa} \, V_{;\,ab} \, \dot \chi^a \, n^b  \\
   &=& V_{;\,abc} \, \dot \chi^a \, \dot \chi^b \, \dot \chi^c + \frac{2 \dot U}{\kappa^2}  \,,\nn
\ea
and differentiating equation (\ref{appeq:troughcondition}) yields
\ba
   \label{appeq:Vttt_tilt_derivation}
   0 &=&
%\frac{\exd}{\exd \sigma} \left[ V_{;\,ab} \, n^a \dot\chi^b - \frac{\dot U}{\kappa} \right]  =
 V_{;\,abc} \, n^a \dot\chi^b \dot\chi^c +
  V_{;\,ab} \, \frac{D\,n^a}{\exd \sigma} \, \dot\chi^b
  +  V_{;\,ab} \, n^a \, \frac{D\dot\chi^b}{\exd \sigma} - \frac{\ddot U}{\kappa} + \frac{\dot \kappa \dot U}{\kappa^2}  \nn\\
 &=& V_{;\,abc} \, n^a \dot\chi^b \dot\chi^c - \frac{1}{\kappa}  \, V_{;\,ab} \, \dot \chi^a \, \dot\chi^b + \frac{1}{\kappa}
   \, V_{;\,ab} \, n^a \, n^b - \frac{\ddot U}{\kappa} + \frac{\dot \kappa \dot U}{\kappa^2}  \\
  \label{appeq:deriv_reln_1}
  &=&
  V_{;\,abc} \, n^a \dot\chi^b \dot\chi^c
  - \frac{2\ddot{U}}{\kappa}   + \frac{m^{2}}{\kappa}
     + \frac{\dot \kappa \dot U}{\kappa^2}\,, \nn
\ea
where
\be
 - \frac{\dot \kappa}{\kappa^2} = \frac{\exd }{\exd \sigma}
 \left( \cG_{ab} \, n^a \frac{D \dot \chi^b}{\exd \sigma}
 \right) = \cG_{ab} \, n^a \frac{D^2 \dot \chi^b}{\exd \sigma^2} \,.
\ee

Finally, differentiating equation (\ref{appeq:tilt_mdefn}) and simplifying yields
\be \label{appeq:tilt_vnnc}
  V_{;\, abc} \, n^a \, n^b \, \dot\chi^c = 2m \dot m +
  \frac{2\dot{U}}{\kappa^{2}} \,.
\ee

The generalization of equation (\ref{appeq:v3_curv}) to tilted troughs becomes
\ba
  V_{; \, abc} \, \dot\chi^a \, \dot\chi^b \, n^c
  - V_{; \, abc} \, \dot\chi^a \, n^b \, \dot\chi^c &=& 0 \,,
  \nn\\
  \label{appeq:tilt_v3sym}
  V_{; \, abc} \, n^a \, \dot\chi^b \, n^c
  - V_{; \, abc} \, n^a \, n^b \, \dot\chi^c &=& \dot{U}/2\rho^{2} \,.
\ea
The first equation of (\ref{appeq:tilt_v3sym}) implies
that when $V_{;\, abc}$ is contracted with two $\dot\chi$'s and one $n$,
the ordering of indices does not matter. The second equation of (\ref{appeq:tilt_v3sym})
implies that all three possible contractions of $V_{;\, abc}$ with two
$n$'s and one $\dot\chi$ can be related to each other. To make this
relation simple and explicit, we introduce the following notation for
symmetrized derivatives:
\be
  V_{ttn} \equiv V_{; \, (abc)} \dot\chi^a \, \dot\chi^b \, n^{c} \,, \qquad
  V_{tnn} \equiv V_{; \, (abc)} \dot\chi^a \, n^b \, n^c \,, \qquad  {\rm etc}.
\ee
Also, we introduce the following notation for non-symmetrized (NS) derivatives:
\be
  V_{tnn}^{\rm NS} \equiv V_{; \, abc} \dot\chi^a \, n^b \, n^c \,, \qquad
  V_{ntn}^{\rm NS} \equiv V_{; \, abc} n^a \, \dot\chi^b \, n^c \,, \qquad {\rm etc}.
\ee
In this new notation, we have
\ba
V_{t} = \dot{U} \,, \qquad V_{n} = 0 \,,
\nn \\
V_{tt} = \ddot{U} \,, \qquad V_{tn} = \frac{\dot{U}}{\kappa} \,,
\qquad
V_{nn} = m^{2} \,.
\ea
For the third derivatives, equation (\ref{appeq:tilt_vccc}) implies
\be
V_{ttt} = \dddot{U} - \frac{2\dot{U}}{\kappa^{2}} \,,
\ee
while equation (\ref{appeq:Vttt_tilt_derivation}), in combination with the first equation of (\ref{appeq:tilt_v3sym}), implies
\be
V_{ntt} = V_{ntt}^{\rm NS} = V_{tnt}^{\rm NS} = V_{ttn}^{\rm NS} =
- \frac{m^{2}}{\kappa} + \frac{2\ddot{U}}{\kappa} - \frac{\dot{U}\dot{\kappa}}{\kappa^{2}} \,.
\ee
Also, equation (\ref{appeq:tilt_vnnc}) implies
\be
  \label{appeq:tilt_vnnc_newnot}
  V_{nnt}^{\rm NS} = 2m \dot m +
  \frac{2\dot{U}}{\kappa^{2}} \,.
\ee
Now, to find explicit expressions for $V_{nnt}$ and $V_{tnn}^{\rm NS}=V_{ntn}^{\rm NS}$, write
\be \label{appeq:vlhh_newnot}
%3V_{tnn} = 2V_{; \, abc} \, \dot\chi^a \, n^b \, n^c +
%V_{; \, abc} \, n^a \, n^b \, \dot\chi^c \,.
3V_{tnn} = 2V_{tnn}^{\rm NS} +
V_{nnt}^{\rm NS} \,.
\ee
Combining equation (\ref{appeq:vlhh_newnot}) and the second equation of  (\ref{appeq:tilt_v3sym}) yields
\ba
  \label{appeq:tilt_vcnn_sym}
  %V_{; \, abc} \, \dot\chi^a \, n^b \, n^c &=& V_{t n n } + \dot{U} / 6\rho^{2} \,,
  V_{tnn}^{\rm NS} &=& V_{t n n } + \dot{U} / 6\rho^{2} \,,
  \nn\\
  \label{appeq:tilt_vnnc_sym}
  %V_{; \, abc} \, n^a \, n^b \, \dot\chi^c &=& V_{t n n }- \dot{U}/3\rho^{2} \,.
  V_{nnt}^{\rm NS} &=& V_{t n n }- \dot{U}/3\rho^{2} \,.
\ea
Combining the second equation of (\ref{appeq:tilt_vnnc_sym}) with (\ref{appeq:tilt_vnnc_newnot}) yields
\be
  \label{appeq:tilt_vtnn_sym}
  V_{t n n } = 2m \dot m +  2 \dot{U}/\kappa^{2} + \dot{U} / 3\rho^{2} \,.
\ee
Therefore, the first equation of (\ref{appeq:tilt_vcnn_sym}) finally becomes
\be
  %V_{; \, abc} \, \dot\chi^a \, n^b \, n^c = 2m \dot m + 2\dot{U}/\kappa^{2}
  %+ \dot{U} / 2\rho^{2} \,.
  V_{tnn}^{\rm NS} = 2m \dot m + 2\dot{U}/\kappa^{2}
  + \dot{U} / 2\rho^{2} \,.
\ee
In summary, we have obtained formulas for all possible contractions of
third covariant derivatives of $V$ with $\dot\chi^a$ or $n^a$, in terms of $m$, $\dot m$, $\kappa$, $\dot{\kappa}$, $\dot{U}$, $\ddot{U}$, $\dddot{U}$, $\rho$, and $V_{n n n}$.

Now, let us define
\be
  V_{\e \e h} \equiv V_{; \, (abc)} e_{-}^{a} \, e_{-}^{b} \, e_{+}^{c} \,, \qquad
  V_{\e h h} \equiv V_{; \, (abc)} e_{-}^{a} \, e_{+}^{b} \, e_{+}^{c} \,, \qquad  {\rm etc}.
\ee
These quantities are important, because they appear in the low-energy effective Lagrangian.
To relate them to $V_{\cdots t , n \cdots }$, we
use:
\ba
  V_{\e} &=&
  %V_{,a}(\dot\chi^a \cos \theta - n^a \sin \theta) =
  V_{t} \cos \theta - V_{n} \sin \theta = \dot{U}\cos\theta
  = \dot{U} + \mathcal{O}(1/m^{4}) \,,
  \nn\\
  V_{h} &=&
  %V_{,a}(n^a \cos \theta + \dot\chi^a \sin \theta) =
  V_{n} \cos \theta + V_{t} \sin \theta = \dot{U}\sin\theta
  = \dot{U}^{2}/m^{2}\kappa + \mathcal{O}(1/m^{4}) \,.
\ea
For the second derivatives, we have, by construction,
\ba
  V_{\e \e} &=& M_{-}^{2} = \ddot{U} - \frac{\dot{U}^{2}}{m^{2}\kappa^{2}}
  + \mathcal{O}\left(\frac{1}{m^{4}}\right) \,, \\
  V_{h h } &=& M_{+}^{2} = m^{2} + \frac{\dot{U}^{2}}{m^{2}\kappa^{2}}
  + \mathcal{O}\left(\frac{1}{m^{4}}\right)\,, \\
  V_{\e h} &=& 0 \,.
\ea
And for the third derivatives,
\ba
  V_{\e \e \e} &=&
  V_{t t t} \cos^{3} \theta
  - 3 V_{n t t} \cos^{2} \theta \sin \theta
  + 3 V_{n n t} \cos \theta \sin^{2} \theta
  - V_{ n n n } \sin^{3} \theta \,,
  \nn \\
  V_{\e \e h} &=&
  V_{ttn}\cos^{3}\theta
  +( V_{ttt} - 2V_{tnn})\cos^{2}\theta \sin \theta
  + (V_{nnn} - 2V_{ttn}) \cos \theta \sin^{2} \theta
  \nn \\
  &&+ V_{tnn} \sin^{3} \theta \,,
  \nn \\
  V_{\e h h} &=&
  V_{tnn} \cos^{3} \theta + (2V_{ttn} - V_{nnn})\cos^{2} \theta \sin \theta
  + (V_{ttt}-2V_{nnt})\cos \theta \sin^{2} \theta
  \\
  && - V_{ttn} \sin^{3} \theta \,,
  \nn \\
  V_{h h h} &=&
  V_{n n n} \cos^{3} \theta + 3V_{nnt} \cos^{2} \theta \sin \theta
  + 3V_{ttn} \cos \theta \sin^{2} \theta + V_{ttt} \sin^{3} \theta \,.
  \nn
\ea
It is useful to expand these complicated expressions in inverse powers of $m$. We find
\ba
  V_{\e \e \e} &=& \dddot{U} + \frac{\dot{U}}{\kappa^{2}}
  -\frac{3 \dot{U}}{m^{2}\kappa^{2}}
  \left[ \ddot{U} -\dot{U}
  \left( \frac{2 \dot{m}}{m} + \frac{\dot{\kappa}}{\kappa}  \right) \right]
  + \mathcal{O}\left(\frac{1}{m^{4}} \right) \,,
  \nn \\
  V_{\e \e h} &=& - \frac{m^{2}}{\kappa}
  + \frac{2 \ddot{U}}{\kappa}
  - \frac{\dot{U}}{\kappa} \left( \frac{4\dot{m}}{m} +\frac{\dot{\kappa}}{\kappa} \right)
  + \frac{\dot{U}}{m^{2} \kappa}
  \left[
  \dddot{U} - \frac{4\ddot{U} \dot{m}}{m}  -  \frac{\dot{U}}{\kappa^{2}} \left(\frac{5}{2}-\lambda_{nnn}\right)
  - \frac{2\dot{U}}{3\rho^{2}}
  \right]
  \nn \\ &&
  + \mathcal{O}\left(\frac{1}{m^{4}} \right) \,,
  \nn \\
  V_{\e h h} &=& 2 m \dot{m} - \dot{U} \left(   \frac{\lambda_{nnn}}{\kappa^{2}} -\frac{1}{3 \rho^{2}} \right)
  + \frac{\dot{U}}{m^{2}\kappa^{2}}
  \left[
  \ddot{U}\Bigl( 2-\lambda_{nnn} \Bigr) - \dot{U} \left(  \frac{7\dot{m}}{m} + \frac{2\dot{\kappa}}{\kappa} \right)
  \right]
  \label{appeq:tilt_v3_lh}
  \\ &&
  + \mathcal{O}\left(\frac{1}{m^{4}} \right) \,,
  \nn \\
  V_{h h h} &=& \frac{m^{2}}{\kappa}\lambda_{nnn} +  \frac{6 \dot{U}\dot{m}}{\kappa m}
  + \frac{\dot{U}}{m^{2}\kappa}
  \left[ \frac{ 6 \ddot{U} \dot{m}}{m}
  + \frac{3\dot{U}}{\kappa^{2}} \left( 1 - \frac{1}{2}\lambda_{nnn}\right)
  + \frac{\dot{U}}{\rho^{2}}
  \right]
  + \mathcal{O}\left(\frac{1}{m^{4}} \right) \,,
  \nn
\ea
where we define
\be \label{appVnnnbardef}
 V_{nnn}= \left( \frac{m^{2}}{\kappa} \right) \, \lambda_{nnn} \,.
\ee

It is sometimes convenient to expand in inverse powers of $M^2 \equiv M^2_{+} \approx m^2 + \dot U^2/\kappa^2 m^2$, the
physical mass of the heavy field, rather than $m^2$. These are related by
\ba
  m^{2} &=& M^{2} - \frac{\dot{U}^{2}}{M^{2}\kappa^{2}} + \mathcal{O}\left( \frac{1}{M^{4}}\right) \,,
  \nn\\
  \frac{1}{m^{2}} &=& \frac{1}{M^{2}} + \frac{\dot{U}^{2}}{M^{6}\kappa^{2}} + \mathcal{O}\left( \frac{1}{M^{8}}\right) \,,
\ea
and
\ba
  2m\dot{m} &=& 2M\dot{M} - \frac{2\dot{U}}{M^{2}\kappa^{2}} \left[
  \ddot{U} - \dot{U}\left( \frac{\dot{M}}{M} + \frac{\dot{\kappa}}{\kappa}\right)
  \right]
  + \mathcal{O}\left( \frac{1}{M^{4}}\right) \,,
  \nn \\
  \frac{\dot{m}}{m} &=& \frac{\dot{M}}{M} - \frac{\dot{U}}{M^{4}\kappa^{2}}
  \left[
  \ddot{U} - \dot{U} \left( 2\frac{\dot{M}}{M} + \frac{\dot{\kappa}}{\kappa}\right)
  \right]
  + \mathcal{O}\left( \frac{1}{M^{6}}\right) \,,
  \nn \\
  2(\dot{m}^{2} + m \ddot{m}) &=&
  2(\dot{M}^{2} + M\ddot{M})
  \\
  && - \frac{2}{M^{2} \kappa^{2}} \Biggl[
  \ddot{U}^{2} + \dot{U}\dddot{U} -4\dot{U}\ddot{U}\left( \frac{\dot{M}}{M} + \frac{\dot{\kappa}}{\kappa}\right)
  \nn \\
  &&
  \phantom{- \frac{2}{M^{2} \kappa^{2}} \Biggl[ }
  +\dot{U}^{2} \left( 3\frac{\dot{M}^{2}}{M^{2}} - \frac{\ddot{M}}{M} + 3\frac{\dot{\kappa}^{2}}{\kappa^{2}} - \frac{\ddot{\kappa}}{\kappa}
  + 4\frac{\dot{M}}{M}\frac{\dot{\kappa}}{\kappa}\right)
  \Biggr]
   + \mathcal{O}\left( \frac{1}{M^{4}}\right) \,.
   \nn
\ea
Employing the above relations to re-express (\ref{appeq:tilt_v3_lh}) in terms of $M$ yields
\ba
  V_{\e \e \e} &=& \dddot{U} + \frac{\dot{U}}{\kappa^{2}}
  -\frac{3 \dot{U}}{M^{2}\kappa^{2}}
  \left[
   \ddot{U}
  -\dot{U}
  \left( 2\frac{\dot{M}}{M} + \frac{\dot{\kappa}}{\kappa}  \right)
  \right]
  + \mathcal{O}\left(\frac{1}{M^{4}} \right) \,,
  \nn \\
  V_{\e \e h} &=& - \frac{M^{2}}{\kappa}
  + \frac{2 \ddot{U}}{\kappa}
  - \frac{\dot{U}}{\kappa} \left(  \frac{4 \dot{M}}{M} +\frac{\dot{\kappa}}{\kappa} \right) \nn \\
  && \qquad\qquad  + \frac{\dot{U}}{M^{2} \kappa}
  \left[
  \dddot{U} - \frac{4\ddot{U}  \dot{M}}{M}  -  \frac{\dot{U}}{\kappa^{2}} \left( \frac{3}{2}-\lambda_{nnn} \right) - \frac{2\dot{U}}{3\rho^{2}}
  \right]
  + \mathcal{O}\left(\frac{1}{M^{4}} \right) \,,
  \nn \\
  V_{\e h h} &=& 2 M \dot{M} - \dot{U} \left[   \frac{\lambda_{nnn}}{\kappa^{2}} -\frac{1}{3 \rho^{2}} \right]
  - \frac{\dot{U}}{M^{2}\kappa^{2}}
  \left(
  \ddot{U}\lambda_{nnn} + 5\dot{U}\frac{\dot{M}}{M}
  \right)
  \label{appeq:tilt_v3_lh_M}
  + \mathcal{O}\left(\frac{1}{M^{4}} \right) \,,
  \\
  V_{h h h} &=& \frac{M^{2}}{\kappa}\lambda_{nnn} + 6 \frac{\dot{U}}{\kappa} \frac{\dot{M}}{M}
  + \frac{\dot{U}}{M^{2}\kappa}
  \left(
  6 \ddot{U} \frac{\dot{M}}{M}
  + \frac{\dot{U}}{\kappa^{2}} \left[3 - \frac{5}{2}\lambda_{nnn}\right]
  + \frac{\dot{U}}{\rho^{2}}
  \right)
  + \mathcal{O}\left(\frac{1}{M^{4}} \right) \,.
  \nn
\ea
The first and second derivatives of $V$, when expanded in inverse powers of $M$, simply become
\ba
  V_{\e} = \dot{U} + \mathcal{O}\left( \frac{1}{M^{4}} \right) \,, \qquad
  V_{h} = \frac{\dot{U}^{2}}{M^{2}\kappa} + \mathcal{O}\left( \frac{1}{M^{4}} \right) \,,
  \nn \\
  V_{\e \e} = \ddot{U} - \frac{\dot{U}^{2}}{M^{2}\kappa^{2}} + \mathcal{O}\left( \frac{1}{M^{4}} \right) \,, \qquad
  V_{\e h} = 0 \,, \qquad
  V_{h h} = M^{2} \,.
\ea

Before going on to calculate the fourth derivatives, we make a remark about dimensional analysis, which
becomes useful due to the proliferation of terms as one takes more derivatives. We use canonical
relativistic units, in which $\hbar = c = 1$. Also, for simplicity, we take $d=4$. Then
$[\ell] = [h] = \mathfrak{M}$, and $[\mathcal{L}] = [V] = \mathfrak{M}^{4}$, and $[V_{i_{1} \cdots i_{k}}] = \mathfrak{M}^{4-k}$, where $\mathfrak{M}$ denotes `mass dimension', and the indices $i_{j}$ can be either $\ell,h$ or $t,n$. In particular, $[m^{2}] = [V_{nn}] = \mathfrak{M}^{2}$, as one would intuitively expect. Moreover, we have $[D / \exd \sigma ] = \mathfrak{M}^{-1}$, and $[\dot{\chi}^{a}] = [n^{a}]= \mathfrak{M}^{0}$, and thus $[\kappa] = \mathfrak{M}$. Also, from the commutator formulas it follows that $[\rho] = \mathfrak{M}$, so both $\kappa$ and $\rho$ share the dimension of the field-space coordinates, $\ell$ and $h$.

Now we calculate the fourth covariant derivatives of $V$. Differentiating
equation (\ref{appeq:tilt_vccc}) and simplifying yields
\be
%  V_{; \, abcd} \, \dot\chi^a \, \dot\chi^b \, \dot\chi^c \, \dot\chi^d =
  V_{tttt}=
  \frac{3m^{2}}{\kappa^{2}}  +\ddddot{U} - \frac{8 \ddot{U}}{\kappa^{2}}
  + \frac{7\dot{U}\dot{\kappa}}{\kappa^{3}}
   \,.
\ee
Differentiating
equation (\ref{appeq:deriv_reln_1}) and simplifying yields
\be \label{appeq:tilt_vnccc}
%  V_{; \, abcd} \, n^a \, \dot\chi^b \, \dot\chi^c \, \dot\chi^d =
  V_{nttt}^{\rm NS}=
   - \frac{m^{2}}{\kappa}\left( 6\frac{\dot{m}}{m} - \frac{\dot{\kappa}}{\kappa} \right)
%  + \frac{m^{2}\dot{\kappa}}{\kappa^{2}}
%  - \frac{6m \dot{m}}{\kappa}
  +\frac{3 \dddot{U}}{\kappa}
  - \frac{3\ddot{U}\dot{\kappa}}{\kappa^{2}}
  - \frac{\dot{U}}{\kappa} \left(
  \frac{6}{\kappa^{2}} -2 \frac{\dot{\kappa}^{2}}{\kappa^{2}} + \frac{\ddot{\kappa}}{\kappa}
  + \frac{1}{2\rho^{2}}
  \right)
%
%  - \frac{6\dot{U}}{\kappa^{3}}
%  - \frac{\dot{U}}{2 \kappa \rho^{2}}
%  - \frac{\dot{U}\ddot{\kappa}}{\kappa^{2}}
%  + \frac{2\dot{U}\dot{\kappa}^{2}}{\kappa^{3}}
  \,.
\ee
Differentiating equation (\ref{appeq:tilt_vnnc}) and simplifying yields
\be
  \label{appeq:tilt_Vnnchichi}
%  V_{; \, abcd} \, n^a \, n^b \, \dot\chi^c \, \dot\chi^d
  V_{nntt}^{\rm NS}
  =
m^{2} \left(
2 \frac{\dot{m}^{2}}{m^{2}} + 2 \frac{\ddot{m}}{m} - \frac{2+\lambda_{nnn}}{\kappa^{2}}
\right)
%   - \frac{V_{nnn}}{\kappa}
%   +2\dot{m}^{2} + 2m \ddot{m}
  + \frac{6\ddot{U}}{\kappa^{2}}
  - \frac{6\dot{U}\dot{\kappa}}{\kappa^{3}} \,,
%  - \frac{2m^{2}}{\kappa^{2}} \,.
\ee
where we have written $V_{nnn} = (m^{2} / \kappa) \lambda_{nnn}$.
Finally, differentiating the definition $V_{nnn} \equiv V_{;\,abc}n^a n^b n^c$ and using (\ref{appeq:tilt_vtnn_sym}) yields
\be
\label{appeq:tilt_Vnnnt_nosym}
%  V_{; \, abcd} \, n^a \, n^b \, n^c \, \dot\chi^d =
  V_{nnnt}^{\rm NS} =
%  \dot{V}_{nnn} + \frac{6m\dot{m}}{\kappa} + \frac{6\dot{U}}{\kappa^{3}}
%  + \frac{\dot{U}}{\kappa \rho^{2}} \,.
  \frac{m^{2}}{\kappa} \left( 2(3+\lambda_{nnn})\frac{\dot{m}}{m} - \lambda_{nnn} \frac{\dot{\kappa}}{\kappa} + \dot{\lambda}_{nnn} \right)
  + \frac{\dot{U}}{\kappa} \left( \frac{6}{\kappa^{2}} + \frac{1}{\rho^{2}} \right) \,.
\ee

Now we look at the symmetries of the fourth derivatives. The generalization of
(\ref{appeq:v4sym_1}) to tilted troughs is
\ba
  \label{appeq:tilt_v4sym_1}
%  V_{;\, abcd} \, \dot\chi^a \, \dot\chi^b \, \dot\chi^c \, n^d
   V_{tttn}^{\rm NS}
%  - V_{;\, abcd} \, \dot\chi^a \, \dot\chi^b \, n^c \, \dot\chi^d &=&
   - V_{ttnt}^{\rm NS} &=&
  - \dot{U}/\kappa \rho^{2} \,,
  \nn\\
  \label{appeq:tilt_v4sym_2}
%  V_{;\, abcd} \, n^a \, n^b \, \dot\chi^c \, n^d
   V_{nntn}^{\rm NS}
%  - V_{;\, abcd} \, n^a \, n^b \, n^c \, \dot\chi^d &=&
   - V_{nnnt}^{\rm NS} &=&
  +\dot{U}/\kappa \rho^{2} \,,
  \\
  \label{appeq:tilt_v4sym_3}
%  V_{;\, abcd} \, \dot\chi^a \, n^b \, \dot\chi^c \, n^d
   V_{tntn}^{\rm NS}
%  - V_{;\, abcd} \, \dot\chi^a \, n^b \, n^c \, \dot\chi^d &=&
   - V_{tnnt}^{\rm NS} &=&
   +(\ddot{U}-m^{2})/2\rho^{2} \,,\nn
\ea
and the generalization of (\ref{appeq:v4sym_3}) is:
\ba
  \label{appeq:tilt_v4sym_4}
%  V_{; \, abcd} \, \dot\chi^a \, \dot\chi^b \, n^c \, \dot\chi^d
   V_{ttnt}^{\rm NS}
%  - V_{; \, abcd} \, \dot\chi^a \, n^b \, \dot\chi^d \, \dot\chi^d
   - V_{tntt}^{\rm NS}
  &=& -\dot{U}/2\kappa \rho^{2} \,,\nn \\
  \label{appeq:tilt_v4sym_5}
%  V_{; \, abcd} \, n^a \, \dot\chi^b \, n^c \, \dot\chi^d
   V_{ntnt}^{\rm NS}
%  - V_{; \, abcd} \, n^a \, n^b \, \dot\chi^c \, \dot\chi^d
   - V_{nntt}^{\rm NS}
  &=& -\dot{U}\dot{\rho}/\rho^{3} + \ddot{U}/2\rho^{2} \,, \nn\\
%  V_{; \, abcd} \, n^a \, \dot\chi^b \, n^c \, n^d
   V_{ntnn}^{\rm NS}
%  - V_{; \, abcd} \, n^a \, n^b \, \dot\chi^c \, n^d
   - V_{nntn}^{\rm NS}
  &=& \dot{U}/2\kappa \rho^{2} - \dot{U} \rho_{,n} / \rho^{3} \,, \\
%  V_{; \, abcd} \, \dot\chi^a \, \dot\chi^b \, n^c \, n^d
   V_{ttnn}^{\rm NS}
%  - V_{; \, abcd} \, \dot\chi^a \, n^b \, \dot\chi^c \, n^d
   - V_{tntn}^{\rm NS}
  &=& -m^{2}/2\rho^{2} \,,
  \nn
\ea
where $\rho_{,n} \equiv n^{a} \nabla_{a} \rho$ is the normal derivative
of the target-space curvature radius.
Now, we have
\be
  4V_{t t t n} =
  2V_{ nttt}^{\rm NS}
  + V_{ ttnt}^{\rm NS}
  + V_{ tttn}^{\rm NS} \,.
\ee
Combining this equation with the first equation of (\ref{appeq:tilt_v4sym_1}) and
the first equation of (\ref{appeq:tilt_v4sym_4}) yields
\ba
  V_{t t t n} &=&
  V_{nttt}^{\rm NS}
  - \frac{\dot{U}}{2\kappa \rho^{2}}
  \nn \\
  &=&
    \frac{m^{2}}{\kappa}\left(  \frac{\dot{\kappa}}{\kappa}
   - \frac{6 \dot{m}}{m} \right)
   + \frac{3 \dddot{U}}{\kappa}
   - \frac{3\ddot{U}\dot{\kappa}}{\kappa^{2}}
   - \frac{\dot{U}}{\kappa} \left(
   \frac{6}{\kappa^{2}}
   - \frac{2\dot{\kappa}^{2}}{\kappa^{2}}
   + \frac{\ddot{\kappa}}{\kappa}
   + \frac{1}{ \rho^{2}}
   \right) \,,
\ea
where in the second equality we have used the expression for $V_{nttt}^{\rm NS}$ given by equation (\ref{appeq:tilt_vnccc}).

Next, let's calculate
\ba
  6 V_{ t t n n } &=& V_{ttnn}^{\rm NS}   + V_{nntt}^{\rm NS}
  + 2V_{tntn}^{\rm NS}
  + 2V_{tnnt}^{\rm NS} \,.
\ea
Using the third equation of (\ref{appeq:tilt_v4sym_1}), and the second and fourth equations
of (\ref{appeq:tilt_v4sym_4}), we may write this as
\ba
%6V_{ttnn} = 5 V_{; \, abcd} \, n^a \, n^b \, \dot\chi^c \, \dot\chi^d
%+V_{; \,abcd} \, \dot\chi^a \, \dot\chi^b \, n^c \, n^d
%-\frac{4\dot{U} \dot{\rho}}{\rho^{3}} +\frac{3\ddot{U}}{\rho^{2}}
%-\frac{m^{2}}{\rho^{2}} \,.
V_{ttnn} &=& V_{nntt}^{\rm NS}
+ \frac{2\ddot{U}}{3\rho^{2}} - \frac{5\dot{U}\dot{\rho}}{6\rho^{3}} - \frac{m^{2}}{3\rho^{2}}
\nn \\
%%&=&
%%  - \frac{V_{nnn}}{\kappa}  +2\dot{m}^{2} + 2m \ddot{m} + \frac{6\ddot{U}}{\kappa^{2}}
%%  - \frac{6\dot{U}\dot{\kappa}}{\kappa^{3}}
%%  - \frac{2m^{2}}{\kappa^{2}}
%%+ \frac{2\ddot{U}}{3\rho^{2}} - \frac{5\dot{U}\dot{\rho}}{6\rho^{3}} - \frac{m^{2}}{3\rho^{2}} \,,
%%\\
&=&
2m^{2} \left[ \frac{\ddot{m}}{m} + \left( \frac{\dot{m}}{m}\right)^{2}\right]
- \frac{m^{2}}{\kappa^{2}}(2+\lambda_{nnn})
- \frac{m^{2}}{3\rho^{2}}
+2 \ddot{U} \left( \frac{3}{\kappa^{2}} + \frac{1}{3\rho^{2}}\right)
\\
&&
-\dot{U} \left( \frac{6\dot{\kappa}}{\kappa^{3}} + \frac{5\dot{\rho}}{6\rho^{3}}\right)
\nn
\ea
where in the second equality we have used the expression for $V_{nntt}^{\rm NS}$ given by equation (\ref{appeq:tilt_Vnnchichi}), and
recall that $V_{nnn} = (m^{2}/\kappa)\lambda_{nnn}$.
%
%%Finally, using equation (\ref{appeq:tilt_Vnnchichi}), and defining
%%%
%%\be
%%V_{t t n n }^{\rm NS} \equiv V_{; \, abcd} \dot\chi^a \, \dot\chi^b \, n^c \, n^d \,,
%%\ee
%%%
%%(where `NS' stands for `no symmetrization'), we obtain
%%%
%%\be
%%V_{ttnn} = \frac{V_{ttnn}^{\rm NS}}{6}
%%- \frac{5V_{nnn}}{6\kappa}
%%-\frac{m^{2}}{6\rho^{2}}
%%-\frac{5m^{2}}{3\kappa^{2}}
%%+ \frac{5\dot{m}^{2}}{3}
%%+\frac{5m\ddot{m}}{3}
%%+ \frac{5\ddot{U}}{\kappa^{2}}
%%- \frac{5\dot{U}\dot{\kappa}}{\kappa^{3}}
%%+ \frac{\ddot{U}}{2\rho^{2}}
%%- \frac{2\dot{\rho}\dot{U}}{3\rho^{3}} \,.
%%\ee
%%%
%
Finally, let's calculate
\be
  4V_{tnnn} = 2V_{tnnn}^{\rm NS} + V_{nntn}^{\rm NS} + V_{nnnt}^{\rm NS} \,.
\ee
Using the second equation of (\ref{appeq:tilt_v4sym_1}) and the third equation of (\ref{appeq:tilt_v4sym_4}), we may write this as
\ba
V_{tnnn} &=& V_{nnnt}^{\rm NS} + \frac{\dot{U}}{\kappa \rho^{2}} - \frac{\dot{U}\rho_{,n}}{2\rho^{3}}
\nn \\
&=&
\dot{V}_{nnn} + \frac{6m\dot{m}}{\kappa} + \frac{6\dot{U}}{\kappa^{3}} + \frac{2\dot{U}}{\kappa \rho^{2}}
 - \frac{\dot{U}\rho_{,n}}{2\rho^{3}}
 \,,
\\
&=&
\frac{m^{2}}{\kappa} \left(2 \frac{\dot{m}}{m}[3+\lambda_{nnn}] - \frac{\dot{\kappa}}{\kappa}\lambda_{nnn} + \dot{\lambda}_{nnn} \right)
+ \dot{U} \left( \frac{6}{\kappa^{3}} + \frac{2}{\kappa \rho^{2}} - \frac{\rho_{,n}}{2\rho^{3}} \right) \,,
\nn
\ea
where in the second equality we have used the expression for $V_{nnnt}^{\rm NS}$ given by equation (\ref{appeq:tilt_Vnnnt_nosym}).

The next step, is to calculate the fourth derivatives in the light and heavy directions. They are given by
\ba
V_{\e \e \e \e} &=& V_{tttt}\cos^{4}\theta - 4V_{tttn}\cos^{3}\theta \sin \theta
+6 V_{ttnn} \cos^{2} \theta \sin^{2} \theta
\nn \\
&& - 4V_{tnnn} \cos \theta \sin^{3} \theta + V_{nnnn} \sin^{4} \theta \,,
\nn \\
V_{\e \e \e h} &=& V_{tttn}\cos^{4}\theta +
(V_{tttt}-3V_{ttnn})\cos^{3}\theta \sin \theta
\nn \\
&& +3(V_{nnnt}-V_{tttn})\cos^{2}\theta \sin^{2} \theta
\nn \\
&& + (3V_{ttnn}-V_{nnnn})\cos \theta \sin^{3} \theta
-V_{nnnt}\sin^{4} \theta   \,,
\nn \\
V_{\e \e h h} &=& V_{ttnn}\cos^{4} \theta +2(V_{tttn}-V_{nnnt})\cos^{3}\theta \sin \theta
\nn \\
&& + (V_{nnnn}+V_{tttt}-4V_{ttnn})\cos^{2}\theta \sin^{2}\theta
\nn \\
&& +2(V_{nnnt}-V_{tttn})\cos \theta \sin^{3} \theta + V_{nntt}\sin^{4} \theta \,,
\\
V_{\e h h h} &=& V_{tnnn}\cos^{4}\theta + (3V_{ttnn}-V_{nnnn})\cos^{3}\theta \sin \theta
\nn \\
&& +3(V_{tttn}-V_{nnnt})\cos^{2}\theta \sin^{2}\theta
\nn \\
&& + (V_{tttt}-3V_{nntt})\cos \theta \sin^{3} \theta
-V_{nttt}\sin^{4} \theta  \,,
\nn \\
V_{h h h h} &=& V_{nnnn}\cos^{4} \theta + 4V_{nnnt}\cos^{3}\theta \sin \theta + 6 V_{nntt} \cos^{2} \theta \sin^{2} \theta
\nn \\
&& + 4 V_{nttt} \cos \theta \sin^{3} \theta + V_{tttt} \sin^{4} \theta  \,.
\nn
\ea
Expanding these in inverse powers of $m^{2}$ yields
\ba
V_{\e \e \e \e} &=& \frac{3m^{2}}{\kappa^{2}} + \ddddot{U} - \frac{8\ddot{U}}{\kappa^{2}}
+\frac{3 \dot{U}}{\kappa^{2}} \left( \frac{\dot{\kappa}}{\kappa} + 8 \frac{\dot{m}}{m}\right)
\nn
\\
&& - \frac{2\dot{U}}{m^{2}\kappa^{2}} \biggl(
6\dddot{U} - 4\ddot{U} \left[ \frac{\dot{\kappa}}{\kappa} + 3 \frac{\dot{m}}{m}\right]
\\
&&
\phantom{- \frac{2\dot{U}}{m^{2}\kappa^{2}} \biggl(}
- \dot{U} \biggl[
\frac{3(1-\lambda_{nnn})}{\kappa^{2}} -4\frac{\dot{\kappa}^{2}}{\kappa^{2}} +2 \frac{\ddot{\kappa}}{\kappa}
+6\frac{\dot{m}^{2}}{m^{2}} + 6\frac{\ddot{m}}{m} + \frac{1}{\rho^{2}} \biggr]
\biggr)
+ \mathcal{O}\left(\frac{1}{m^{4}} \right) \,,
\nn \\
V_{\e \e \e h} &=& -\frac{m^{2}}{\kappa} \left( 6\frac{\dot{m}}{m} - \frac{\dot{\kappa}}{\kappa} \right)
+ \frac{3\dddot{U}}{\kappa} - \frac{3\ddot{U}\dot{\kappa}}{\kappa^{2}}
\nn \\
&&
- \frac{\dot{U}}{\kappa} \left( 6\frac{\dot{m}^{2}}{m^{2}} + 6 \frac{\ddot{m}}{m}
-\frac{3(1+\lambda_{nnn})}{\kappa^{2}} -2\frac{\dot{\kappa}^{2}}{\kappa^{2}} + \frac{\ddot{\kappa}}{\kappa}\right)
\nn \\
&&
+ \frac{\dot{U}}{m^{2}\kappa} \biggl(
\ddddot{U} - \ddot{U}
\left[
6\frac{\dot{m}^{2}}{m^{2}} + 6 \frac{\ddot{m}}{m} + \frac{17-3\lambda_{nnn}}{\kappa^{2}}
+ \frac{1}{\rho^{2}}
\right]
\nn
\\
&&
\phantom{+ \frac{\dot{U}}{m^{2}\kappa} \biggl(}
+\dot{U}
\left[
\frac{1}{\kappa^{2}} \left( 6(8+\lambda_{nnn})\frac{\dot{m}}{m} + (20-3\lambda_{nnn})\frac{\dot{\kappa}}{\kappa} + 3\dot{\lambda}_{nnn} \right)
+ \frac{5\dot{\rho}}{2\rho^{3}}
\right]
\biggr)
\nn \\
&& + \mathcal{O}\left(\frac{1}{m^{4}} \right) \,,
\nn \\
V_{\e \e h h} &=&
m^{2} \left( 2 \frac{\dot{m}^{2}}{m^{2}} + 2\frac{\ddot{m}}{m} -\frac{2+\lambda_{nnn}}{\kappa^{2}} - \frac{1}{3\rho^{2}}\right)
+2 \ddot{U} \left( \frac{3}{\kappa^{2}} + \frac{1}{3\rho^{2}}\right)
\nn \\
&& - 2\frac{\dot{U}}{\kappa^{2}} \left( 1 + \frac{\ddot{U}}{m^{2}}\right) \left[
2(6+\lambda_{nnn})\frac{\dot{m}}{m} + (2 - \lambda_{nnn})\frac{\dot{\kappa}}{\kappa} + \dot{\lambda}_{nnn}
\right]
- \frac{5 \dot{U} \dot{\rho}}{6 \rho^{3}}
\nn \\
&&
+ \frac{\dot{U}}{m^{2}\kappa}\biggl(
6\frac{\dddot{U}}{\kappa}
- \dot{U} \left[ \frac{1}{\kappa}
\left(
12 \frac{\dot{m}^{2}}{m^{2}} + 12 \frac{\ddot{m}}{m}
+ \frac{9-6\lambda_{nnn}-\lambda_{nnnn}}{\kappa^{2}} - 4 \frac{\dot{\kappa}^{2}}{\kappa^{2}} + 2 \frac{\ddot{\kappa}}{\kappa} + \frac{4}{\rho^{2}}
\right)
 - \frac{\rho_{,n}}{\rho^{3}}\right]
\biggr)
\nn \\
&& + \mathcal{O}\left(\frac{1}{m^{4}} \right) \,,
\nn \\
V_{\e h h h} &=&
\frac{m^{2}}{\kappa}
\left(
2(3+\lambda_{nnn})\frac{\dot{m}}{m} - \frac{\dot{\kappa}}{\kappa} \lambda_{nnn} + \dot{\lambda}_{nnn}
\right)
\nn \\
&&
+ \frac{\dot{U}}{\kappa} \left( 1 + \frac{\ddot{U}}{m^{2}} \right)
\left(
6 \frac{\dot{m}^{2}}{m^{2}} + 6 \frac{\ddot{m}}{m} - \frac{3\lambda_{nnn} + \lambda_{nnnn}}{\kappa^{2}} + \frac{1}{\rho^{2}}
\right)
- \frac{\dot{U}\rho_{,n}}{2\rho^{3}}
\nn \\
&&
+ \frac{\dot{U}}{m^{2} \kappa}
\left(
12\frac{\ddot{U}}{\kappa^{2}}
- \dot{U} \left[
\frac{1}{\kappa^{2}}
\left(
2(24+5\lambda_{nnn})\frac{\dot{m}}{m} + 5(3-\lambda_{nnn})\frac{\dot{\kappa}}{\kappa}
+ 5 \dot{\lambda}_{nnn}
\right)
+ \frac{5\dot{\rho}}{2\rho^{3}}
\right]
\right)
\nn \\
&&
 + \mathcal{O}\left(\frac{1}{m^{4}} \right) \,,
\nn \\
V_{h h h h} &=&
\frac{m^{2}}{\kappa^{2}} \lambda_{nnnn}
+ \frac{4\dot{U}}{\kappa^{2}}
\left( 1 + \frac{\ddot{U}}{m^{2}} \right)
\left(
2(3+\lambda_{nnn})\frac{\dot{m}}{m} - \lambda_{nnn} \frac{\dot{\kappa}}{\kappa} + \dot{\lambda}_{nnn}
\right)
\nn \\
&&
+ \frac{2\dot{U}^{2}}{m^{2} \kappa} \left(
\frac{1}{\kappa} \left[
6\frac{\dot{m}^{2}}{m^{2}} + 6\frac{\ddot{m}}{m} + \frac{6-3\lambda_{nnn}-\lambda_{nnnn}}{\kappa^{2}} + \frac{3}{\rho^{2}}
\right]
- \frac{\rho_{,n}}{\rho^{3}}
 \right) + \mathcal{O}\left(\frac{1}{m^{4}} \right) \,,
\ea
where we have written $V_{nnnn} = (m^{2}/\kappa^{2})\lambda_{nnnn}$.

Now the final step is to replace the $1/m^{2}$ expansion with a $1/M^{2}$ expansion. Doing so yields
\ba
V_{\e \e \e \e} &=& \frac{3M^{2}}{\kappa^{2}} + \ddddot{U} - \frac{8\ddot{U}}{\kappa^{2}}
+\frac{3 \dot{U}}{\kappa^{2}} \left( \frac{\dot{\kappa}}{\kappa} + 8 \frac{\dot{M}}{M}\right)
\nn
\\
&& - \frac{2\dot{U}}{M^{2}\kappa^{2}} \biggl(
6\dddot{U} - 4\ddot{U} \left[ \frac{\dot{\kappa}}{\kappa} + 3 \frac{\dot{M}}{M}\right]
\\
&&
\phantom{- \frac{2\dot{U}}{M^{2}\kappa^{2}} \biggl(}
- \dot{U} \biggl[
\frac{3(1-2\lambda_{nnn})}{2\kappa^{2}} -4\frac{\dot{\kappa}^{2}}{\kappa^{2}} +2 \frac{\ddot{\kappa}}{\kappa}
+6\frac{\dot{M}^{2}}{M^{2}} + 6\frac{\ddot{M}}{M} + \frac{1}{\rho^{2}} \biggr]
\biggr)
+ \mathcal{O}\left(\frac{1}{M^{4}} \right) \,,
\nn
\\
V_{\e \e \e h} &=& -\frac{M^{2}}{\kappa} \left( 6\frac{\dot{M}}{M} - \frac{\dot{\kappa}}{\kappa} \right)
+ \frac{3\dddot{U}}{\kappa} - \frac{3\ddot{U}\dot{\kappa}}{\kappa^{2}}
\nn \\
&&
- \frac{\dot{U}}{\kappa} \left( 6\frac{\dot{M}^{2}}{M^{2}} + 6 \frac{\ddot{M}}{M}
-\frac{3(1+\lambda_{nnn})}{\kappa^{2}} -2\frac{\dot{\kappa}^{2}}{\kappa^{2}} + \frac{\ddot{\kappa}}{\kappa}\right)
\nn \\
&&
+ \frac{\dot{U}}{M^{2}\kappa} \biggl(
\ddddot{U} - \ddot{U}
\left[
6\frac{\dot{M}^{2}}{M^{2}} + 6 \frac{\ddot{M}}{M} + \frac{11-3\lambda_{nnn}}{\kappa^{2}}
+ \frac{1}{\rho^{2}}
\right]
\nn
\\
&&
\phantom{+ \frac{\dot{U}}{M^{2}\kappa} \biggl(}
+\dot{U}
\left[
\frac{1}{\kappa^{2}} \left( 6(7+\lambda_{nnn})\frac{\dot{M}}{M} + (13-3\lambda_{nnn})\frac{\dot{\kappa}}{\kappa} + 3\dot{\lambda}_{nnn} \right)
+ \frac{5\dot{\rho}}{2\rho^{3}}
\right]
\biggr)
\nn \\
&& + \mathcal{O}\left(\frac{1}{M^{4}} \right) \,,
\nn \\
V_{\e \e h h} &=&
M^{2} \left( 2 \frac{\dot{M}^{2}}{M^{2}} + 2\frac{\ddot{M}}{M} -\frac{2+\lambda_{nnn}}{\kappa^{2}} - \frac{1}{3\rho^{2}}\right)
+2 \ddot{U} \left( \frac{3}{\kappa^{2}} + \frac{1}{3\rho^{2}}\right)
\nn \\
&& - 2\frac{\dot{U}}{\kappa^{2}}  \left[
2(6+\lambda_{nnn})\frac{\dot{M}}{M} + (2 - \lambda_{nnn})\frac{\dot{\kappa}}{\kappa} + \dot{\lambda}_{nnn}
\right]
- \frac{5 \dot{U} \dot{\rho}}{6 \rho^{3}}
- \frac{2 \ddot{U}^{2}}{M^{2}\kappa^{2}}
\nn \\
&&
+ \frac{\dot{U}}{M^{2}\kappa}\Biggl(
4\frac{\dddot{U}}{\kappa}
-2 \frac{\ddot{U}}{\kappa} \left[
2(4+\lambda_{nnn})\frac{\dot{M}}{M} - (2 + \lambda_{nnn})\frac{\dot{\kappa}}{\kappa} + \dot{\lambda}_{nnn}
\right]
\nn \\
&&
\phantom{+ \frac{\dot{U}}{M^{2}\kappa}\Biggl(}
-  \frac{\dot{U}}{\kappa}
\left[
18 \frac{\dot{M}^{2}}{M^{2}} + 10 \frac{\ddot{M}}{M}
+ \frac{7-7\lambda_{nnn}-\lambda_{nnnn}}{\kappa^{2}} +2 \frac{\dot{\kappa}^{2}}{\kappa^{2}} +8\frac{\dot{M}}{M}\frac{\dot{\kappa}}{\kappa}+\frac{11}{3\rho^{2}}
\right]
 + \dot{U}\frac{\rho_{,n}}{\rho^{3}}
\Biggr)
\nn \\
&& + \mathcal{O}\left(\frac{1}{M^{4}} \right) \,,
\nn \\
V_{\e h h h} &=&
\frac{M^{2}}{\kappa}
\left(
2(3+\lambda_{nnn})\frac{\dot{M}}{M} - \frac{\dot{\kappa}}{\kappa} \lambda_{nnn} + \dot{\lambda}_{nnn}
\right)
\nn \\
&&
+ \frac{\dot{U}}{\kappa}
\left(
6 \frac{\dot{M}^{2}}{M^{2}} + 6 \frac{\ddot{M}}{M} - \frac{3\lambda_{nnn} + \lambda_{nnnn}}{\kappa^{2}} + \frac{1}{\rho^{2}}
\right)
- \frac{\dot{U}\rho_{,n}}{2\rho^{3}}
\nn \\
&&
+ \frac{\dot{U}}{M^{2} \kappa}
\Biggl(
\ddot{U}\left[ 6 \frac{\dot{M}^{2}}{M^{2}} + 6 \frac{\ddot{M}}{M} + \frac{6-5\lambda_{nnn} - \lambda_{nnnn}}{\kappa^{2}} + \frac{1}{\rho^{2}}  \right]
\nn \\
&&
\phantom{+ \frac{\dot{U}}{M^{2} \kappa}
\biggl(}
- \dot{U} \left[
\frac{1}{\kappa^{2}}
\left(
2(21+4\lambda_{nnn})\frac{\dot{M}}{M} + (9-8\lambda_{nnn})\frac{\dot{\kappa}}{\kappa}
+ 6 \dot{\lambda}_{nnn}
\right)
+ \frac{5\dot{\rho}}{2\rho^{3}}
\right]
\Biggr)
\nn \\
&&
 + \mathcal{O}\left(\frac{1}{M^{4}} \right) \,,
\nn \\
V_{h h h h} &=&
\frac{M^{2}}{\kappa^{2}} \lambda_{nnnn}
+ \frac{4\dot{U}}{\kappa^{2}}
\left( 1 + \frac{\ddot{U}}{M^{2}} \right)
\left(
2(3+\lambda_{nnn})\frac{\dot{M}}{M} - \lambda_{nnn} \frac{\dot{\kappa}}{\kappa} + \dot{\lambda}_{nnn}
\right)
\nn \\
&&
+ \frac{2\dot{U}^{2}}{M^{2} \kappa} \left(
\frac{1}{\kappa} \left[
6\frac{\dot{M}^{2}}{M^{2}} + 6\frac{\ddot{M}}{M} + \frac{3(4-2\lambda_{nnn}-\lambda_{nnnn})}{2\kappa^{2}} + \frac{3}{\rho^{2}}
\right]
- \frac{\rho_{,n}}{\rho^{3}}
 \right) + \mathcal{O}\left(\frac{1}{M^{4}} \right) \,.
\nn
\ea

\section{Integrating out the heavy fields}
\label{app:integratingout}

This appendix provides the details of the classical integration over the heavy field to obtain the low-energy effective theory of the light field along the bottom of the trough. The plan is to use $S_{\rm eff}(\e) = S[\e, h(\e)]$, where $h(\e)$ satisfies $\delta S/\delta h = 0$ \cite{EFTrevs}. For these purposes it is useful to integrate by parts in order to write the classical action as follows,
\ba \label{app_act}
 \frac{\mathcal L}{\sqrt{-g}} &=& - \frac12 \, \partial_\mu \e \, \partial^\mu \e - V_{\rm tr}(\p, \e) + \frac12 \, h \, \Delta_h h - J_{(1)} h - \frac13 \, J_{(3)} h^3 - \frac14 \, J_{(4)} h^4 \nn\\
 &=& \frac{{\mathcal L}_{\rm tr}}{\sqrt{-g}} + \frac12 \, h \, \Delta_h h - J_{(1)} h - \frac13 \, J_{(3)} h^3 - \frac14 \, J_{(4)} h^4\,,
\ea
where the truncated potential is
\be
 V_{\rm tr} := \Bigl. V \Bigr|_{h=0} = U(\p) + (j + V_\e) \e + \frac{\mu^2}{2} \, \e^2 + \frac16 \, V_{\e\e \e} \, \e^3 + \frac{1}{24} \, V_{\e\e\e\e} \, \e^4
\ee
and we couple an external current, $j$, to the light field, $\e$. The kinetic operator for $h$ is $\Delta_h := \Omega - \cM^2$, where
\begin{eqnarray}  \label{appeq:otherdefs}
 \Omega &:=& \left( 1 - \frac{\e^2}{6 \rho^2} \right) \square \nn\\
 \hbox{and} \quad
 \cM^2 &:=& M^2 + V_{hh\e} \, \e + \frac12 \, V_{\e\e h h} \, \e^2 - \frac{1}{2 \rho^2} \, \partial_\mu\e \partial^\mu \e - \frac{1}{3\rho^2} \, \e \, \square \e \,.
\end{eqnarray}
Recall here that $\rho$ denotes the target-space radius of curvature as defined using the Ricci scalar constructed from the target-space metric, $\cG_{ab}$, and the last two terms of eq.~\pref{appeq:otherdefs} are obtained by repeatedly integrating by parts the two-derivative interactions
\be \label{lke}
 \cL_{\rm 2\,deriv} = \frac{1}{12 \rho^2} \Bigl( \e^2 \, \partial_\mu h \, \partial^\mu h + h^2 \, \partial_\mu \e \, \partial^\mu \e - 2 h\e \, \partial_\mu \e \, \partial^\mu h\Bigr) \,.
\ee
Finally, the $J_{(i)}$ are given by
\bea
 J_{(1)} &:=& V_h + \frac12 \, V_{h\e\e} \, \e^2 + \frac16 \, V_{\e\e\e h} \, \e^3 \nn\\
 J_{(3)} &:=& \frac12 \, V_{hhh} + \frac12 \, V_{\e h h h} \, \e \\
 \hbox{and} \quad J_{(4)} &:=& \frac16 \, V_{hhhh} \,. \nn
\eea

\subsection*{Integrating out $h$}

To integrate out the $h$ field we compute
\ba \label{int}
 e^{iS_{\rm eff}[\ell\,]} &=& e^{iS_{\rm tr}[\e\,]} \left[ e^{ -\frac{i}{3} \int J_{(3)} \left(i \frac{\delta}{\delta \cJ} \right)^3} \, e^{-\frac{i}{4} \int J_{(4)} \left(i \frac{\delta}{\delta \cJ} \right)^4} \int \mathcal \cD h \; e^{\frac{i}{2} \int h \Delta_h h - i \int h[J_{(1)} + \cJ]} \right]_{\cJ = 0} \\
 &=& e^{iS_{\rm tr}[\e\,]} \left[ e^{ -\frac{i}{3} \int J_{(3)} \left( i \frac{\delta}{\delta \cJ} \right)^3} \, e^{-\frac{i}{4} \int J_{(4)} \left( i \frac{\delta}{\delta \cJ} \right)^4} \left( e^{-\frac{i}{2}\int [J_{(1)} + \cJ] \Delta_h^{-1} [J_{(1)} + \cJ ]} \; \left[ \det \,\Delta_h \right]^{-1/2} \right) \right]_{\cJ = 0} \,, \nn
\ea
in the classical approximation (for which the determinant may be neglected). Evaluating the derivatives and taking the logarithm (and so dropping disconnected terms) then gives
\ba \label{eap}
 \frac{\cL_{\rm eff}[\ell\,]}{\sqrt{-g}} &=& \frac{\cL_{\rm tr}[\ell\,]}{\sqrt{-g}} - \frac{1}{2} \, J_{(1)} \, \Delta_h^{-1} J_{(1)} - \frac{1}{3} \, J_{(3)} \left[ \Delta_h^{-1} J_{(1)} \right]^3 - \frac{1}{4} \, J_{(4)} \left[ \Delta_h^{-1} J_{(1)} \right]^4 \,.
\ea

Expanding up to and including order $\cM^{-6}$ then gives
\begin{eqnarray} \label{app_eap}
 \frac{\cL_{\rm eff}}{\sqrt{-g}} &=& \frac{\cL_{\rm tr}}{\sqrt{-g}} - \frac12 \, J_{(1)} \left( \Omega - \cM^2 \right)^{-1} J_{(1)} - \frac{J_{(3)}}{3} \Bigl[ \left( \Omega - \cM^2 \right)^{-1} J_{(1)} \Bigr]^3
 + \cdots  \,,
\end{eqnarray}
where
\be \label{app_kerexp}
 \Delta_h^{-1} = (\Omega - \cM^2)^{-1} = -\frac{1}{\cM^2}\sum_{n=0}^\infty \Bigl(\Omega \, \frac{1}{\cM^2}\Bigr)^n \,,
\ee
and so
\be \label{appeq:leff_series}
  \mathcal{L}_{\rm eff} =
  \mathcal{L}_{\rm trunc} + \delta \mathcal{L}^E_{(1)} + \delta \mathcal{L}^E_{(2)} + \delta \mathcal{L}^E_{(3)} + \mathcal{O} \left(\frac{1}{\cM^{8}}\right) \,,
\ee
with
\be \label{app_dl1}
 \frac{\delta\mathcal L^E_{(1)}}{\sqrt{-g}} = \frac{1}{2 \cM^2} \left( V_h + \frac12 \, V_{h\e\e} \, \e^2 + \frac16 \, V_{\e\e\e h} \, \e^3 \right)^2 \,,
\ee
\be \label{app_dl2}
 \frac{\delta\mathcal L^E_{(2)}}{\sqrt{-g}} = \frac{1}{2 \cM^2} \left( V_h + \frac12 \, V_{h\e\e} \, \e^2 + \frac16 \, V_{\e\e\e h} \, \e^3 \right) \Omega \, \frac{1}{\cM^2} \left( V_h + \frac12 \, V_{h\e\e} \, \e^2 + \frac16 \, V_{\e\e\e h}\, \e^3 \right) \,,
\ee
and
\begin{eqnarray}
 \frac{\delta\mathcal L^E_{(3)}}{\sqrt{-g}} &=&  \frac{1}{2 \cM^2} \left( V_h + \frac12 \, V_{h\e\e} \, \e^2 + \frac16 \, V_{\e\e\e h} \, \e^3 \right) \Omega \, \frac{1}{\cM^2} \, \Omega \, \frac{1}{\cM^2} \left( V_h + \frac12 \, V_{h\e\e} \, \e^2 + \frac16 \, V_{\e\e\e h} \, \e^3 \right) \nn\\
 && \qquad + \frac{1}{3 \cM^6} \left( \frac12 \, V_{h h h} + \frac12 \, V_{\e h h h} \, \e \right) \left( V_h + \frac12 \, V_{h\e\e} \, \e^2 + \frac16 \, V_{\e\e\e h} \, \e^3 \right)^3 \,.
\end{eqnarray}
Here the superscript $E$ indicates that this is an expansion\footnote{Notice that this expansion, and the effective field theory to which it leads, would break down if the terms in $\cM^2$ were to cancel one another so that $\cM^2$ were small.} in powers of $\cM^{-1}$ (as opposed to our later expansions in inverse powers of $M$).

We next assume that the scale $M^2$ dominates all of the others in $\cM^2$, and gather terms that are suppressed by a fixed power of $1/M^2$. This leads to
\be \label{appeq:leff_series v2}
  \mathcal{L}_{\rm eff} =
  \mathcal{L}_{(0)} + \frac{ \mathcal{L}_{(1)}}{M^2} + \frac{ \mathcal{L}_{(2)}}{M^4} + \frac{ \mathcal{L}_{(3)}}{M^6} + \mathcal{O} \left(\frac{1}{M^{8}}\right) \,,
\ee
where
\be \label{app_Dl1}
 \frac{ \mathcal{L}_{(1)}}{M^2} = \frac{1}{2 M^2} \left( V_h + \frac12 \, V_{h\e\e} \, \e^2 + \frac16 \, V_{\e\e\e h} \, \e^3 \right)^2
\ee
\begin{eqnarray} \label{app_Dl2}
 \frac{ \mathcal{L}_{(2)}}{M^4} &=& - \frac{1}{2M^4} \left( V_h + \frac12 \, V_{h\e\e} \, \e^2 + \frac16 \, V_{\e\e\e h} \, \e^3 \right)^2 \left( V_{hh\e} \, \e + \frac12 \, V_{\e\e hh} \, \e^2 - \frac{1}{2 \rho^2} \, (\partial \e)^2 -\frac{1}{3 \rho^2} \, \e \, \square\e \right) \nn\\
  && \quad + \frac{1}{2M^4} \left( V_h + \frac12 \, V_{h\e\e} \, \e^2 + \frac16 \, V_{\e\e\e h} \, \e^3 \right) \left(1 - \frac{\e^2}{6 \rho^2} \right) \square \left( V_h + \frac12 \, V_{h\e\e} \, \e^2 + \frac16 \, V_{\e\e\e h} \, \e^3 \right) \nn\\
  &=& - \frac{1}{2M^4} \left( V_h + \frac{1}{2} \, V_{h\e\e} \, \e^2 + \frac{1}{6} \, V_{\e\e\e h} \, \e^3 \right)^2 \left( V_{hh\e} \, \e + \frac{1}{2} \, V_{\e\e hh} \, \e^2 \right) \\ && \qquad - \frac{\e^2}{2M^4} \, (\partial\e)^2 \left( V_{h\e\e} + \frac{1}{2} \, V_{\e\e\e h} \, \e \right)^2 + \frac{1}{12 \rho^2 M^4} \, (\partial\e)^2 \left( V_h - \frac{1}{2} \, V_{h\e\e} \, \e^2 - \frac{1}{3} \, V_{\e\e\e h} \, \e^3 \right)^2 \nn
\end{eqnarray}
and
\ba \label{app_Dl3} %\label{o3c}
 \frac{ \mathcal{L}_{(3)}}{M^6} &=&  \frac{1}{2M^6}\left(V_h + \frac{\e^2}{2}V_{\e\e h} \right)^2\left(V_{\e h h}\e + \frac{\e^2}{2}V_{\e\e h h}\right)^2 \nn\\
 &+& \frac{1}{6M^6}\left(V_{hhh} + V_{\e hhh}\e\right)\left(V_h + \frac{\e^2}{2}V_{\e\e h} + \frac{\e^3}{6}V_{\e\e\e h}\right)^3 \nn \\
 &+& \frac{V_h^2}{2M^6}\left(\frac{1}{4\rho^4}(\partial\e)^4 + \frac{1}{9\rho^4}\e^2\square\e\square\e + \frac{1}{3\rho^4}(\partial\e)^2\e\square\e\right) \nn\\
 &+& \frac{V_h^2}{2\rho^2 M^6} (\partial\e)^2 \left(\frac{\e}{3}V_{\e h h} + \frac{\e^2}{2}V_{\e\e h h}\right) + \frac{V_h}{2M^6}V_{\e hh}V_{\e\e\e h}\e^2(\partial\e)^2 \\
 &+& \frac{(\partial\e)^2}{M^6}\left(V_hV_{\e\e h} - \frac{V_h^2}{6\rho^2}\right)\left(V_{\e h h }\e + V_{\e\e h h}\e^2\right) \nn\\
 &+& \frac{1}{\rho^2M^6}\left(V_hV_{\e\e h} - \frac{V_h^2}{6\rho^2}\right)\left(\frac{(\partial\e)^4}{2} + \frac{\e^2}{3}(\square\e)^2 + \frac{5}{6}\e\square\e(\partial\e)^2\right) \nn\\
 &+& \frac{1}{2M^6}\left(V^2_{\e\e h} - \frac{V_hV_{\e\e h}}{3\rho^2}\right)\left((\partial\e)^4 + \e^2(\square\e)^2 + 2\e\square\e(\partial\e)^2\right) \,, \nn
\ea
and so on. Some algebra and integration by parts (checked numerically using {\em Mathematica}) is used above to obtain the expressions for $\cL_{(3)}$ and the second equality for $\cL_{(2)}$.

\subsection*{Expressions in terms of $U$, $\kappa$ and $\rho$}

The final step is to trade symmetrized derivatives like $V_\e$, $V_h$, $V_{\e\e\e}$, as well as the mass eigenvalues $M^2 = M^2_+$ and $\mu^2 = M^2_-$, for $U$, $m$, $\kappa$ and $\rho$ and their derivatives. This step is a crucial one because some of the interactions --- like $V_{\e \e h}$ in eq.~\pref{Vllh} or the quartic interactions $V_{\e\e hh}$ and $V_{\e\e\e\e}$ computed in appendix \ref{appssec:troughgeometry} --- contain terms proportional to a positive power of $m^2$, allowing them to contribute to higher order in the $1/m^2$ expansion than naively expected. The formulae relevant for performing this replacement are given in earlier sections and the Appendices, but are reproduced here for convenience of reference:
\ba
 && \qquad\qquad V_h \simeq \frac{\dot U^2}{\kappa m^2} \,,
 \qquad
 M^2 \simeq  m^2 + \frac{{\dot{U}^{2}}}{{\kappa^{2} m^{2}}} \,, \qquad
 \mu^2 \simeq  \ddot U - \frac{{\dot{U}^{2}}}{{\kappa^{2} m^{2}}} \,, \nn\\
 &&
 V_{h \e \e} \simeq  - \frac{m^{2}}{\kappa} + \frac{2\, \ddot{U}}{\kappa} - \frac{\dot{U}}{\kappa} \left( \frac{4 \, \dot m}{ m} + \frac{\dot \kappa}{\kappa}  \right) + \cO \left( \frac{1}{m^2} \right)\,, \qquad
 V_{\e\e\e} \simeq \dddot U + \frac{\dot U}{\kappa^2} + \cO \left( \frac{1}{m^2} \right) \nn\\
 &&\qquad\qquad V_{\e\e\e \e} \simeq   \frac{3m^{2}}{\kappa^{2}} +\ddddot{U} - \frac{8 \ddot{U}}{\kappa^{2}}
  + \frac{3\dot{U}}{\kappa^{2}} \left( \frac{8\, \dot{m}}{m} +  \frac{\dot{\kappa}}{\kappa}
   \right) + \cO \left( \frac{1}{m^2} \right)
\ea
and
%%% V_{\e\e\e h} \simeq \frac{3\dddot{U}}{\kappa} + \frac{2\dot{U}}{\kappa^{3}}
%%% - \frac{6m\dot{m}}{\kappa} -\frac{\dot{U}}{2\kappa \rho^{2}}
%%% - \frac{3\ddot{U}\dot{\kappa}}{\kappa^{2}} + \frac{m^{2}\dot{\kappa}}{\kappa^{2}}
%%% \nn \\
%%% && \qquad\qquad\qquad
%%% - \frac{\dot{U}\ddot{\kappa}}{\kappa^{2}} + \frac{2\dot{U}\dot{\kappa}^{2}}{\kappa^{3}}
%%% -\frac{5\dot{U}\dot{m}^{2}}{\kappa m^{2}} - \frac{5\dot{U}\ddot{m}}{\kappa m}
%%% + \frac{5V_{nnn}\dot{U}}{2 \kappa^{2} m^{2}} - \frac{V_{ttnn}^{\rm NS}\dot{U}}{2\kappa m^{2}}
%%% + \mathcal{O}\left( \frac{1}{m^{2}}\right) \,, \nn
\ba
  V_{\e \e \e h} & \simeq &
  -\frac{m^{2}}{\kappa} \left( \frac{6\, \dot{m}}{m} - \frac{\dot{\kappa}}{\kappa} \right)
  + \frac{3\dddot{U}}{\kappa} - \frac{3\ddot{U}\dot{\kappa}}{\kappa^{2}}
  \nn \\
  && \qquad
  - \frac{\dot{U}}{\kappa} \left[ \frac{6 \, \dot{m}^{2}}{m^{2}} + \frac{6 \, \ddot{m}}{m}
  -\frac{3(1+\lambda_{nnn})}{\kappa^{2}} - \frac{2\, \dot{\kappa}^{2}}{\kappa^{2}} + \frac{\ddot{\kappa}}{\kappa}\right] + \cO \left( \frac{1}{m^2} \right)\,.
\ea

Inserting these into the effective Lagrangian leads to the following, intermediate, form for the action out to four-derivative order
\be \label{eaint}
 \mathcal L_{\rm eff} = - V_{\rm eff}(\e) -\frac{1}{2} \, \hat G(\e) \, (\partial\e)^2  + H(\e) \, (\partial\e)^4 + K_1(\e) \, (\partial\e)^2 \square\e + K_2(\e) \, \square\e \, \square\e \,,
\ee
where
\be \label{appeq:geffp}
 \hat G(\e) \simeq 1 + \frac{\e^2}{\kappa^2} \left(1 + \frac{2 \dot U \dot\kappa}{\kappa m^2} - \frac{4 \ddot U}{m^2}+ \frac{8 \dot U \dot m}{m^3} \right) + \cO(\ell^3)\,,
\ee
and
\ba \label{appeq:veffp}
 V_{\rm eff}(\e) &=& U(\varphi + \e) + \frac{\e^3}{6} \left( \frac{\dot U}{\kappa ^2} \right) \left( 1 + \frac{2 \dot U \dot \kappa}{\kappa m^2} -  \frac{3 \ddot U}{m^2} + \frac{6 \dot U \dot m}{m^3} \right) \nn\\
 &+& \e^4\left( -\frac{ \dot U \dot \kappa }{8\kappa ^3}+\frac{ \ddot U}{6\kappa ^2} + \frac{2\dot U \ddot U \dot m}{\kappa^2 m^3} - \frac{2\dot U^2 \dot \kappa \dot m}{3\kappa^3 m^3} - \frac{2\dot U^2}{\kappa^2 m^2}\frac{\dot m^2}{m^2} \right) \\
 &+& \frac{\e^4}{24 \kappa^4 m^2} \Bigl[ -12 \kappa ^2 \ddot U^2+\dot U^2 \left(6+3 \lambda_{nnn}-11 \dot \kappa^2+4 \kappa  \ddot \kappa \right)+\dot U \left(20 \kappa  \dot\kappa  \ddot U-6 \kappa ^2 \dddot U\right)\Bigr] + \cO(\ell^5)\,, \nn
\ea
while
\be \label{appeq:Heffp}
 H(\e) = \frac{1}{2\kappa^2m^2} + \cO(\ell) \,.
\ee
In all of these expressions we keep only sufficient powers of the light field to track the action out to quartic order in $\ell$. The detailed form of the two functions $K_1$ and $K_2$ is less important for later purposes, but they are formally given by
\ba \label{k1}
 K_1(\e) &=& \frac{V_h^2}{6\rho^4 M^6}\e +\frac{5\e}{6\rho^2M^6}\left(V_hV_{\e\e h} - \frac{V_h^2}{6\rho^2}\right) + \frac{\e}{M^6}\left(V_{\e\e h}^2 - \frac{V_h V_{\e\e h}}{3\rho^2}\right) \\
 \label{k2}
 K_2(\e) &=& \frac{V_h^2}{18\rho^4 M^6}\e^2 +\frac{\e^2}{3\rho^2M^6}\left(V_hV_{\e\e h} - \frac{V_h^2}{6\rho^2}\right) + \frac{\e^2}{2M^6}\left(V_{\e\e h}^2 - \frac{V_h V_{\e\e h}}{3\rho^2}\right) \,.
\ea

The action quoted above is only `intermediate' because the terms involving $\Box \ell$ can be absorbed into the others by making the field redefinition \cite{EFTrevs}
\be \label{fr}
 \e \to \e + \Delta(\e) \,,
\ee
which changes the action by a term
\ba \label{ac2}
 \Delta\mathcal L_{\rm eff} &=& \left( \hat G \square\e - V_{\rm eff}'  + \frac{\hat G^{'}}{2}(\partial\e)^2\right) \Delta(\e) \\
 && \qquad + \frac{1}{2}\Delta(\e)\left(\frac{1}{2} \, \hat G^{''}(\partial\e)^2 + \hat G^{'} \square\e + \hat G^{'} \partial_\mu\e \partial^\mu + \hat G \square - V_{\rm eff}''\right) \Delta(\e) \,, \nn\\
 &\simeq& \left( \hat G \, \square\e - V_{\rm eff}' \right) \Delta(\e) - \frac{1}{2} \, \Delta(\e)^2 V_{\rm eff}'' \,,
\ea
where primes denote differentiation with respect to $\e$, and the approximate equality assumes $\Delta(\e)$ is at least quadratic in $\e$ and drops terms in $\Delta \cL$ that involve more than four derivatives or four powers of $\ell$.

The choice
\be \label{fredp}
 \Delta(\e) = - \left( \frac{K_1 (\partial\e)^2 + K_2\square\e }{\hat G} \right) \,,
\ee
implies $\Delta(\e)^2$ is at least sixth order in $\e$ (and so for our purposes can be neglected) and leads to
\be \label{eaint2}
 \mathcal L_{\rm eff} = - V_{\rm eff}(\e) -\frac{1}{2} \, \hat G(\e) (\partial\e)^2 + H(\e)(\partial\e)^4 + K_1(\e) \, (\partial\e)^2 \left( \frac{V_{\rm eff}'}{\hat G} \right) + K_2(\e) \, \square\e \left( \frac{V_{\rm eff}'}{\hat G} \right),
\ee
which is the same as making the replacement
\be \label{osr}
 \square\e \to \frac{V_{\rm eff}'}{\hat G} - \frac{\hat G^{'}}{2 \hat G} \, (\partial\e)^2
\ee
which amounts to eliminating $\Box \ell$ using its equation of motion. Since eq.~(\ref{appeq:geffp}) implies the second term in (\ref{osr}) is at least cubic in $\e$, it can be dropped to the extent that we follow only terms out to order $\ell^4$ in $\cL_{\rm eff}$. Similarly, eqs.~(\ref{eaint}), (\ref{k1}) and (\ref{k2}) ensure we need only evaluate $\square\e$ to at most quadratic order in $\e$. Doing so gives
\be \label{blsol}
 \square\e = \frac{V_{\rm eff}'}{\hat G}= \dot U + \ell \ddot U + \ell^2 \left[ - \frac{\dot U}{2 \kappa ^2} - \frac{5 \dot m \dot U^2}{m^3 \kappa ^2} + \frac{\dot U }{m^2 \kappa ^2}\left(-\frac{\dot U \dot \kappa}{\kappa }+\frac{5 \dot U}{2}\right) +\frac{\dddot U}{2}\right] \,,
\ee
and so after integrating the last term in (\ref{eaint2}) by parts, one arrives at the effective action
\be \label{eaintf}
 \mathcal L_{\rm eff} =  - V_{\rm eff}(\e) -\frac{1}{2} \, G(\e)(\partial\e)^2 + H(\e)(\partial\e)^4 \,,
\ee
with
\ba \label{app:Gf}
 G(\e) &=& \hat G  - \frac{2 V_{\rm eff}' K_1}{\hat G} + 2\left(\frac{K_2 V_{\rm eff}'}{\hat G}\right)' \nn\\
 &\simeq& 1 + \frac{\e^2}{\kappa^2}\left(1 + \frac{2\dot U\dot\kappa}{\kappa m^2} - \frac{3\ddot U}{m^2} + \frac{8\dot U \dot m}{m^3} \right) \,,
\ea
and $V_{\rm eff}$ and $H$ given by their earlier expressions, eqs.~(\ref{appeq:veffp}) and (\ref{appeq:Heffp}) respectively. This is the expression used in the main text.

\end{document}